\documentclass[11pt,a4paper]{article}
\pdfoutput=1
\usepackage{jheppub}
\usepackage{graphicx,psfrag}
\usepackage{bm}
\usepackage{mathbbol}
\usepackage{slashed}
\usepackage{lscape}
\usepackage{graphics}
\usepackage{xcolor,ulem}
\usepackage{array,longtable}
\usepackage{multirow}
\usepackage{makecell,multirow}
\allowdisplaybreaks

\graphicspath{{./figures/}}

\newcolumntype{C}{>{$}c<{$}}  

\definecolor{greeen}{rgb}{0.03,0.84,0.13}
\definecolor{test}{rgb}{0.03,0.74,0.33}
\definecolor{viol}{rgb}{0.44,0,0.94}
\definecolor{or}{rgb}{0.95,0.65,0}

\newcommand{\specialcell}[2][c]{%
  \begin{tabular}[#1]{@{}c@{}}#2\end{tabular}}

\title{Leptonic Scalars and Collider Signatures in a UV-complete Model}

\author[a]{P. S. Bhupal Dev,}
\author[b]{Bhaskar Dutta,}
\author[c,d,e]{Tathagata Ghosh,}
\author[e]{Tao Han,}
\author[e]{Han Qin,}
\author[f]{Yongchao Zhang}

\affiliation[a]{Department of Physics and McDonnell Center for the Space Sciences,  Washington University, St. Louis, MO 63130, USA}
\affiliation[b]{Mitchell Institute for Fundamental Physics and Astronomy, Department of Physics and Astronomy, Texas A\&M University, College Station, TX 77843, USA}
\affiliation[c]{Regional Centre for Accelerator-based Particle Physics, Harish-Chandra Research Institute, HBNI, Chhatnag Road, Jhunsi,
Prayagraj (Allahabad) 211019,
India }
\affiliation[d]{Instituto de F\'{i}sica, Universidade de S\~{a}o Paulo, R. do Matao 1371, S\~{a}o Paulo, SP 05508-090, Brazil}
\affiliation[e]{PITT PACC, Department of Physics and Astronomy, University of Pittsburgh,
3941 O'Hara St., Pittsburgh, PA 15260, USA}
\affiliation[f]{School of Physics, Southeast University, Nanjing 211189, China}

\emailAdd{bdev@wustl.edu, dutta@tamu.edu, tathagataghosh@hri.res.in,  than@pitt.edu, han.qin@pitt.edu, zhangyongchao@seu.edu.cn}
\preprint{PITT-PACC-2108, MI-TH-2112, HRI-RECAPP-2021-007}

\date{\today}

\begin{document}

\abstract{We study the non-standard interactions of neutrinos with light leptonic scalars ($\phi$) in a global $(B-L)$-conserved ultraviolet (UV)-complete model. The model utilizes Type-II seesaw motivated neutrino interactions with an $SU(2)_L$-triplet scalar, along with an additional singlet in the scalar sector.
This UV-completion
leads to an enriched spectrum and consequently new observable signatures. We examine the low-energy lepton flavor violation constraints, as well as the perturbativity and unitarity constraints on the model parameters. Then we lay out a search strategy for the unique signature of the model resulting from the leptonic scalars at the hadron colliders via the processes $H^{\pm\pm} \to W^\pm W^\pm \phi$ and $H^\pm \to W^\pm \phi$ for both small and large leptonic Yukawa coupling cases.
We find that via these associated production processes at the HL-LHC, the prospects of doubly-charged scalar $H^{\pm\pm}$ can reach up to 800 (500) GeV and 1.1 (0.8) TeV at the  $2\sigma \ (5\sigma)$ significance for small and large Yukawa couplings, respectively.
A future 100 TeV hadron collider will further increase  the mass reaches up to 3.8 (2.6) TeV and 4 (2.7) TeV, at the  $2\sigma \ (5\sigma)$  significance, respectively.
We also demonstrate that the mass of $\phi$
can be determined at about 10\% accuracy at the LHC for the large Yukawa coupling case even though it escapes as missing energy from the detectors. }

\maketitle

\section{Introduction}
Explanation of tiny but non-zero masses of neutrinos, as confirmed in various experiments over the past two decades~\cite{Zyla:2020zbs}, requires new physics beyond the Standard Model (SM).
In addition to the origin of their masses and mixing, neutrinos pose many more unanswered questions. 
For example, we still do not know whether the neutrino masses are of Dirac-type or Majorana-type; see Ref.~\cite{Bilenky:2020wjn} for a recent review. We would also like to understand whether the neutrino sector contains  new interactions beyond those allowed by the SM gauge structure, {\it i.e.}~the so-called non-standard interactions (NSIs); see Ref.~\cite{Dev:2019qno} for a recent status report.
Just like neutrinos, the origin of dark matter (DM) is also a puzzle and it is conceivable that these two puzzles could be somehow correlated at a fundamental level
\cite{Lattanzi:2014mia}.
We also wonder whether the leptonic sector breaks CP-symmetry and whether it is responsible for the observed matter-antimatter asymmetry in the Universe~\cite{Hagedorn:2017wjy}. In order to address these outstanding puzzles, construction of  neutrino models and investigation of their predictions at  various  experiments are highly motivated.


If neutrinos are Majorana particles, lepton number $L$, which is an accidental global symmetry of the SM Lagrangian, must be broken either at tree-level or loop-level. On the other hand, if neutrinos are Dirac particles, lepton number (or some non-anomalous symmetry that contains $L$, such as $B-L$) remains a good symmetry of the Lagrangian. We will focus on this latter case, assuming that $B-L$ is conserved even in presence of higher-dimensional operators. Thus, any new, additional degrees of freedom must be charged appropriately under global $B-L$~\cite{Berryman:2018ogk}.
In a recent paper \cite{deGouvea:2019qaz}, motivated by certain observational considerations at the LHC and beyond,
we considered the possibility that Dirac neutrinos could exhibit NSIs with a new (light) scalar field $\phi$ which has a $B-L$ charge of $+2$ but is a singlet under the SM gauge group. These were dubbed as ``leptonic scalars", which can only couple to right-handed neutrinos $(\nu_R)$ (or left-handed anti-neutrinos) like $\nu_R^TC\nu_R\phi$ at the renormalizable level. Then the question arises as to how these leptonic scalars couple to the SM fields. At the dimension-6 level, we can write an effective coupling of the form 
\begin{eqnarray}
\label{eqn:operator}
\frac{1}{\Lambda^2} (LH)(LH)\phi \,,
\end{eqnarray}
where $L$ and $H$ are the SM lepton and Higgs doublets, respectively, and $\Lambda$ is the new physics scale. After electroweak (EW) symmetry breaking, the operator~\eqref{eqn:operator} yields flavor-dependent NSIs of neutrinos with the leptonic scalar of the form $\lambda_{\alpha \beta} \phi \nu_\alpha \nu_\beta$. Furthermore, at energy scales below the mass of $\phi$, this leads to an effective non-standard neutrino self-interaction, which could have observable cosmological consequences~\cite{Kreisch:2019yzn, Blinov:2019gcj,DeGouvea:2019wpf, Lyu:2020lps,Kelly:2020aks, Das:2020xke}.

Our goal in this paper is to find an ultraviolet (UV)-completion of the operator~\eqref{eqn:operator} and to test the model at the ongoing LHC and future 100 TeV colliders, such as the Future Circular Collider (FCC-hh) at CERN~\cite{FCC:2018vvp} and the Super Proton-Proton Collider (SPPC) in China~\cite{Tang:2015qga}.
To be concrete, we adopt a 
Type-II seesaw motivated neutrino mass model~\cite{Konetschny:1977bn, Magg:1980ut, Schechter:1980gr, Cheng:1980qt, Mohapatra:1980yp, Lazarides:1980nt}, which can also account for the baryon asymmetry in the Universe~\cite{Barrie:2021mwi}.
In our model the neutral component of the triplet scalar field $\Delta$ does not acquire a vacuum expectation value (VEV),  which keeps the custodial symmetry intact. The lepton number is not broken and the neutrinos are Dirac-type in this model. We also add a SM-singlet complex scalar field $\Phi$, which gives rise to the leptonic scalar $\phi$ in the model.
Beyond the NSIs between the active neutrinos and the  leptonic scalar, the particle spectrum
and new interactions in this model  lead to rich phenomenology and consequently new observable signatures. In some other UV-complete models, the effective interactions of $\phi$ with the SM neutrinos stemming from Eq.~(\ref{eqn:operator}) might also be relevant to DM phenomenology~\cite{Pospelov:2007mp, Kelly:2019wow, DeGouvea:2019wpf, Du:2020avz,Kelly:2020aks}.

In this paper we will show that the distinguishing features of the signatures of our UV-complete model compared to the standard Type-II seesaw model is due to the new sources of missing energy carried away by $\phi$, which would help the model to be detected at the ongoing LHC and future higher-energy colliders.
After taking into account the current limits from the low-energy lepton flavor violating (LFV) constraints (cf. Table~\ref{tab:limits} and Fig.~\ref{fig:limits}) and the theoretical limits from perturbativity and unitarity (see Fig.~\ref{fig:pert}), we consider three scenarios with respectively small, large and intermediate Yukawa couplings of the leptonic scalar $\phi$. In all these scenarios, $\phi$ can be produced either from the doubly-charged scalar $H^{\pm\pm} \to W^\pm W^\pm \phi$ or from the singly-charged scalar $H^\pm \to W^\pm \phi$ $-$ channels which are unique and absent in the standard Type-II seesaw. As the leptonic scalar $\phi$ decays exclusively into neutrinos, these new channels will lead to same-sign dilepton plus  missing transverse energy plus jets signal at the hadron colliders.
Detailed cut-based analysis  is carried out for both scenarios, and the technique of Boosted Decision Tree (BDT)~\cite{Roe:2004na} is also utilized to improve the observational significance (see Tables~\ref{tab:events} and \ref{tab:ap_events}). We find that the mass of doubly-charged scalars in the small and large Yukawa coupling scenarios can be probed up to respectively 800 GeV and 1.1 TeV at the $2\sigma$ significance, corresponding to a $95\%$ confidence level, in the new channels at the high-luminosity LHC (HL-LHC) with integrated luminosity of 3 ab$^{-1}$, and can be improved up to 3.8 TeV and 4 TeV respectively at future 100 TeV colliders with luminosity of 30 ab$^{-1}$.
This can be further improved in the intermediate Yukawa coupling case, with the help of increasing leptonic decay channel of the doubly-charged scalar.
We also show that since in the large Yukawa coupling case, the missing energy is completely from the leptonic scalar in the associate production channel $pp \to H^{\pm\pm} H^\mp$, its mass can be determined with an accuracy of about $10\%$ at the HL-LHC.


The rest of the paper is organized as follows.
In Section \ref{sec:Model}, we present the model details and lay out relevant experimental and theoretical constraints, including the key parameters and resultant main decay channels of $H^{\pm\pm}$ and $H^\pm$ in Section~\ref{sec:interactions}, the current LFV constraints on $H^{\pm\pm}$ in Section~\ref{sec:Limits}, and the high-energy limits from perturbativity and unitarity in Section~\ref{sec:Limits2}.
In Section \ref{sec:LHC}, we discuss our search strategy at the LHC and future 100 TeV hadron colliders, presenting the small Yukawa coupling case in Section~\ref{sec:small}, large Yukawa coupling scenario in Section~\ref{sec:large}, and the intermediate Yukawa coupling case in Section~\ref{sec:medium-yukawa}. We show the discovery potential by utilizing the cut-based analysis and the BDT techniques, and obtain the prospect for determining the mass of $\phi$ in the large Yukawa coupling case even though the scalar $\phi$ escapes from the detectors as missing energy.
The main results are summarized in Section~\ref{sec:Sum}. For the sake of completeness, the complete set of Feynman rules for the model are listed in  Appendix~\ref{sec:AppA}. The functions $G$ and  ${\cal F}$ for some three-body decays are given in Appendix~\ref{sec:F}. The renormalization group equations (RGEs) for the couplings are detailed in Appendix~\ref{sec:AppB}. The perturbativity limits
are analytically derived in Appendix~\ref{sec:AppC}, and the unitarity limits are described in Appendix~\ref{sec:AppD}.

\section{The model}
\label{sec:Model}

In this section, we present a global $(B-L)$-conserved UV-complete model of a leptonic scalar, which is motivated by the well-known Type-{II} seesaw model~\cite{Konetschny:1977bn, Magg:1980ut, Schechter:1980gr, Cheng:1980qt, Mohapatra:1980yp, Lazarides:1980nt}.
The enlarged particle content of the model includes a leptonic complex scalar $\Phi$, which is a singlet under the SM gauge groups and carries a $B-L$ charge of $+2$. The model contains also an $SU(2)_L$ triplet scalar $\Delta$ with hypercharge +1 and $B-L$ charge +2:
\begin{equation}
\Delta = \begin{pmatrix}
\frac{1}{\sqrt{2}} \delta^+ & \delta^{++} \\
\delta^0 & -\frac{1}{\sqrt{2}} \delta^+
\end{pmatrix}
\end{equation}
and three SM-singlet $B-L=-1$ right-handed neutrino fields $\nu_{R_i} \,(i=1,2,3)$. 


The allowed Yukawa interactions in the model are given by
\begin{equation}
-\mathcal{L}_Y = y_{\nu,\,\alpha\beta} \overline{L}_\alpha H \nu_{R_\beta} + Y_{\alpha\beta} L_\alpha^{\sf T} C i \sigma_2 \Delta L_\beta + \tilde{y}_{\nu,\,\alpha\beta} \nu_{R_\alpha}^{\sf T} C \nu_{R_\beta} \Phi  + {\rm H.c.} \, ,
\label{eq:LYuk}
\end{equation}
where $\alpha,\; \beta = e,\; \mu,\; \tau$ are the lepton flavor indices, $C$ is the charge-conjugation operator, $\sigma_2$ is the second Pauli matrix, $y_{\nu}$ are the SM-like Yukawa couplings of the neutrinos, $Y_{\alpha \beta}$ are the new leptonic Yukawa couplings of the triplet that govern the heavy scalar phenomenology, and $\tilde y_{\nu}$ are the Yukawa couplings of the leptonic scalar $\Phi$ to the right-handed neutrinos.
In a $(B-L)$-conserved theory where $\Delta$ and $\Phi$ do not acquire any VEV, neutrinos are Dirac fermions and non-zero neutrino masses can be generated after the EW symmetry breaking from the first term of the Yukawa Lagrangian given in Eq.~(\ref{eq:LYuk}), just like the other fermions in the SM. However, one requires $y_{\nu} \lesssim 10^{-12}$ in order to satisfy the absolute neutrino mass constraints~\cite{Planck:2018vyg, KATRIN:2019yun}. 

The kinetic and potential terms of the scalar sector are given by
\begin{eqnarray}
{\cal{L}}_{\rm Scalar}  =
(D_\mu H)^{\dagger}(D^\mu H)+{\rm Tr}[(D_\mu \Delta)^{\dagger}(D^\mu \Delta)]+(\partial_\mu \Phi)^\dagger(\partial^\mu \Phi)-V(H,\Delta,\Phi),
\end{eqnarray}
where the covariant derivatives are given by
\begin{align}
\label{eqn:DH}
D_\mu H &= \partial_\mu H- i\frac{g_L}{2}W^a_\mu \sigma_a H-i\frac{g_Y}{2}B_\mu H \,,     \\
\label{eqn:DDelta}
D_\mu \Delta &=\partial_\mu \Delta - i\frac{g_L}{2}[W^a_\mu \sigma_a,\Delta] - i g_Y B_\mu \Delta \,,
\end{align}
{with $g_L$ and $g_Y$ respectively the gauge couplings for the SM gauge groups $SU(2)_L$ and $U(1)_Y$,
and $\sigma_a$ ($a = 1,\, 2,\, 3$) the Pauli matrices. } The most general renormalizable potential involving the scalar fields of the model is given by
\begin{eqnarray}
\label{V_Delta_phi}
V(H,\Delta,\Phi) =
& & - m^2_H + \frac{\lambda}{4}(H^\dagger H)^2 + M^2_\Delta {\rm Tr}(\Delta^\dagger \Delta) + M^2_\Phi \Phi^\dagger \Phi \nonumber \\
&& + \lambda_1(H^\dagger H){\rm Tr}(\Delta^\dagger \Delta) + \lambda_2 [{\rm Tr}(\Delta^\dagger\Delta)]^2 + \lambda_3 {\rm Tr}[(\Delta^\dagger \Delta)^2] + \lambda_4 (H^\dagger \Delta) (\Delta^\dagger H) \nonumber \\
& & + \lambda_5(\Phi^\dagger \Phi)^2 + \lambda_6 (\Phi^\dagger \Phi)(H^\dagger H) + \lambda_7(\Phi^\dagger \Phi){\rm Tr}(\Delta^\dagger \Delta)
\nonumber \\
& &
+ \lambda_8(i\Phi H^{\sf T} \sigma_2 \Delta^\dagger H + {\rm H.c.}) \,,
\end{eqnarray}
where all the mass parameters $m_H^2$, $M_{\Delta}^2$, $M_\Phi^2$ and the quartic couplings $\lambda$ and $\lambda_i$ are assumed to be real.
The scalar $\Delta$ in our model carries the same $SU(3)_C\times SU(2)_L\times U(1)_Y$ charges ({\bf 1},{\bf 3},1) as in the Type-II seesaw model. However, the presence of a $(B-L)$-charged $\Phi$ and the $B-L$ conservation in our model have important phenomenological consequences associated with the triplet $\Delta$, which is different from that in the Type-II seesaw scenario. In the Type-II seesaw model, the EW symmetry breaking induces a non-vanishing VEV for the triplet $\Delta$ via the cubic term $H^{\sf T} i\sigma_2  \Delta^\dagger H$. However, due to the $B-L$ conservation  such a cubic term does not exist in our model, and as a result the triplet $\Delta$ does not develop a VEV in our model. As we will see in Section~\ref{sec:LHC}, this leads to very interesting
signatures at the LHC and future 100 TeV colliders, which are key to distinguish our model from the Type-II seesaw.

After the EW symmetry breaking, the Higgs doublet $H$ develops a VEV $v = (\sqrt2 G_F)^{-1/2}$ with $G_F$ being the Fermi constant, and the mass matrix of the  CP-even neutral components in the $\{h, \delta^{0r}, \Phi^r\}$ basis (here $X^r$ refers to the real component of the field $X$) is
\begin{align}
\label{eqn:matrix}
M^2_{\rm CP-even}=\begin{pmatrix}
\frac{1}{2} \lambda v^2 & 0 & 0\\
0 & M^2_{\Delta}+\frac{1}{2} (\lambda_1+\lambda_4) v^2  & -\frac{1}{2} \lambda_8 v^2 \\
0 & -\frac{1}{2} \lambda_8 v^2  & M^2_{\Phi} + \frac{1}{2} \lambda_6 v^2
\end{pmatrix} \, .
\end{align}
As the singlet and triplet scalars do not have VEVs, the component $h$ from the SM doublet $H$ does not mix with other neutral scalars, as can be seen from Eq.~(\ref{eqn:matrix}). Then $h$ can be readily identified as the 125 GeV Higgs boson observed at the LHC~\cite{ATLAS:2012yve,CMS:2012qbp}, and the quartic coupling $\lambda$ can be identified as the SM quartic coupling.
The two remaining physical CP-even scalar eigenstates are from mixing of the components $\Phi^r$ and $\delta^{0r}$ of the leptonic fields $\Phi$ and $\Delta$ with $B-L$ charge of $+2$, and thus are both physical leptonic scalars. Denoting $H_1$ as the lighter one and $H_2$ as the heavier one, they
can be obtained by the following rotation
\begin{align}
\label{eqn:mixing}
\begin{pmatrix}
H_1 \\ H_2
\end{pmatrix} = \begin{pmatrix} \cos \theta & -\sin \theta \\  \sin \theta & \cos \theta \end{pmatrix} \begin{pmatrix}
\Phi^{r} \\ \delta^{0r}
\end{pmatrix} \, ,
\end{align}
where the mixing angle $\theta$ is given by
\begin{equation}
\tan 2\theta = \frac{\lambda_8 v^2}{M^2_{\Delta}+v^2 (\lambda_1+\lambda_4 -\lambda_6)/2 - M^2_\Phi} \,,
\label{theta}\end{equation}
and the two eigenvalue masses are
\begin{align}
M^2_{H_{1,\,2}} = \ & \frac{1}{2} \left( M^2_\Delta + M^2_\Phi \right) + \frac{1}{4} (\lambda_1 + \lambda_4 +\lambda_6) v^2 \nonumber \\
& \mp \frac{1}{4} \sqrt{\left[ 2M^2_\Delta - 2M^2_\Phi + (\lambda_1+\lambda_4 -\lambda_6) v^2 \right]^2 + 4 \lambda_8^2 v^4_H} \,. 
\end{align}
Similarly, the two CP-odd leptonic scalars ($A_1, A_2$) from the imaginary components  $\Phi^i,\delta^{0i}$ have exactly the same masses as the CP-even scalars, {\it i.e.}
\begin{eqnarray}
M_{A_1} = M_{H_1} \,, \quad M_{A_2} = M_{H_2} \,.
\end{eqnarray}
For the sake of illustration, we choose to work in the regime where the leptonic scalars ($A_1, H_1$) are in the mass range $M_h/2<M_{H_1, A_1}\lesssim {\cal O}(100)$ GeV. The lower mass bound  is to avoid the invisible decay of the SM Higgs $h \to H_1 H_1,\; A_1 A_1 \to \nu \nu \bar{\nu} \bar{\nu}$,
while the upper bound is mainly motivated from our previous collider study~\cite{deGouvea:2019qaz}, where the sensitivity in the vector boson fusion (VBF) channel was found to drop exponentially beyond 100 GeV or so.
In order to keep the two leptonic scalars ($A_1, H_1$) light, we choose the simplest scenario $\lambda_6 = 0$. There is also a pair of heavy leptonic scalars $H_2$ and $A_2$, which can either decay into neutrinos or cascade decay into gauge bosons and lighter scalars. For simplicity,  we just assume $(H_2, A_2)$ to be heavier than the EW scale such that they are not relevant for our consideration here, and a detailed collider study of their phenomenology is deferred to future work.
Finally, it is trivial to get the masses of the singly- and doubly-charged scalars, which are respectively given  by
\begin{align}
M^2_{H^{\pm}} &= M^2_\Delta + \frac14 (2\lambda_1 + \lambda_4) v^2 \,,  \\
M^2_{H^{\pm\pm}} &= M^2_\Delta + \frac12 \lambda_1 v^2 \,.
\end{align}
Depending on the sign of $\lambda_4$, $H^\pm$ can be lighter or heavier than  $H^{\pm\pm}$.

\subsection{Key parameters and decay channels of $H^{\pm\pm}$ and $H^\pm$}
\label{sec:interactions}

\begin{table}[!t]
\centering
\caption{Important couplings for the neutral scalars $H_1$, $A_1$, the singly-charged scalar $H^\pm$ and the doubly-charged scalar $H^{\pm\pm}$. Here $e$ is the electric charge, $s_W \equiv \sin\theta_W$ and $c_W \equiv \cos\theta_W$ the sine and cosine of the Weinberg angle $\theta_W$,  $p_{1,\,2}$ the momenta for the first and second particles in the vertices, and $P_L = \frac{1}{2}(1-\gamma_5)$ the left-handed projection operator. See Appendix.~\ref{sec:AppA} for the full set of Feynman rules.}
\label{keyvertices1}
\vspace{5pt}
\begin{tabular}{lc}
\hline\hline
Vertices & Couplings \\ \hline
$H_1 \nu_\alpha \nu_\beta$ & $ -i\, \sqrt{2} Y_{\alpha\beta} \sin \theta \, P_L$ \\
$A_1 \nu_\alpha \nu_\beta$ & $  \sqrt{2} Y_{\alpha\beta} \sin \theta \, P_L$ \\ \hline
$H^+ H^- \gamma_\mu$ & $i\, e (p_1 - p_2)_\mu$ \\
$H^+ H^- Z_\mu$ & $-i\, e \dfrac{s_W}{c_W} (p_1 - p_2)_\mu$ \\
$H^{+} \ell^-_\alpha \nu_\beta$ & $\sqrt{2} i \, Y_{\alpha\beta} \, P_L$ \\
$H^{+} H_1 W^-_\mu$  &  $i\dfrac{g_L}{\sqrt{2}} \, (p_1 -p_2)_\mu \, \sin\theta$ \\
$H^{+} A_1 W^-_\mu$  &  $\dfrac{g_L}{\sqrt{2}} \, (p_1 -p_2)_\mu \, \sin \theta$ \\ \hline
$H^{++} H^{--} \gamma_\mu$ &  $2i\, e (p_1 -p_2)_\mu$  \\ 
$H^{++} H^{--} Z_\mu$ & $i\, e \dfrac{c^2_W - s^2_W}{c_W s_W} (p_1 -p_2)_\mu$ \\ 
$H^{++} \ell^-_\alpha \ell^-_\beta$ & $2 i \, Y_{\alpha\beta} \, P_L$ \\ 
$H^{++} H^- W^-_\mu$ & $-i g_L \, (p_1-p_2)_\mu$ \\ 
$H^{++}W^-_\mu W^-_\nu H_1$ & $-i\sqrt{2}g_L^2 \, \sin\theta  \, g_{\mu\nu}$ \\ 
$H^{++}W^-_\mu W^-_\nu A_1$ & $-\sqrt{2}g_L^2 \, \sin\theta  \, g_{\mu\nu}$ \\
\hline\hline
\end{tabular}
\end{table}


The interactions of the new scalars with the SM fields are generated through the gauge couplings in Eqs.~(\ref{eqn:DH}) and (\ref{eqn:DDelta}), the scalar couplings in Eq.~(\ref{V_Delta_phi}) and  the Yukawa interactions in Eq.~(\ref{eq:LYuk}) including potential scalar mixing in Eq.~(\ref{eqn:mixing}). All the key interactions of the neutral scalars $H_1$, $A_1$, the singly-charged scalar $H^\pm$ and the doubly-charged scalar $H^{\pm\pm}$ for the hadron collider analysis below are collected in Table~\ref{keyvertices1}.
For the sake of completeness, we have listed the complete set of Feynman rules in Tables~\ref{table:couplings:trilinear} to \ref{table:couplings:Yukawa} in  Appendix~\ref{sec:AppA}.


The gauge interactions of $H^\pm$ and $H^{\pm\pm}$ with the SM photon, $W$ and $Z$ bosons in Table~\ref{keyvertices1} are relevant for the pair production $H^{++}H^{--}$ and the associated production $H^{\pm\pm}H^{\mp}$ of the doubly-charged scalar at hadron colliders, as in the Type-II seesaw case~\cite{Chakrabarti:1998qy, Chun:2003ej, Akeroyd:2005gt,  FileviezPerez:2008jbu, delAguila:2008cj,Akeroyd:2011zza, Melfo:2011nx, Aoki:2011pz,  Chiang:2012dk, Han:2015hba, Babu:2016rcr, Ghosh:2017pxl, Dev:2018kpa, Du:2018eaw, Antusch:2018svb, Primulando:2019evb, deMelo:2019asm, Padhan:2019jlc, Ashanujjaman:2021txz}.   
The remaining couplings in Table~\ref{keyvertices1} are relevant to the decays of $H^\pm$ and $H^{\pm\pm}$. For the singly-charged scalar $H^\pm$, besides the leptonic final states, it can decay into a light neutral scalar $H_1$ or $A_1$ and a $W$ boson, which is absent in the Type-II seesaw model. The corresponding partial decay widths are respectively
\begin{align}
\Gamma(H^\pm \rightarrow \ell^\pm_\alpha v_\beta) & \ = \ \frac{ Y^2_{\alpha\beta} M_{H^\pm}}{8\pi} \,, \\
\Gamma(H^\pm \rightarrow W^\pm H_1) & \ = \ \Gamma(H^\pm \rightarrow W^\pm A_1)
\ = \ \frac{G_F \sin^2\theta M^3_{H^\pm}}{4\sqrt2 \pi} \lambda^{3/2} \left( \frac{M^2_W}{M^2_{H^\pm}}, \frac{M^2_{H_1}}{M^2_{H^\pm}} \right) \,,
\end{align}
where the function
\begin{eqnarray}
\lambda(x,y)  \ \equiv \ 1+x^2+y^2-2xy-2x-2y \,.
\end{eqnarray}

As in the standard Type-II seesaw, the singly-charged scalar $H^\pm$ can decay into a heavy scalar $H_2$ or $A_2$ and a $W$ boson. However, the mass splitting between the triplet scalar components is severely constrained by the EW precision data (EWPT), in terms of the oblique $S$ and $T$ parameters~\cite{Peskin:1990zt, Peskin:1991sw}: depending on the triplet scalar masses, it is required that the mass splitting $\Delta M \lesssim 50$ GeV~\cite{Melfo:2011nx, Kanemura:2012rs, Chun:2012jw, Primulando:2019evb}. Therefore the $W$ boson is always off-shell, {\it i.e.} $H^{\pm} \to W^{\pm \ast} H_2,\; W^{\pm \ast} A_2$ (the corresponding interaction can be found in Table~\ref{table:couplings:trilinear2}), and the corresponding widths are given by
\begin{eqnarray}
\label{eqn:Width:HpWH2}
\Gamma(H^\pm \rightarrow W^{\pm *} H_2) \ = \
\Gamma(H^\pm \rightarrow W^{\pm *} A_2) \ = \
\frac{9 g_L^4 \cos^2\theta M_{H^\pm}}{256 \pi^3} G\left(\frac{M^2_{H_2}}{M^2_{H^\pm}},\frac{M^2_W}{M^2_{H^\pm}}\right) ,~
\end{eqnarray}
where the function $G(x,y)$ is explicitly given in Appendix~\ref{sec:F}.
This channel is highly suppressed by the off-shell $W^\ast$ boson, and will be neglected in the following sections.

In our model, the doubly-charged scalar $H^{\pm\pm}$ can decay into same-sign dilepton pairs 
and the three-body final state $W^\pm W^\pm H_1$ and $W^\pm W^\pm A_1$. The partial widths are given respectively by
\begin{align}
\Gamma(H^{\pm\pm} \rightarrow \ell^\pm_\alpha \ell^\pm_\beta) & \ = \ \frac{S_{\alpha\beta} Y^2_{\alpha\beta} M_{H^{\pm\pm}}}{4\pi} \,,\\
\label{eqn:decayWWphi}
\Gamma(H^{\pm\pm} \rightarrow W^\pm W^\pm H_1) & \ = \
\Gamma(H^{\pm\pm} \rightarrow W^\pm W^\pm A_1)
= \frac{g_L^4\sin^2\theta}{512 \pi^3 M_{H^{\pm\pm}}^3} \int {\cal F} dm^2_{12} dm^2_{23} \,,
\end{align}
where $S_{\alpha\beta} = 1/2$ (1) for $\alpha \neq \beta$ ($\alpha = \beta$) is a symmetry factor, and the dimensionless lengthy function ${\cal F}$ is put in Appendix~\ref{sec:F}, which is a function of $m_{12}^2$ and $m_{23}^2$.
The phase space is integrated over the Dalitz plot where the ranges for $m^2_{12}$ and $m^2_{23}$ are respectively $[4M^2_W, (M_{H^{\pm\pm}}-M_{H_1})^2]$ and $[(M_W+ M_{H_1})^2, (M_{H^{\pm\pm}}-M_{W})^2]$. There is also a two-body bosonic channel, with the partial width
\begin{align}
\Gamma(H^{\pm\pm} \rightarrow W^{\pm *} H^\pm)&=\frac{9 g_L^4 M_{H^{\pm\pm}}}{128 \pi^3} G \left( \frac{M^2_{H^+}}{M^2_{H^{\pm\pm}}},\frac{M^2_W}{M^2_{H^{\pm\pm}}} \right) \,.
\label{eq:w}
\end{align}
with the function $G(x,y)$ defined in Appendix~\ref{sec:F}.
As for the singly-charged scalar in Eq.~(\ref{eqn:Width:HpWH2}), this channel is highly suppressed by the off-shell $W$ boson, and will be neglected in the following analysis. Since the masses and decay properties of $H_1$ and $A_1$ are the same in our model, we henceforth collectively
use $\phi$
to denote both the leptonic scalars $H_1$ and $A_1$, {\it i.e.} $\phi = H_1,\, A_1$.

In the standard Type-II seesaw, there is also the cascade decay channel for the doubly-charged scalar~\cite{ FileviezPerez:2008jbu, Melfo:2011nx}:
\begin{eqnarray}
\label{eqn:cascade}
H^{\pm\pm} \to H^\pm W^{\pm \ast} \to H_2 W^{\pm \ast} W^{\pm \ast} \,. 
\end{eqnarray}
In a large region of parameter space, the dilepton channels $H^{\pm\pm} \to \ell^\pm \ell^\pm$ and diboson channel $H^{\pm\pm} \to W^\pm W^\pm$ are highly suppressed respectively by the small Yukawa couplings $Y_{\alpha\beta}$ and the small VEV $v_\Delta$ of the triplet, and the doubly-charged scalar $H^{\pm\pm}$ decays mostly via the cascade channel above. When the mixing of $H_2$ with the SM Higgs is small, the neutral scalar $H_2$ decays mostly further into neutrinos via the Yukawa coupling $Y_{\alpha\beta}$. If the cascade channel dominates, the current direct LHC constraints on $M_{H^{\pm\pm}}$ in the $\ell^\pm \ell^\pm$~\cite{Aaboud:2017qph, CMS:2017pet} and $W^\pm W^\pm$~\cite{ATLAS:2018ceg, Aad:2021lzu} channels will be largely weakened. Then a relatively light $H^{\pm\pm}$ implies that the neutral scalar $H_2$ may also be light. This makes the decay channel of (\ref{eqn:cascade}) in the standard Type-II seesaw to some extent similar to our case in Eq.~(\ref{eqn:decayWWphi}), both leading to the signal of same-sign dilepton plus missing transverse energy (assuming $W$ boson decaying leptonically). However, as a result of the severe EWPT constraint on the mass splitting  $\Delta M$ of the triplet scalars~\cite{Melfo:2011nx, Kanemura:2012rs, Chun:2012jw, Primulando:2019evb},
the two $W$ bosons are both off-shell in the cascade decay in Eq.~(\ref{eqn:cascade}), which is very different from the on-shell $W$ bosons in Eq~(\ref{eqn:decayWWphi}) in our case.

Similarly, in the standard Type-II seesaw model the singly-charged scalar $H^\pm$ can decay into $\ell^\pm \nu$ and $h W^\pm$, $Z W^\pm$, $t\bar{b}$, which are respectively proportional to the couplings $Y_{\alpha\beta}$ and $v_\Delta$~\cite{FileviezPerez:2008jbu}. 
When both $Y_{\alpha\beta}$ and $v_\Delta$ are relatively small, the decay of $H^\pm$ will be dominated by
\begin{eqnarray}
\label{eqn:decay:HpH2W}
H^\pm \to H_2 W^{\pm \ast} \,,
\end{eqnarray}
where the $W$ boson is again off-shell as a result of the EWPT limit on the triplet scalar mass splitting. As in the doubly-charged scalar case, the decay $H^\pm \to H_2 W^{\pm \ast}$ with a light $H_2$ in the Type-II seesaw is very similar to the channel $H^\pm \to W^\pm \phi$ in our model, except for the off-shell $W$ boson.

Therefore, the new decay channels $H^{\pm\pm} \to W^\pm W^\pm \phi$ and $H^\pm \to W^\pm \phi$ make our model very different from the standard Type-II seesaw in the following aspects, which can be used to distinguish the two models at the high-energy colliders:
\begin{itemize}
    \item The $W^\pm W^\pm \phi$ final state from the $H^{\pm\pm}$ decay is absent in the standard Type-II seesaw model, where the $W$ bosons in the decays in Eqs.~(\ref{eqn:decayWWphi}) and (\ref{eqn:decay:HpH2W}) are off-shell.
    \item Another distinguishing feature of this model is that the decays $H^{\pm\pm}\to W^\pm W^\pm \phi$ and $H^\pm \to W^\pm \phi$ does not necessarily correspond to the compressed mass gaps among different particle states of the triplet $\Delta$, whereas in the standard Type-II seesaw model the decays in Eqs.~(\ref{eqn:decayWWphi}) and (\ref{eqn:decay:HpH2W}) are very sensitive to the mass splitting $\Delta M$ of the triplet scalars.
\end{itemize}

\begin{figure}[!t]
\centering
\includegraphics[width=0.45\textwidth]{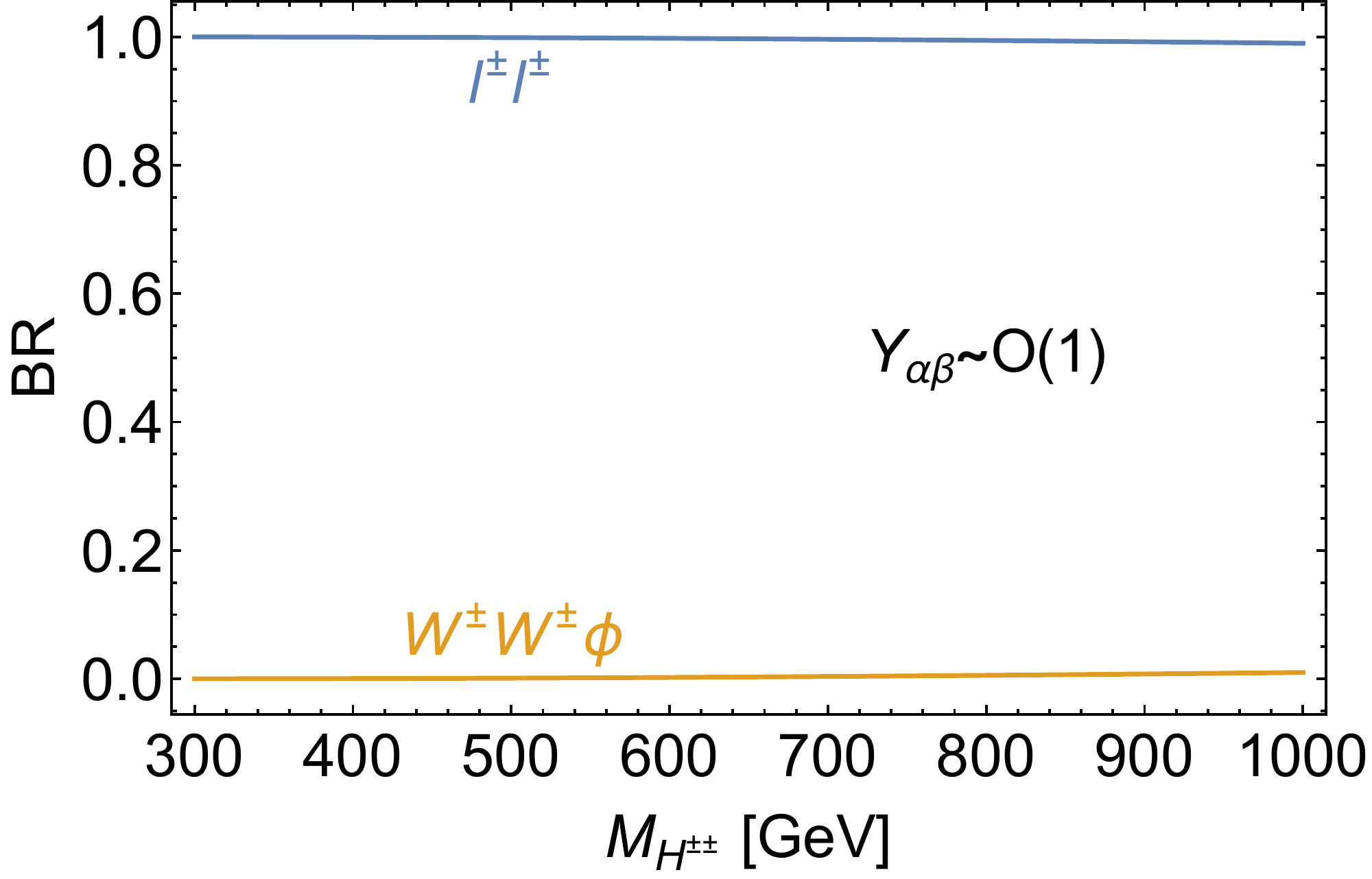}
\includegraphics[width=0.45\textwidth]{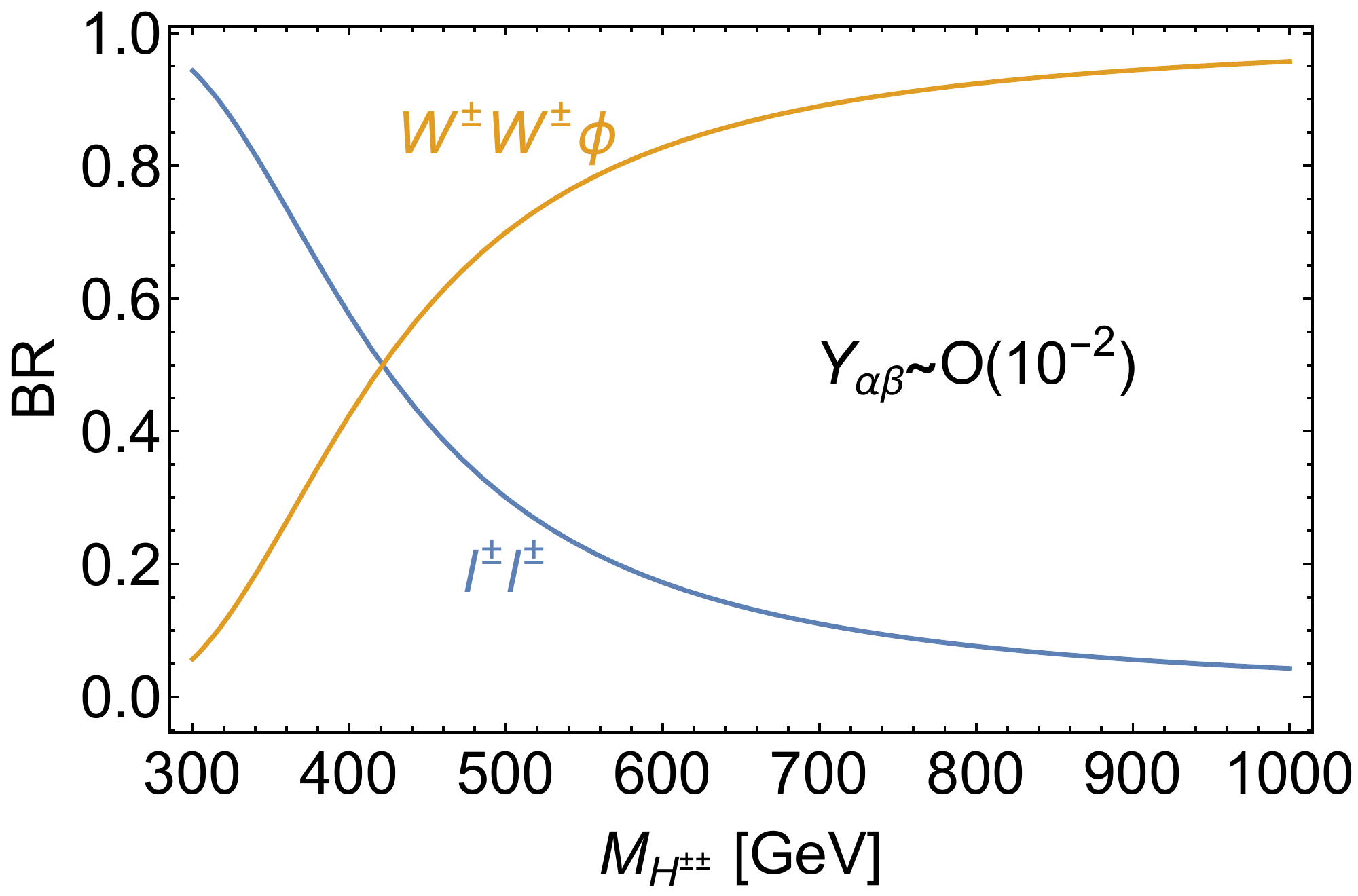}

\vspace{5mm}

\centering
\includegraphics[width=0.45\textwidth]{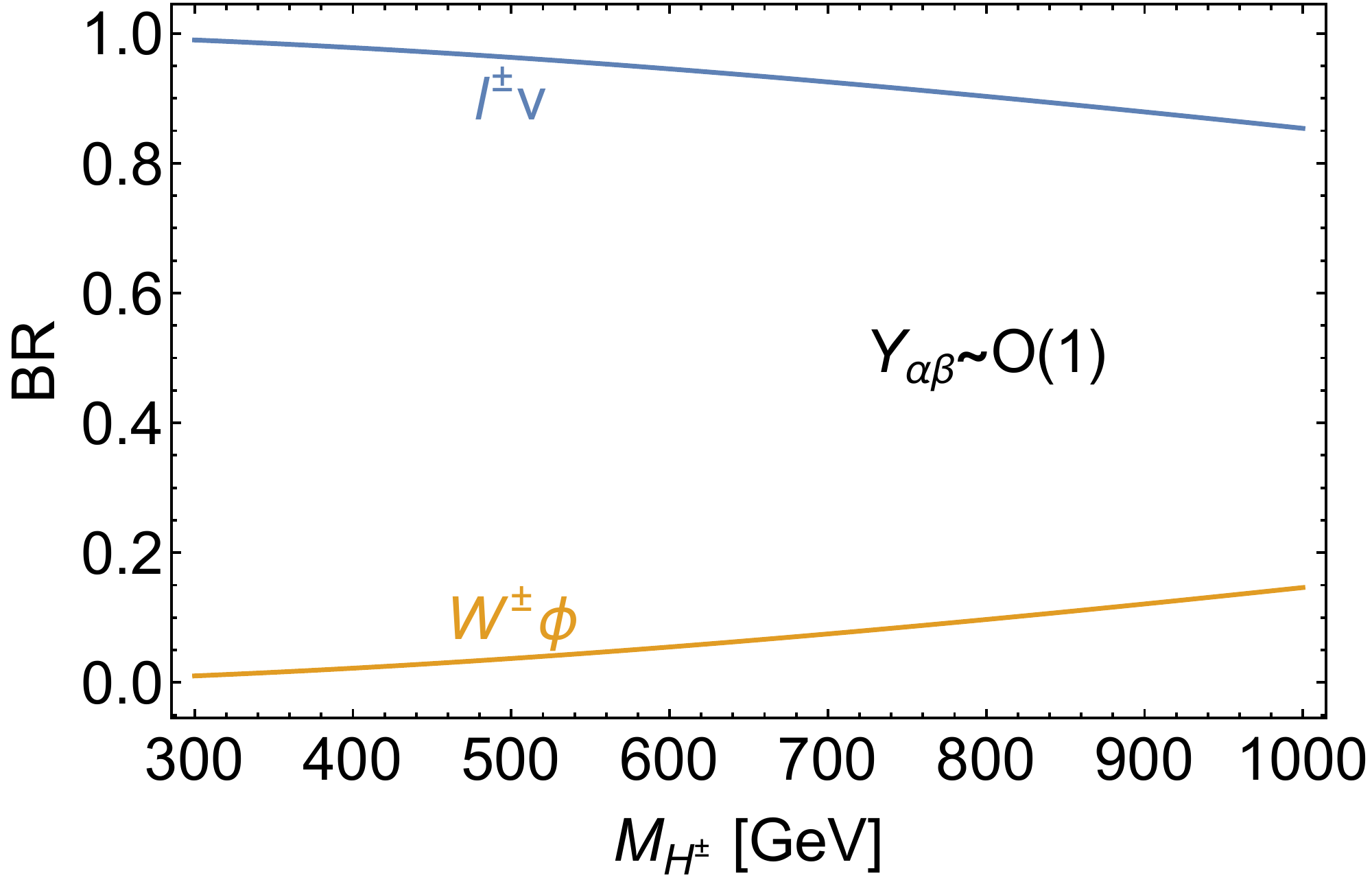}
\includegraphics[width=0.45\textwidth]{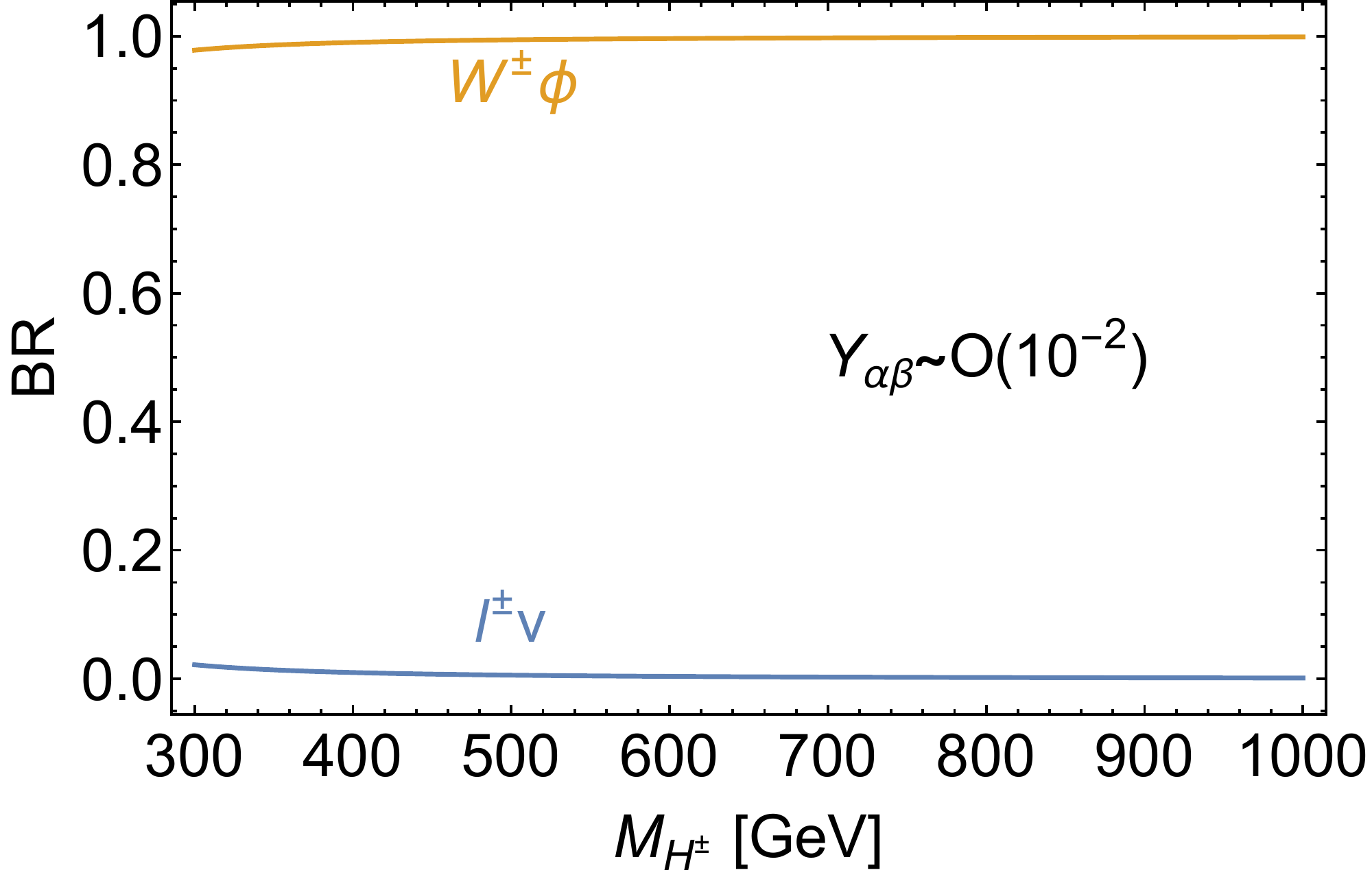}
\caption{Branching fractions of $H^{\pm\pm}$ decay (upper panels)  and $H^\pm$ decay (lower panels)  as a function of their masses. The left and right panels are for the large and small Yukawa coupling scenarios, respectively. Here $\phi$ denotes a leptonic scalar $H_1/A_1$.}
\label{fig:br_Hpp}
\end{figure}

Depending on the value of the Yukawa couplings $Y_{\alpha\beta}$, there are two distinct scenarios for the decays of $H^{\pm\pm}$ and $H^{\pm}$:
\begin{itemize}
    \item Large Yukawa coupling scenario with $Y_{\alpha\beta}\sim {\cal O}(1)$. In this case the leptonic channels $H^{\pm\pm}\rightarrow \ell^\pm \ell^\pm$ and $H^\pm \rightarrow \ell^\pm \nu$ dominate, which are from the Yukawa interactions $Y_{\alpha\beta}$.
    \item Small  Yukawa coupling scenario with $Y_{\alpha\beta}\lesssim {\cal O}(10^{-2})$. In this case the bosonic channels $H^{\pm\pm}\rightarrow W^\pm W^\pm \phi$ and $H^\pm \rightarrow W^\pm \phi$ dominate, which originate from the gauge couplings in Eqs.~(\ref{eqn:DH}) and (\ref{eqn:DDelta}).
\end{itemize}
For simplicity, we will not consider the intermediate scenarios, where the branching fractions (BRs) of bosonic and fermionic decay channels above are comparable.
The $W$-dominated final states  for small Yukawa couplings $Y_{\alpha\beta}$ depend on the scalar mixing angle $\sin\theta$, which in turn depends on $\lambda_8$ as shown in Eq.~\eqref{theta}, where we find that $\lambda_8$ needs to be ${\cal O}(1)$ in order to have a sizable $\sin \theta$.
The decay branching fractions of $H^{\pm\pm}$ and $H^{\pm}$ are shown respectively in the upper and lower panels of Fig.~\ref{fig:br_Hpp} as a function of their masses. The left and right panels are respectively for the large and small Yukawa coupling scenarios. As shown in the bottom left panel, if the Yukawa couplings 
are of order one, the dominant decay channels of $H^\pm$ will be $\ell^\pm \nu $, but the bosonic channel $W^\pm \phi$ is still feasible in the high mass regime with a branching fraction  around $10\%$. For small Yukawa couplings of order ${\cal O}(10^{-2})$, the singly-charged scalar $H^\pm$ decays predominantly into $W^\pm \phi$, as demonstrated in the bottom right panel. On the other hand, as shown in the top left panel, the doubly-charged scalar $H^{\pm\pm}$ will decay mostly to $\ell^\pm \ell^\pm$ if the Yukawa couplings are large, while the $W^\pm W^\pm \phi$ channel is dominant for small Yukawa couplings although a crossover happens for low $M_{H^{\pm\pm}}$, as shown in the top right panel.

\subsection{LFV constraints}
\label{sec:Limits}

\begin{figure*}[!t]
\centering
\includegraphics[width=0.55\textwidth]{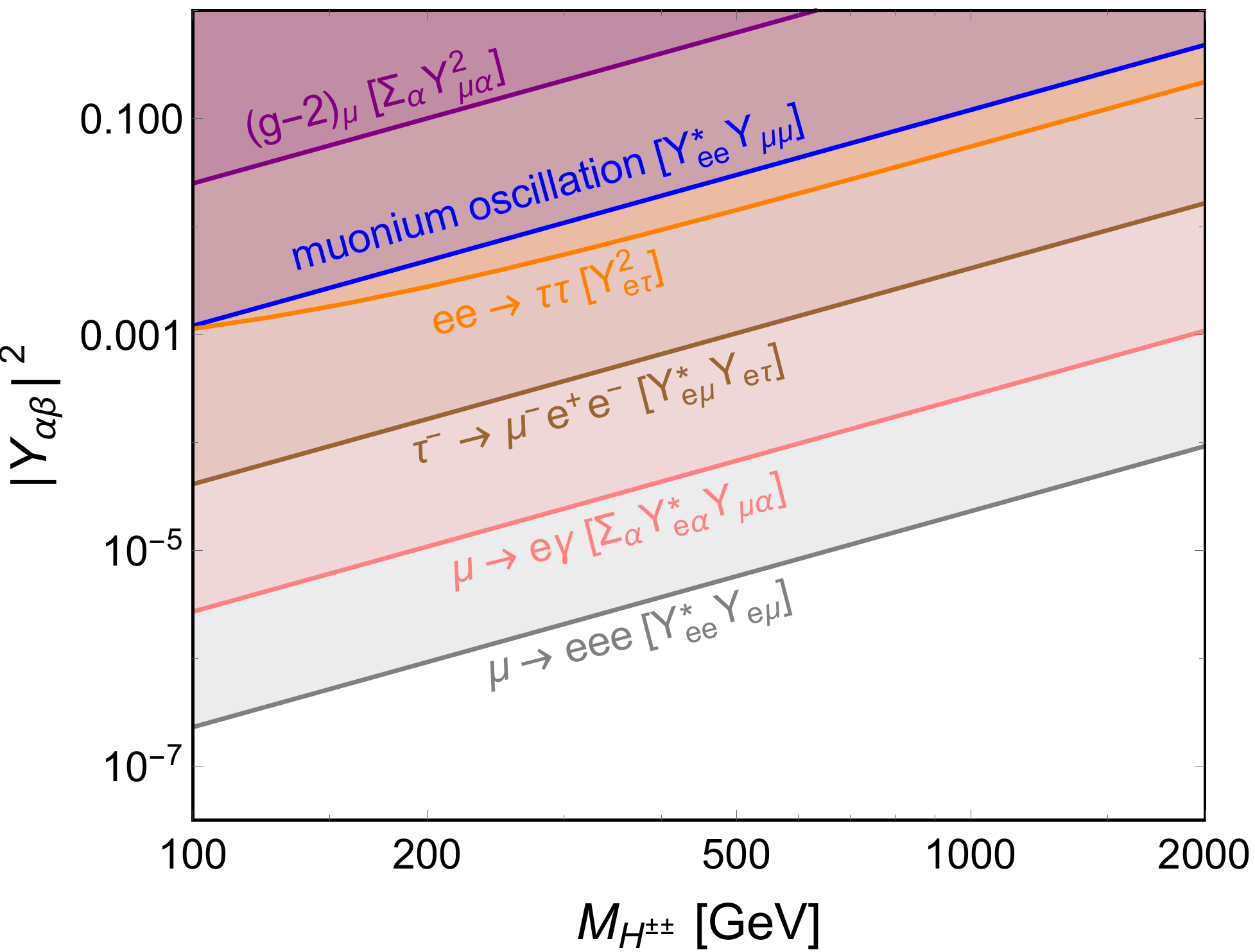}
\caption{LFV limits on the Yukawa couplings $|Y_{\alpha\beta}|^2$ as a function of the doubly-charged scalar mass $M_{H^{\pm\pm}}$. The shaded regions are excluded. See text and Table~\ref{tab:limits} for more details. }
\label{fig:limits}
\end{figure*}

There exist numerous constraints on the charged Higgs sector from the low-energy flavor data, such as those from the LFV decays $\ell_\alpha \to \ell_\beta \ell_\gamma \ell_\delta$,\  $\ell_\alpha \to \ell_\beta \gamma$~\cite{Zyla:2020zbs, Amhis:2016xyh}, anomalous electron~\cite{Hanneke:2008tm} and muon~\cite{Bennett:2006fi, Muong-2:2021ojo} magnetic moments, muonium oscillation~\cite{Willmann:1998gd}, and the LEP $e^+e^- \to \ell^+\ell^-$ data~\cite{Abdallah:2005ph}. Following  Ref.~\cite{Dev:2018kpa}, the updated LFV limits on the Yukawa couplings $Y_{\alpha\beta}$ are collected in Table~\ref{tab:limits}, and the most stringent ones are shown in Fig.~\ref{fig:limits}, as a function of the doubly-charged scalar mass $M_{H^{\pm\pm}}$. We see that the products involving two flavor transitions are highly constrained, while the bounds on an individual coupling are much weaker, especially for the tau flavor.

It should be noted that the contributions of $H^{\pm\pm}$ to the electron and muon $g-2$ are always negative~\cite{Lindner:2016bgg}. Therefore, the recent measurement of muon $g-2$ at Fermilab~\cite{Muong-2:2021ojo} cannot be interpreted as the effect of $H^{\pm\pm}$ in our model. On the other hand, we can use the reported measurement of Ref.~\cite{Muong-2:2021ojo}
\begin{eqnarray}
\Delta a_\mu = a_\mu^{\rm exp}- a_\mu^{\rm SM} = (251 \pm 59) \times 10^{-11} \,,
\end{eqnarray}
which is $4.2\sigma$ larger than the SM prediction~\cite{Aoyama:2020ynm}, to set limits on the $H^{\pm\pm}$ parameter space. We will use a conservative $5\sigma$ bound, {\it i.e.} require that the magnitude of the new contribution to $(g-2)_\mu$ from $H^{\pm\pm}$ must not exceed $0.8\times 59\times 10^{-11}$.
The corresponding limit on the Yukawa coupling $Y_{\mu\beta}$ is shown by the purple shaded region in Fig.~\ref{fig:limits} and also in Table~\ref{tab:limits}. Note that if a light scalar has an LFV coupling $h_{\mu\tau}$ to muon and tau, it could be a viable candidate to explain the muon $g-2$ anomaly, while satisfying all current constraints~\cite{Dev:2017ftk, BhupalDev:2018vpr, Li:2018cod, Evans:2019xer, Iguro:2020rby, Li:2021lnz, Hou:2021qmf}. Such neutral scalar interpretations of muon $g-2$ anomaly can be definitively tested at a future muon collider~\cite{ Capdevilla:2020qel,Buttazzo:2020eyl, Yin:2020afe, Capdevilla:2021rwo, Haghighat:2021djz}.

\begin{table}[!t]
  \centering
  \small
  \caption[]{Upper limits on the Yukawa couplings $|Y_{\alpha\beta}|^2$ (or $|Y_{\alpha\gamma}^\dagger Y_{\beta\gamma}|$) from the current experimental limits on the LFV branching fractions of $\ell_\alpha \to \ell_\beta \ell_\gamma \ell_\delta$, $\ell_\alpha \to \ell_\beta \gamma$~\cite{Zyla:2020zbs, Amhis:2016xyh}, anomalous electron~\cite{Hanneke:2008tm} and muon~\cite{Bennett:2006fi, Muong-2:2021ojo} magnetic moments, muonium oscillation~\cite{Willmann:1998gd}, and LEP $e^+e^- \to \ell^+\ell^-$ data~\cite{Abdallah:2005ph}. See also Fig.~\ref{fig:limits}.}
  \label{tab:limits}
  \begin{tabular}[t]{ccc}
  \hline\hline

  Process  & Experimental bound &
  Constraint $\times \left(\frac{M_{H^{\pm\pm}}}{100 \, {\rm GeV}} \right)^2$  \\ \hline

  $\mu^- \to e^- e^+ e^-$ & $< 1.0 \times 10^{-12}$ &
  $|Y_{ee}^\dagger Y_{e\mu}|< 2.3 \times 10^{-7}$ \\ \hline

  $\tau^- \to e^- e^+ e^-$ & $< 1.4 \times 10^{-8}$ &
  $|Y_{ee}^\dagger Y_{e\tau}| < 6.5 \times 10^{-5}$  \\
  $\tau^- \to e^- \mu^+ \mu^-$ & $< 1.6 \times 10^{-8}$ &
  $|Y_{e\mu}^\dagger Y_{\mu\tau}| < 4.9 \times 10^{-5}$  \\
  $\tau^- \to \mu^- e^+ \mu^-$ & $< 9.8 \times 10^{-9}$ &
  $|Y_{e\tau}^\dagger Y_{\mu\mu}| < 5.5 \times 10^{-5}$ \\

  $\tau^- \to \mu^- e^+ e^-$ & $< 1.1 \times 10^{-8}$ &
  $|Y_{e\mu}^\dagger Y_{e\tau}| < 4.1 \times 10^{-5}$ \\
  $\tau^- \to e^- \mu^+ e^-$ & $< 8.4 \times 10^{-9}$ &
  $|Y_{ee}^\dagger Y_{\mu\tau}| < 5.1 \times 10^{-5}$ \\
  $\tau^- \to \mu^- \mu^+ \mu^-$ & $< 1.2 \times 10^{-8}$ &
  $|Y_{\mu\mu}^\dagger Y_{\mu\tau}| < 6.1 \times 10^{-5}$ \\ \hline

  $\mu^- \to e^- \gamma$ & $< 4.2 \times 10^{-13}$ &
  $|\sum_\gamma Y_{e\gamma}^\dagger Y_{\mu \gamma}| < 2.7 \times 10^{-6}$  \\
  $\tau^- \to e^- \gamma$ & $< 3.3 \times 10^{-8}$ &
  $|\sum_\gamma Y_{e\gamma}^\dagger Y_{\tau \gamma}| < 1.8 \times 10^{-3}$ \\
  $\tau^- \to \mu^- \gamma$ & $< 4.4 \times 10^{-8}$ &
  $|\sum_\gamma Y_{\mu \gamma}^\dagger Y_{\tau \gamma}| < 2.1 \times 10^{-3}$ \\ \hline

  electron $g-2$ & $< 5.2 \times 10^{-13}$ &
  $\sum_\beta |Y_{e\beta}|^2 < 1.2$ \\
  muon $g-2$ & $< 4.7 \times 10^{-10}$ &
  $\sum_\beta |Y_{\mu \beta}|^2 < 0.025$ \\ \hline

  \makecell{muonium \vspace{-3pt} \\  oscillation} & $<8.2 \times 10^{-11}$ &
  $|Y_{ee}^\dagger Y_{\mu\mu}| < 0.0012$ \\ \hline

  $e^+e^- \to e^+e^-$ & $\Lambda_{\rm eff} > 5.2$ TeV &
  $|Y_{ee}|^2 < 0.0012$  \\
  $e^+e^- \to \mu^+\mu^-$ & $\Lambda_{\rm eff} > 7.0$ TeV &
  $|Y_{e\mu}|^2 < 6.4 \times 10^{-4}$  \\
  $e^+e^- \to \tau^+\tau^-$ & $\Lambda_{\rm eff} > 7.6$ TeV &
  $|Y_{e\tau}|^2 < 5.4 \times 10^{-4}$  \\
  \hline\hline
  \end{tabular}
\end{table}


The doubly-charged scalar $H^{\pm\pm}$ can induce   leptonic decays of SM $Z$ and Higgs boson at 1-loop level. With the coupling $Y_{\alpha\beta}$, the corresponding partial widths are respectively~\cite{Perez:1992hc, Nemevsek:2016enw}
\begin{eqnarray}
\label{eqn:Z_LFV}
\Gamma (Z \to \ell_\alpha^+ \ell_\beta^-) &\simeq&
\frac{g_L^2 M_Z}{144 \pi^4} \left( \frac{\cos2\theta_w}{\cos\theta_w} \right)^2
\left| \frac{\sum_\gamma m_{\ell_\gamma}^2 Y_{\alpha\gamma} Y_{\gamma\beta}^\ast}{M_{H^{\pm\pm}}^2} \right|^2 \,, \\
\label{eqn:h_LFV}
\Gamma (h \to \ell_\alpha^+ \ell_\beta^-) &\simeq&
\frac{M_h (\lambda_1 v)^2}{2^{15} \pi^5}
\left| \frac{\sum_\gamma m_{\ell_\gamma} Y_{\alpha\gamma} Y_{\gamma\beta}^\ast}{M_{H^{\pm\pm}}^2} \right|^2 \left| F \left( \frac{4M_{H^{\pm\pm}}^2}{M_h^2} \right) \right|^2 \,,
\end{eqnarray}
where $M_Z$ is the $Z$ boson mass, $m_{\ell_\gamma}$ is the mass for the charged lepton $\ell_\gamma$, the factor of $\lambda_1 v$ in Eq.~(\ref{eqn:h_LFV}) is from the trilinear scalar coupling $h H^{++} H^{--}$ in Table~\ref{table:couplings:trilinear}, and the loop function $F(x)$ can be found in Eq.~(B.8) of Ref.~\cite{Nemevsek:2016enw}. For the case of $\alpha \neq \beta$, the $H^{\pm\pm}$ induced decays in Eqs.~(\ref{eqn:Z_LFV}) and (\ref{eqn:h_LFV}) are apparently LFV. However, in addition to the loop factor, both the (LFV) decays of SM Higgs and $Z$ bosons above are highly suppressed by powers of the small ratio $m_{\ell_\gamma}/M_{h,\,Z}$. It turns out that the current precision $Z$ and Higgs data~\cite{Zyla:2020zbs} can only exclude $|Y_{\alpha\beta}|^2 \gg 1$ for $M_{H^{\pm\pm}}=1$ TeV, and the corresponding limits are much weaker than those in Table~\ref{tab:limits} and Fig.~\ref{fig:limits}.

Similarly, given the coupling $Y_{\alpha\beta}$, the couplings of the leptonic scalar $\phi$ with neutrinos induce the tree-level invisible decays $Z \to \nu_\alpha \nu_\beta \phi$, $h \to \nu_\alpha \nu_\beta \phi$ and the leptonic decay $W \to \ell_\alpha \nu_\beta \phi$. However, the limits from current precision EW and Higgs data are at most $Y_{\alpha\beta}\gtrsim {\cal O}(1)$~\cite{Berryman:2018ogk, deGouvea:2019qaz}, and therefore, are not shown in Table~\ref{tab:limits} and Fig.~\ref{fig:limits}.

\subsection{High-energy behavior: perturbativity and unitarity limits}
\label{sec:Limits2}

Since larger values of $\lambda_8$ and $Y_{\alpha\beta}$ play important roles for the hadron collider signal of this model, let us first check the largest values of these couplings which can be accommodated at the EW scale without becoming non-perturbative at a higher energy scale. For the purpose of illustration, we set just one Yukawa coupling  $Y_{\mu\mu}$ to be non-vanishing, with all other Yukawa couplings $Y_{\alpha\beta}$ ($\alpha\beta \neq \mu\mu$) to be zero. This choice is compatible with the current limits in Table~\ref{tab:limits}, as the products of the Yukawa couplings must be small due to the existing LFV limits, while a single coupling ($Y_{\mu\mu}$ in our case) can be as large as $Y_{\mu\mu}\sim {\cal O} (1)$ for $M_{H^{\pm\pm}} \sim 1$ TeV.

\begin{figure}[tb]
\centering
\includegraphics[width=0.49\textwidth]{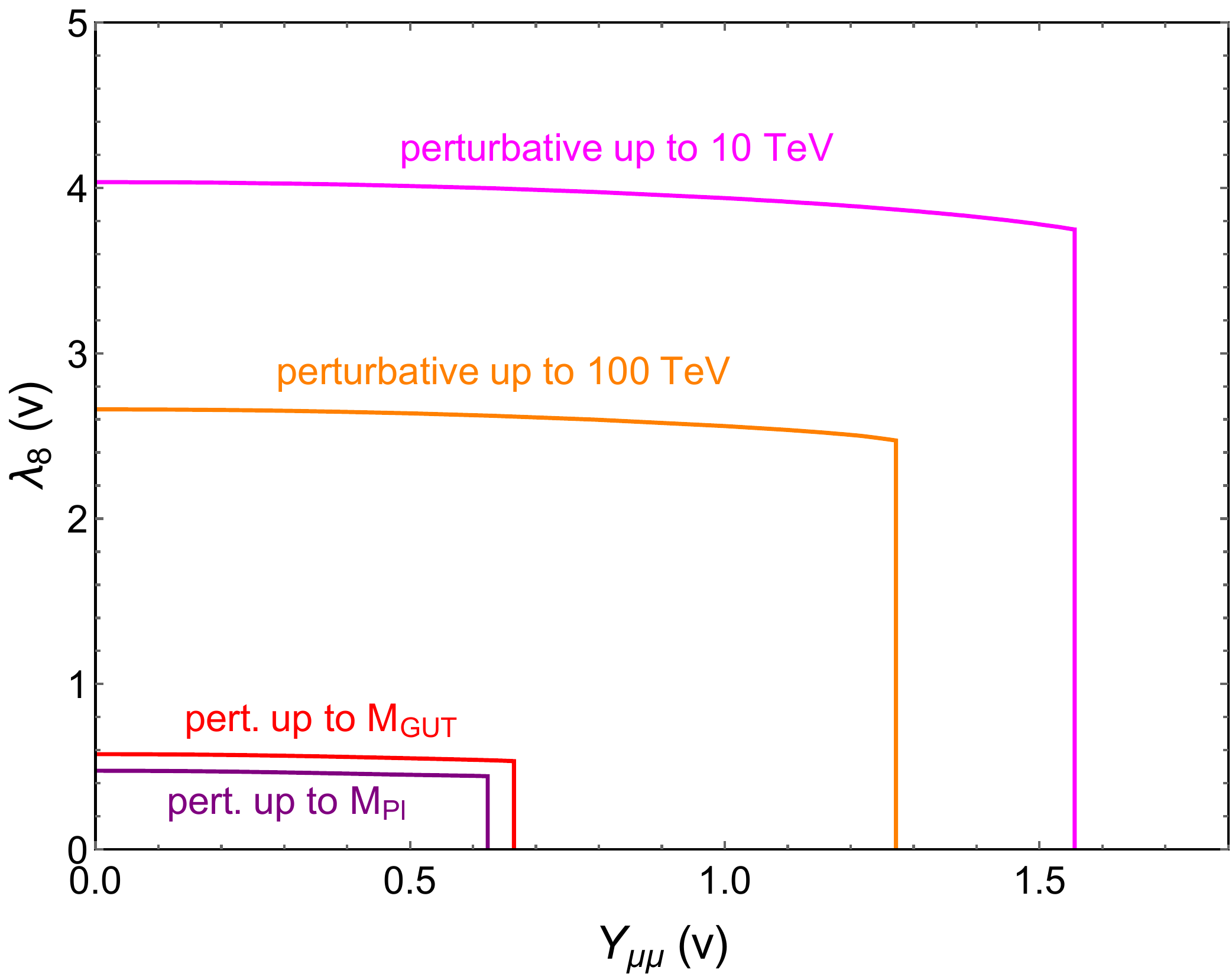}
\includegraphics[width=0.49\textwidth]{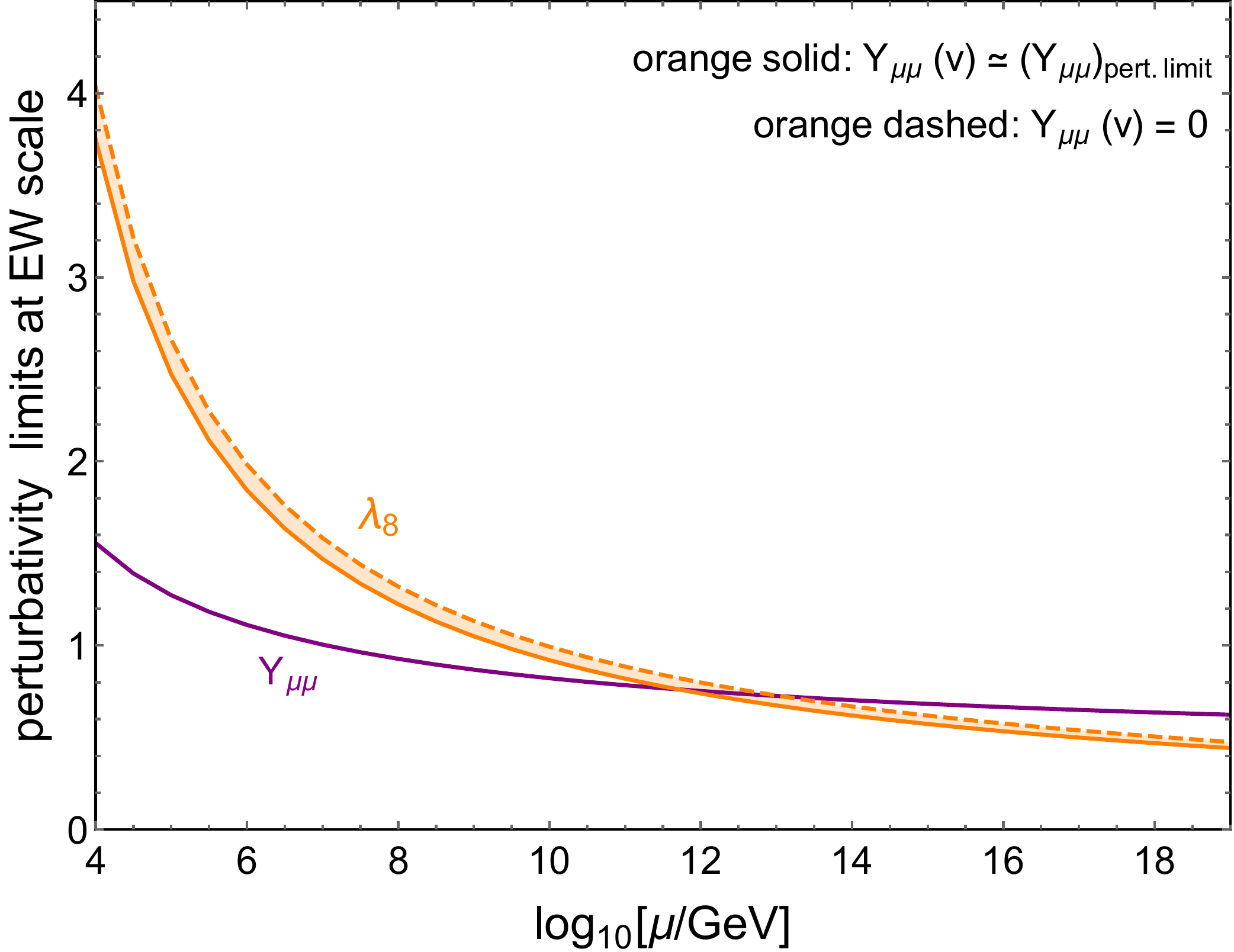}
\caption{
\label{fig:pert}
 {\it Left panel}: Perturbativity limits on $\lambda_8 (v)$ by the Landau pole at a higher scale of 10 TeV (magenta), 100 TeV (orange), the GUT scale (red) and the Planck scale (purple), as function of $Y_{\mu\mu} (v)$. {\it Right panel}: Perturbativity limits on $\lambda_8$ (orange) and $Y_{\mu\mu}$ (purple) at the EW scale, as function of the Landau pole scale $\mu$. For the solid and dashed orange lines, we take $Y_{\mu\mu}$ to be respectively the perturbativity limit and zero at the EW scale.
}
\end{figure}

{To implement the perturbativity limits from the high-energy scale, we use the RGEs  in Appendix~\ref{sec:AppB} for all the gauge, scalar and Yukawa couplings given in Eqs.~(\ref{eqn:DH}), (\ref{eqn:DDelta}), (\ref{eq:LYuk}) and (\ref{V_Delta_phi})}.
From the  RGEs, we find that $\lambda_8$ depends on $Y_{\mu\mu}$ at one-loop level, since both  $\lambda_8$ and $Y_{\mu\mu}$ are associated with the interaction terms which involve the triplet scalars. The dependence of perturbativity limits on $\lambda_8$ on the Yukawa coupling $Y_{\mu\mu} (v)$ at the EW scale is shown in the left panel of Fig.~\ref{fig:pert}, with perturbativity up to Planck scale $M_{\rm Pl}$ and the grand unified theory (GUT) scale $M_{\rm GUT}$ for the purple and red lines, and up to the 100 TeV and 10 TeV scales for the orange and pink lines, respectively. Comparing these lines, we can see that the perturbativity limits on $\lambda_8$ are very sensitive to the value of $Y_{\mu\mu}$ at the EW scale. To have a perturbative $\lambda_8$ at the 10 TeV (100 TeV) scale, it is required that the coupling $Y_{\mu\mu} (v) \lesssim 1.6 \; (1.3)$. For a perturbative theory up to the GUT or Planck scale, the coupling $Y_{\mu\mu}$ needs to be even smaller, {\it i.e.} $Y_{\mu\mu} (v) \lesssim 0.67$. The perturbativity limits on $\lambda_8$ and $Y_{\mu\mu}$ at the EW scale as function of the scale $10 \; {\rm TeV} < \mu < M_{\rm Pl}$ are shown in the right panel of Fig.~\ref{fig:limits}. For the quartic coupling $\lambda_8$, the solid and dashed lines correspond respectively to the cases of $Y_{\mu\mu}$ set at the perturbative limit and $Y_{\mu\mu}=0$ at the EW scale. As shown in both the two panels of Fig.~\ref{fig:limits}, the quartic coupling $\lambda_8$ can be as large as 4 (2.7), with perturbativity holding up to 10 TeV (100 TeV). With the requirement of perturbativity up to the Planck (GUT) scale, we have $\lambda_8 \lesssim 0.48 \; (0.58)$ at the EW scale.


The high-energy behavior of $\lambda_8$, $Y_{\mu\mu}$ and other couplings can be understood analytically from  the solutions of RGEs for these couplings. As a rough approximation, let us first see the analytical solution of $Y_{\mu\mu}$ without including the contributions from the gauge couplings $g_{S,\, L,\, Y}$ for the $SU(3)_C, SU(2)_L, U(1)_Y$ respectively. Defining $\alpha_{\mu} \equiv Y_{\mu\mu}^2/4\pi$,  
it is trivial to get the analytical solution of $\alpha_\mu$ at scale $\mu$ from Eq.~(\ref{eqn:beta:Ymumu}) as
\begin{eqnarray}
\alpha_{\mu} (\mu) = \frac{ \alpha_{\mu} (v) }{ 1 - \frac{4}{\pi} \alpha_{\mu} (v)t } \,,\quad \text{with} \quad t=\ln{\mu \over v} \ .
\end{eqnarray}
It is clear from the above equation that the coupling $Y_{\mu\mu}$ is not asymptotically free and will blow up when the scale  parameter approaches the value of
\begin{eqnarray}
t_c = \ln \left( \frac{\mu_c}{v} \right) = \frac{\pi^2}{Y_{\mu\mu}^2(v)} \,.
\end{eqnarray}
With an initial value of $Y_{\mu\mu} (v) = 1.5$ at the EW scale, we can get the critical value of $t_c \simeq 4.39$, which corresponds to an energy scale of $\mu \simeq 20$ TeV. The full analytic solution of $Y_{\mu\mu}$ including the gauge coupling contributions is shown in Appendix~\ref{sec:AppC}. Following the running of gauge couplings, and taking $g_L (M_Z) = 0.65100$, $g_Y (M_Z) = 0.357254$~\cite{Fusaoka:1998vc, Xing:2007fb, Xing:2011aa, Antusch:2013jca, Huang:2020hdv}, we find that in this case $t_c = 4.67$, which corresponds to $\mu \simeq 26$ TeV.



The contribution of $Y_{\mu\mu}$ to the evolution of $\lambda_8$ can be obtained from the following analytical solution of the RGE for $\lambda_8$ (see Appendix~\ref{sec:AppC} for more details)
\begin{equation}
    \lambda_8(\mu)=\lambda_8(v)\exp \left[ \frac{1}{4\pi^2}\int^\mu_v E_8(\mu)d\mu \right] \, ,
\end{equation}
where $E_8$ depends on $Y_{\mu\mu}$ as well as the couplings $g_{L,\, Y}$ and the top-quark Yukawa coupling $y_t$ and is given in Eq.~(\ref{eqn:E8}). As soon as $Y_{\mu\mu}$ turns non-perturbative, the exponential  becomes very large and  $\lambda_8$ also becomes non-perturbative. 


We have also checked the unitarity constraints on $Y_{\mu\mu}$ and $\lambda_8$, and the details are given in Appendix \ref{sec:AppD}. It is found that the unitarity constraints are much weaker $\lambda_8 <10.0$, compared to the perturbativity constraints obtained here.

%
%

\section{Collider signatures}
\label{sec:LHC}

\begin{figure}[tb]
\centering
\includegraphics[width=0.6\textwidth]{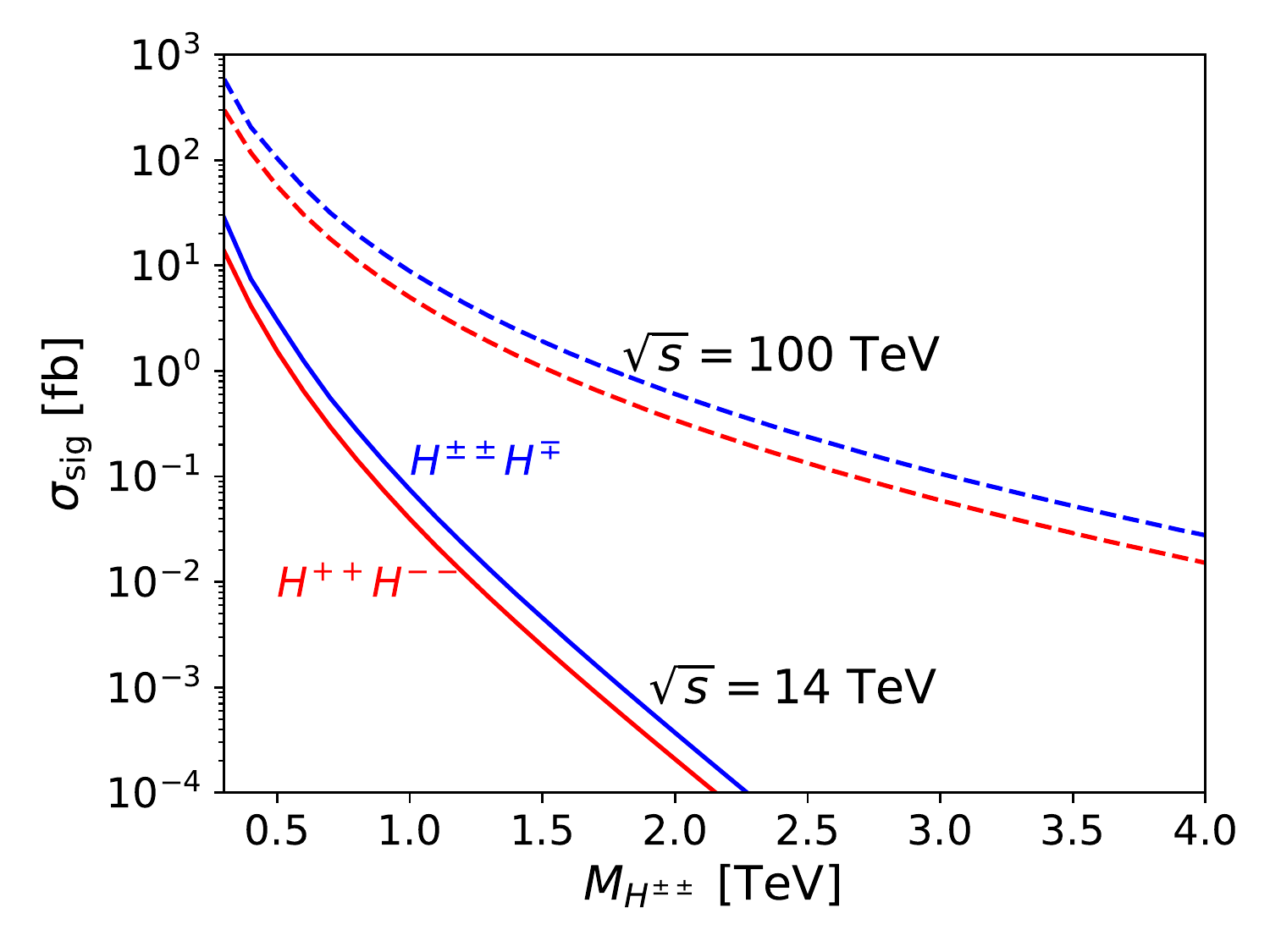}
\caption{\label{fig:cross_section}
Cross sections of $H^{++}H^{--}$ pair (red) and $H^{\pm\pm}H^{\pm}$ associated (blue) production of doubly-charged scalars at $\sqrt{s}=14$ TeV (solid) and $\sqrt{s}=100$ TeV (dashed) $pp$ colliders.  }
\end{figure}

In this section we analyze the striking signatures of this model at the LHC and future 100 TeV hadron colliders. We consider both the pair production and the associated production  channels: 
\begin{eqnarray}
pp \to H^{++} H^{--} ,\; H^{\pm\pm} H^\mp \,.
\end{eqnarray}
The production cross sections in the two channels for the doubly-charged scalar coming from an $SU(2)_L$-triplet $\Delta$ at the 14 TeV  LHC and future 100 TeV colliders have been estimated in Refs.~\cite{Du:2018eaw, Arkani-Hamed:2015vfh}, which are reproduced in Fig.~\ref{fig:cross_section}.  As shown in Section~\ref{sec:interactions},
the final states associated with these production processes depend on the decay  branching fractions of $H^{\pm \pm}$ and $H^{\pm}$. Our model predicts novel decay processes
\begin{equation}
H^{\pm \pm} \rightarrow W^{\pm} W^{\pm} \phi \quad {\rm  and}\quad
H^{\mp} \rightarrow W^{\mp} \phi \,,
\end{equation}
where the light leptonic scalars $\phi = H_1,\; A_1$ will escape from detection and lead to missing momentum. This can be used to distinguish our model from the standard Type-II seesaw. In this paper, we will focus on these novel channels. The prospects of the small Yukawa coupling scenario at future hadron colliders are investigated in Section~\ref{sec:small}, the large Yukawa coupling case is analyzed in Section~\ref{sec:large}, and the intermediate Yukawa coupling case is considered in Section~\ref{sec:medium-yukawa}.


\subsection{Small Yukawa coupling scenario}
\label{sec:small}

One typical choice of parameter is that the Yukawa coupling $Y_{\alpha\beta} \lesssim 10^{-2}$ to satisfy all the low-energy experimental limits in Section~\ref{sec:Limits}. Note that this choice of $Y_{\alpha \beta}$ would result in an effective $\nu_\alpha\nu_\beta\phi$ coupling $\lambda_{\alpha\beta}$ of order $10^{-3}$, which is too small to probe in the VBF channel discussed in Ref.~\cite{deGouvea:2019qaz}, but accessible in our UV-complete model due to the additional interactions, as shown below. In particular, under this choice of small Yukawa coupling, the doubly-charged scalar $H^{\pm\pm}$ will mostly decay to two $W$ bosons and a light neutral leptonic scalar $\phi = H_1, A_1$; cf.~the top right panel of Fig.~\ref{fig:br_Hpp}. With two same-sign $W$ bosons decaying leptonically and the other two decaying hadronically, the final state of our signal features two same-sign leptons ($e$ or $\mu$) plus jets and large missing transverse momentum in the pair production channel, {\it i.e.}
\begin{align*}
p p \rightarrow H^{++}(\rightarrow W^+W^+ \phi) \; H^{--}(\rightarrow W^-W^- \phi) \rightarrow \ell^{\pm} \ell^{\pm}+\text{4 jets} + E_T^{\text{miss}} \,.
\end{align*}
Similarly, we also have the associated production $p p \rightarrow H^{\pm\pm}H^{\mp}$ with $H^{\mp} \rightarrow W^{\mp} \phi$ which also has the same final states. However, due to the presence of less number of $W$'s, the contribution from the associated production is small to our signal.

We use {\tt FeynRules}~\cite{Alloul:2013bka} to define the fields and the Lagrangian of our model, then the resulting UFO model file is fed into {\tt MadGraph5\textunderscore aMC@NLO}~\cite{Alwall:2014hca} to generate the Monte Carlo events where the decay of vector bosons is achieved by the {\tt Madspin}~\cite{Artoisenet:2012st} module integrated within {\tt MadGraph5}. Next-to-leading order corrections are included by a $k$-factor of $1.25$~\cite{Muhlleitner:2003me} for our signal process. The leading SM backgrounds come from $WZ$ and $WW$ productions and the sub-leading ones from $WWW$ and $t \bar{t}W$ processes are also considered. We use {\tt MadGraph5} to generate the background events, and the leading ones are generated with two extra jets to properly account for the jet multiplicity in the final states. The events from the hard processes are showered with {\tt Pythia8}~\cite{Sjostrand:2007gs} and the jets are clustered using {\tt Fastjet}~\cite{Cacciari:2011ma} with the anti-$k_T$ algorithm~\cite{Cacciari:2008gp} and the cone radius $\Delta R=0.4$. All the signal and background events are smeared to simulate the detector effect by our own code
using {\tt Delphes} {\tt CMS\textunderscore PhaseII} cards~\cite{deFavereau:2013fsa}.

Electrons (muons) are selected by requiring that $p_T > 10 \, \text{GeV}$ and $ |\eta| < 2.47 \; (2.5)$, jets are required to have $p_T > 20\, \text{GeV} $ and $ |\eta| < 3$. We adopt the $b$-tagging formula from the {\tt Delphes} default card where the efficiency is $\varepsilon_b = 0.8 \mathrm{tanh} \left(0.003 p_T^{\text{$b$-jet}} \right) \times 30/(1 + 0.086p_T^{\text{$b$-jet}})$ (with $p_T^{\text{$b$-jet}}$ in unit of GeV)~\cite{deFavereau:2013fsa}. We apply some pre-selection cuts before launching 
the carefully designed analysis below. First, all events should have exactly two same-sign leptons and the number of jets should be at least 3: $N_{\text{jet}} \geq 3$. Finally we veto any event with $b$-tagged jet: $N_{\text{$b$-jet}} = 0$.

\subsubsection{Cut-based analysis}
\label{sec:cut}

The same-sign $W$ pair signal from $H^{\pm\pm} \to W^\pm W^\pm$ has  been searched for at the LHC by the ATLAS collaboration~\cite{ATLAS:2018ceg, Aad:2021lzu}. In the searches of same-sign dilepton plus jets plus missing energy, the most stringent lower limit on doubly-charged scalar mass is 350 GeV~\cite{Aad:2021lzu}.
As a case study, we first consider the scenario of $M_{H^{\pm\pm}}=400\,\text{GeV}$, which satisfies the current direct LHC constraints. The kinematic variables we use to distinguish the signal from  backgrounds are the missing  transverse energy $E_T^{\text{miss}}$,  the effective mass $M_{\text{eff}}$ defined as scalar sum of transverse momenta of all reconstructed leptons, jets, and missing energy, the separation $\Delta R_{\ell\ell}$ between two leptons,  the azimuthal angle $\Delta \phi(\ell\ell, E_T^{\text{miss}})$ between the two lepton system and $E_T^{\text{miss}}$,  the invariant mass of all jets $M_{\text{jets}}$, and  the cluster transverse mass from jets and $E_T^{\text{miss}}$ defined as  \cite{Barger:1987re}
\begin{align}
M_T^{\text{jets}} \equiv \left[ \left( \sqrt{M_{\text{jets}}^2+ \bigg| \sum_{j} \overrightarrow{p}^j_T \bigg|^2} + E_T^{\text{miss}} \right)^2 - \Bigg| \sum_{j} \overrightarrow{p}^j_T + \overrightarrow{E}_T^{\text{miss}} \Bigg|^2 \right]^{1/2} \ .
\label{eq:M_T}
\end{align}

\begin{figure}[!t]
\makebox[\linewidth][c]{%
\centering
\includegraphics[width=0.5\textwidth]{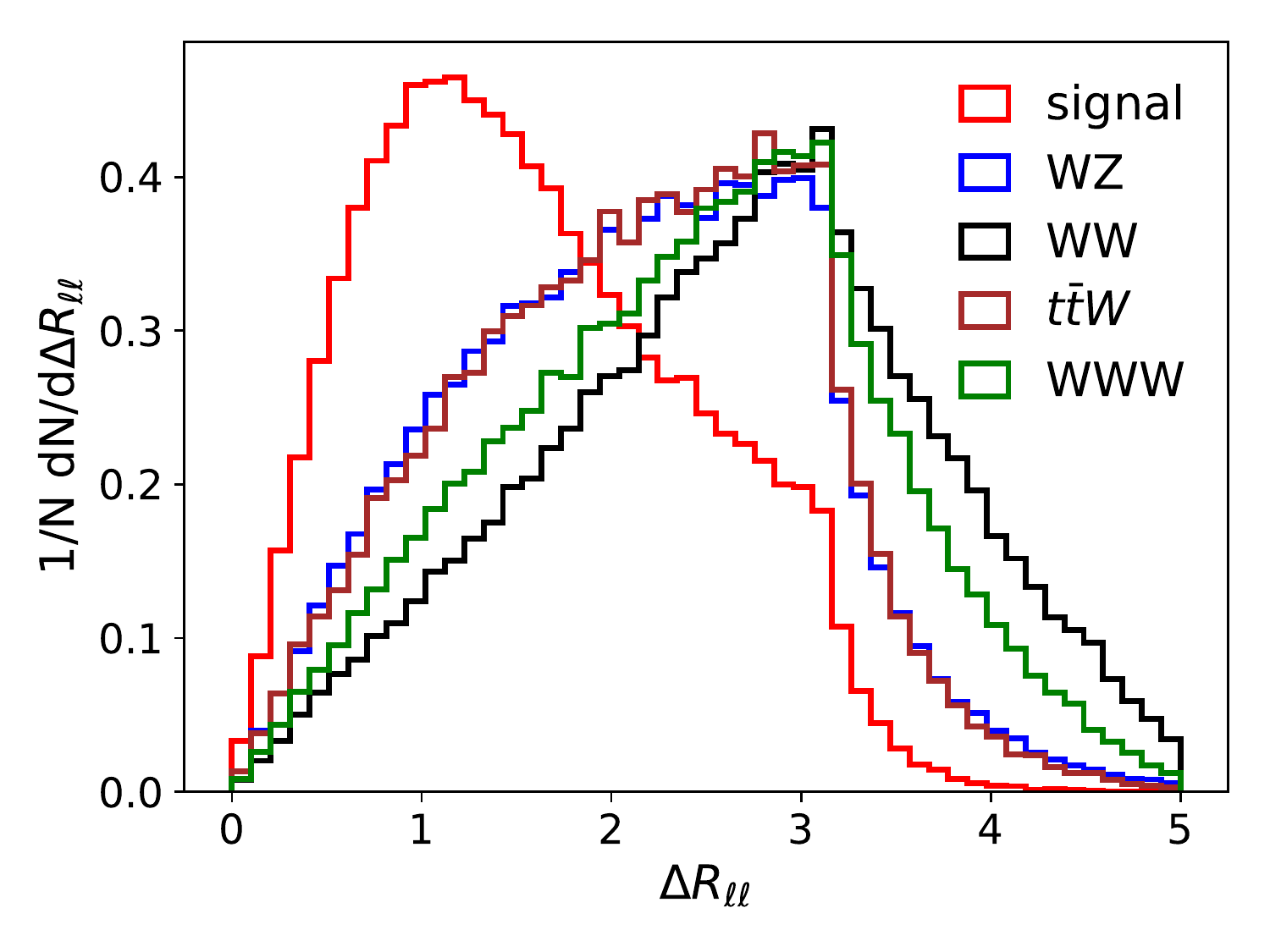}
\includegraphics[width=0.5\textwidth]{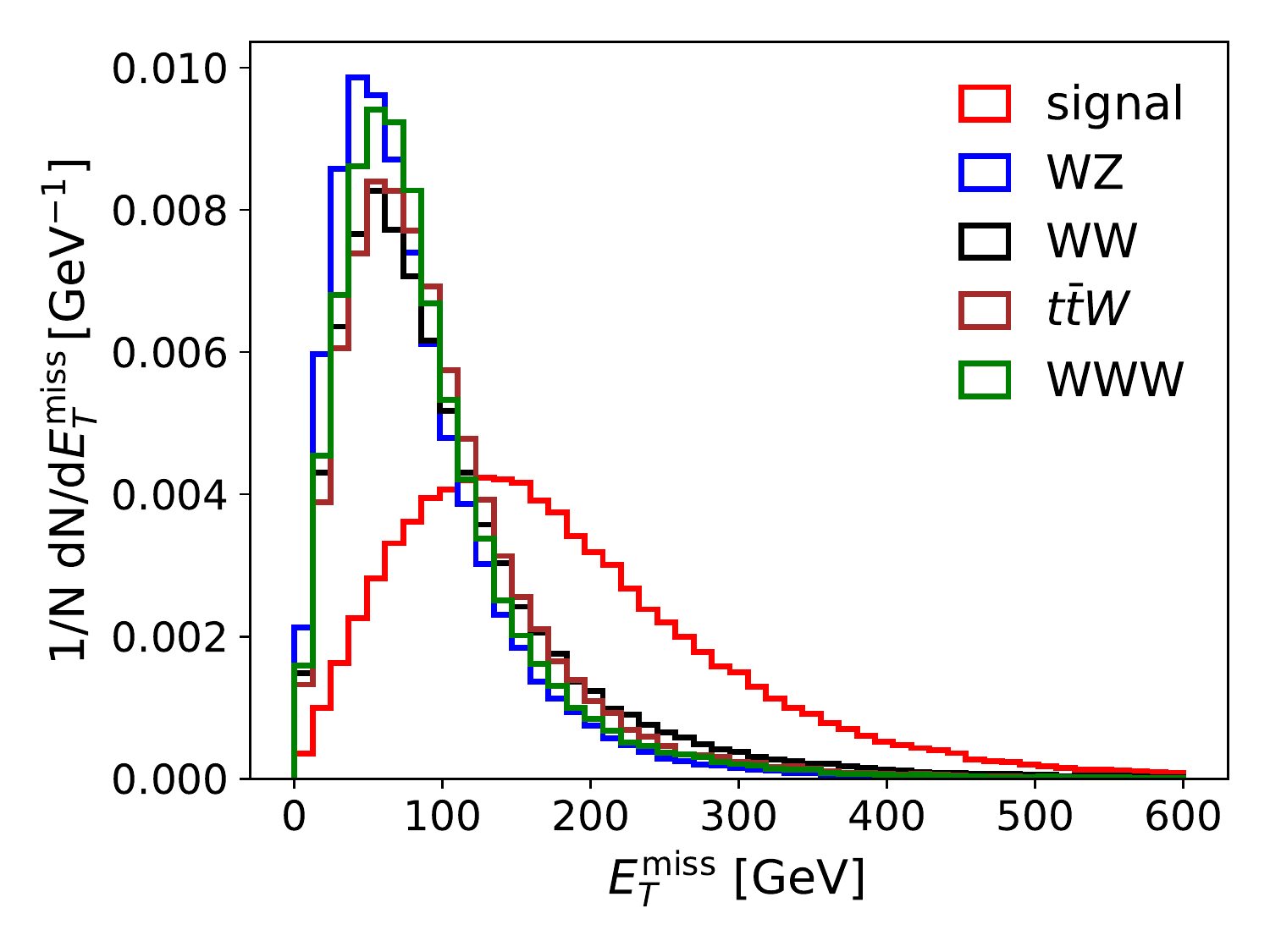}
}
\makebox[\linewidth][c]{%
\centering
\includegraphics[width=0.5\textwidth]{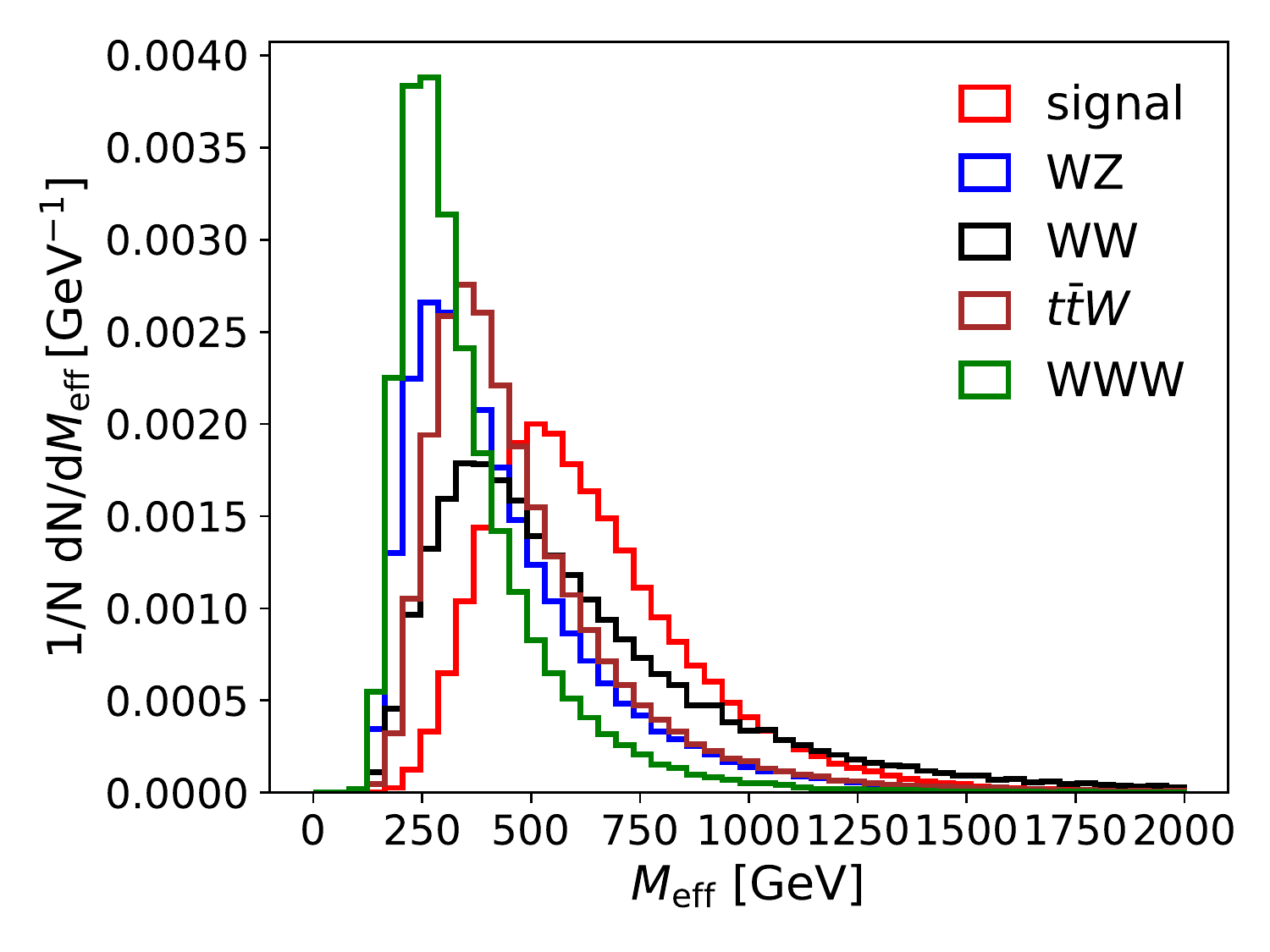}
\includegraphics[width=0.5\textwidth]{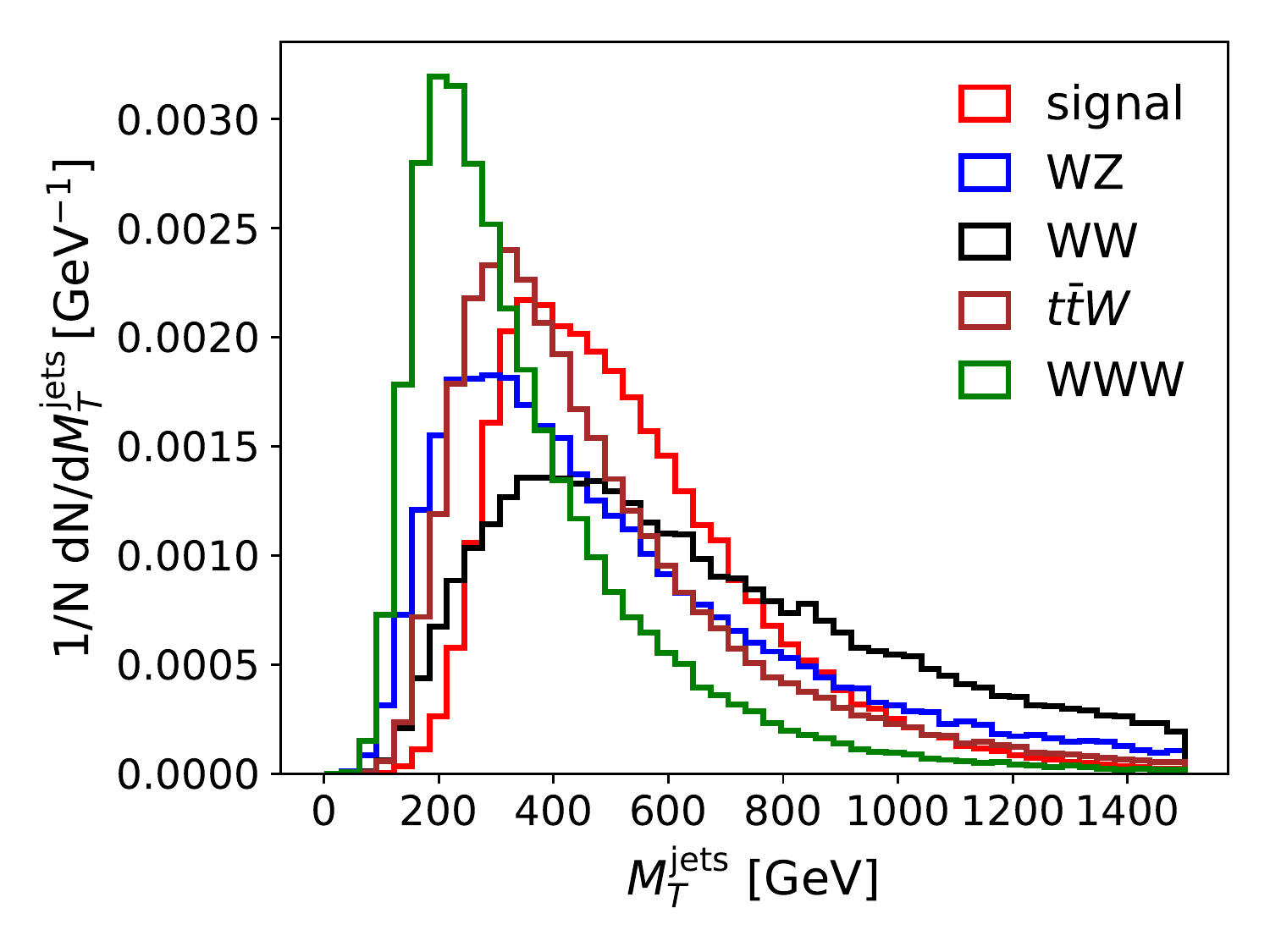}
}
\makebox[\linewidth][c]{%
\centering

\includegraphics[width=0.5\textwidth]{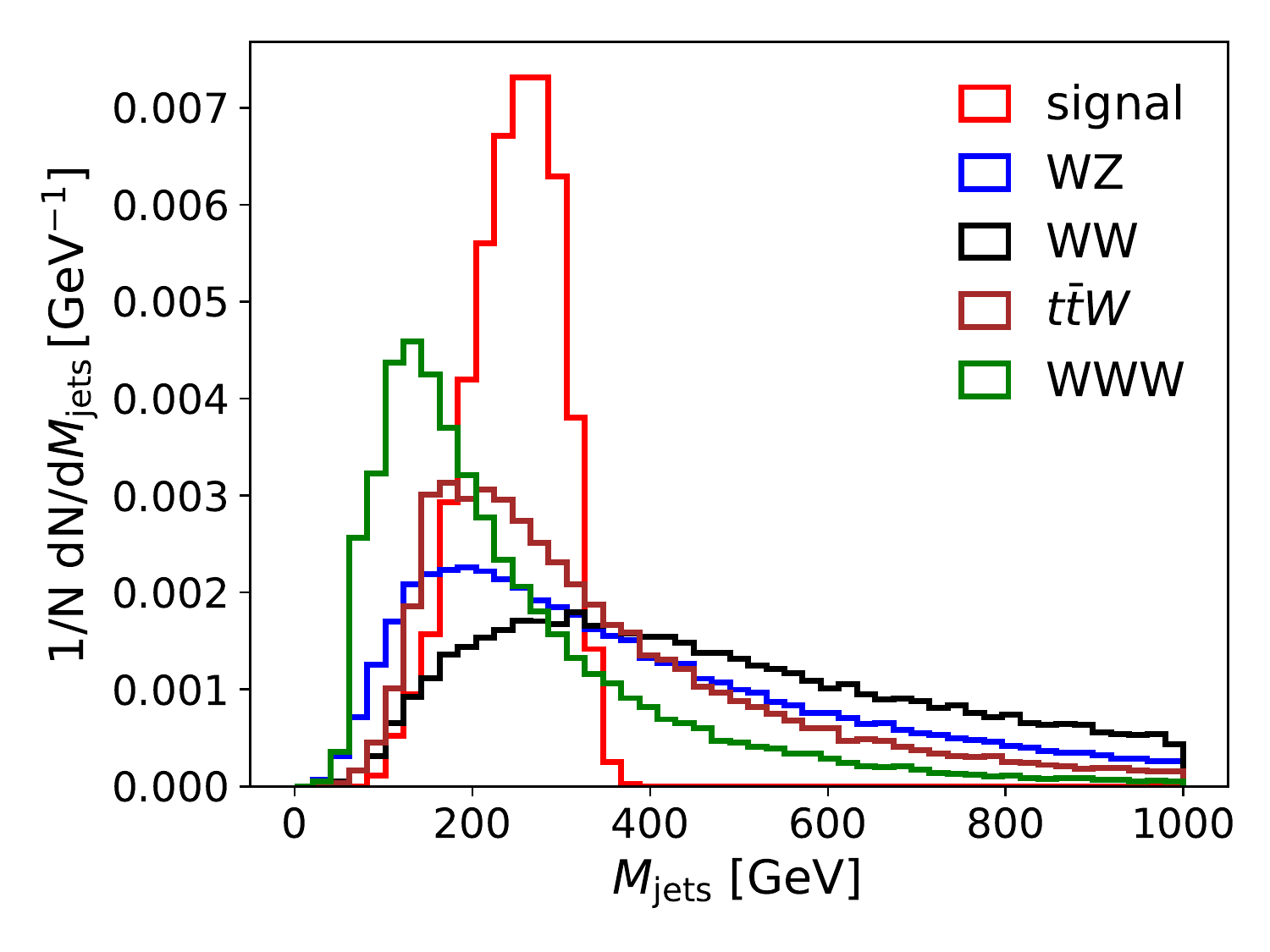}
\includegraphics[width=0.5\textwidth]{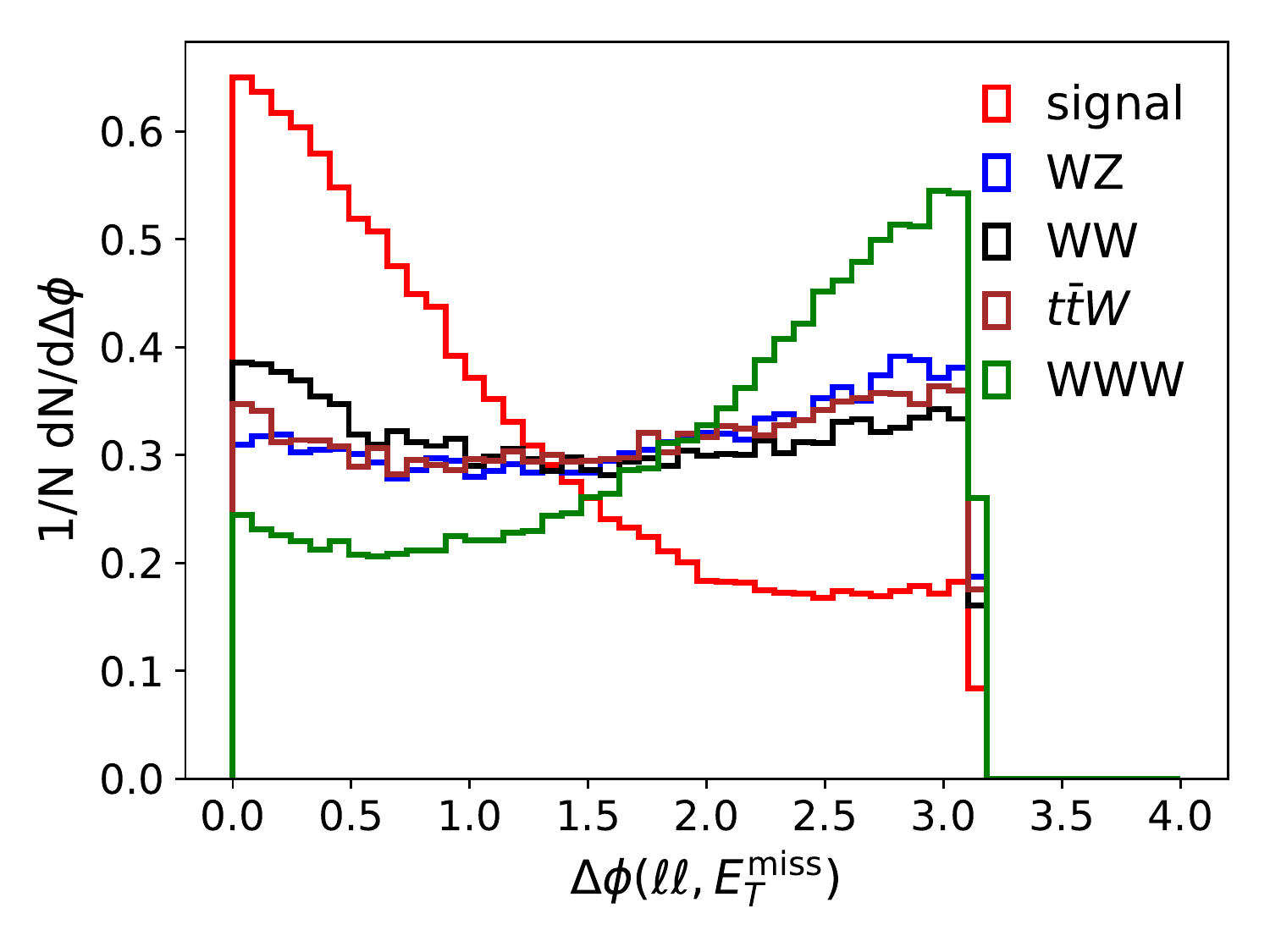}
}
\caption{\label{fig:distribution}
Distributions of observables used in cut-based analysis for the signal $W^\pm W^\pm \phi$ and SM backgrounds $WZ$, $WW$, $t \bar{t} W$, $WWW$: separation of two leptons $\Delta R_{\ell\ell}$ (top left), missing transverse energy $E_T^{\rm miss}$ (top right), effective mass $M_{\rm eff}$ (middle left), transverse mass $M_{T}^{\rm jets}$ of $H^{\pm\pm}$ defined in Eq.~(\ref{eq:M_T}) (middle right), invariant mass $M_{\rm jets}$ of jets (bottom left), and the azimuthal angle $\Delta \phi (\ell\ell,\, E_T^{\rm miss})$ between dilepton and missing energy (bottom right). All the distributions are normalized to be unity.}
\end{figure}

To enhance the signal-to-background ratio, the selection cuts we applied are as follows, and the corresponding cut-flows for the cross sections of signal and backgrounds are collected in Table~\ref{tab:cut_flow}.
\begin{itemize}
\item $ 0.3 < \Delta R_{\ell\ell} < 2.0$. The lower limit of $\Delta R_{\ell\ell}$  separates the leptons for isolation. The leptons in our signal emerge from the decay of two same-sign $W$ bosons which are from the decay of $H^{\pm\pm}$.
However, the leptons associated with the background processes emerge from the decays of $W$ and $Z$ bosons which are well separated. Therefore, the leptons in the signal tend to have smaller $\Delta R_{\ell\ell}$. {The distributions of $\Delta R_{\ell\ell}$ for the signal and backgrounds are presented in the top left panel of Fig.~\ref{fig:distribution}}.

\item $ E_T^{\text{miss}} > 110 \, \text{GeV}$. One of the decay products emerging from $H^{\pm\pm}$ is the light neutral scalar $\phi$ which decays only into neutrinos and appears to be invisible in the detector. Due to the existence of the massive $\phi$ along with the neutrinos from $W$ boson decay, our signal tends to have larger missing transverse energy compared to the  background processes {(see the top right panel of Fig.~\ref{fig:distribution} for distributions)}. Consequently, we choose a high  $ E_T^{\text{miss}} $ threshold to distinguish {the signal from backgrounds}.

\item $ M_{\text{eff}} > 350 \, \text{GeV}$. Borrowed from the SUSY searches~\cite{ATLAS:2020srl, ATLAS:2020ghe}, the effective mass $M_{\rm eff}$ is a measure of the overall activity of the event. It provides a good discrimination especially for  signals with energetic jets. The jets in our signal are from $W$ decay while the jets associated with backgrounds are from the QCD productions, which makes the jets from the signal to be more energetic in general. This can be seen in the middle left panel of Fig.~\ref{fig:distribution}.  Thus the effective mass associated with the signal is distributed at higher values.

\item $ M_T^{\text{jets}} > 300 \, \text{GeV} $. Since the decay products from $H^{\pm\pm}$ contain invisible particles, we cannot fully reconstruct its mass. The transverse mass $M_T^{\text{jets}}$ is an alternative option in this situation. We choose to reconstruct the transverse mass $M_{T}^{\rm jets}$ of $H^{\pm\pm}$ using jets and $ E_T^{\text{miss}} $ in order to reproduce its mass peak as close as possible. From the distributions shown in the middle right panel of Fig.~\ref{fig:distribution}, we can see that the transverse mass for the signal peaks around 400 GeV while for backgrounds it peaks at a smaller value. Consequently, a large $ M_T^{\text{jets}}$ cut can help us to discriminate the signal from backgrounds.

\item $ 150\,\text{GeV} < M_{\text{jets}} < 350\,\text{GeV}$. As mentioned above, the jets in the signal emerge from the  hadronic decays of $W$ boson while the jets associated with the main backgrounds are from QCD production. As a result, the invariant mass of all jets from backgrounds has a broader and flatter distribution, while the distribution for the signal is concentrated in the region between the two $W$ boson mass threshold and the doubly-charged scalar mass, {as shown in the bottom left panel of Fig.~\ref{fig:distribution}}. This provides a good observable to distinguish the signal from backgrounds.

\item $ \Delta \phi(\ell\ell, E_T^{\text{miss}}) < 1.5 $. The contributions to $E_T^{\text{miss}}$ associated with the signal are neutrinos and the light neutral scalar $\phi$ from the decay of $H^{\pm\pm}$. The signal decay products include also same-sign dileptons and,  consequently, the azimuthal angle between the same-sign dilepton and $E_T^{\text{miss}}$ in the signal tends to have a small value. In contrast, the backgrounds do not have such kinematics and thus the distribution of $\Delta \phi(\ell\ell, E_T^{\text{miss}})$ is rather flat for the background processes. The distributions for the signal and backgrounds are shown in the bottom right panel of Fig.~\ref{fig:distribution}.
\end{itemize}

\begin{table}[tb]
\begin{center}
\caption{
\label{tab:cut_flow}
Cut-flow of the cross sections for signal and SM backgrounds $WZ$, $WW$, $t \bar{t} W$, $WWW$ at the HL-LHC with $M_{H^{\pm\pm}}=400\,\text{GeV}$.}
\vspace{5pt}
\begin{tabular}{ |c|c|c c c c| }
 \hline
 Cut Selection& \specialcell{Signal \\ $[\text{fb}]$} &\specialcell{$WZ$ \\ $[\text{fb}]$} & \specialcell{$WW$\\ $[\text{fb}]$}  & \specialcell{$t\bar{t}W$ \\ $[\text{fb}]$} & \specialcell{$WWW$  \\ $[\text{fb}]$}   \\ \hline
 \hline
$ 0.3 < \Delta R_{\ell\ell} < 2.0$ & 0.092 & 4.5 & 1.3 & 0.64 & 0.25 \\ \hline
$ E_T^{\text{miss}} > 110 \, \text{GeV}$ & 0.067 & 1.1 & 0.41 & 0.191& 0.053 \\ \hline
$ M_{\text{eff}} > 350 \, \text{GeV}$ & 0.066 & 0.95 & 0.39 & 0.18 & 0.039 \\ \hline
$ M_T^{\text{jets}} > 300 \, \text{GeV} $ & 0.064 & 0.94 & 0.39 & 0.18 & 0.038 \\ \hline
$ 150\,\text{GeV} < M_{\text{jets}} < 350\,\text{GeV}$ & 0.062 & 0.22 & 0.067 & 0.073 & 0.018 \\ \hline
$ \Delta \phi(\ell\ell, E_T^{\text{miss}}) < 1.5 $ & 0.049 & 0.13 & 0.035 & 0.040 & 0.010 \\ \hline
\end{tabular}
\end{center}
\end{table}

After all the cuts, it is found in Table~\ref{tab:cut_flow} that the cross section for our signal is only a few times smaller than that for the SM backgrounds. To calculate the signal significance, we use the metric $\sigma = S/ \sqrt{S+B}$ where $S$ and $B$ are the numbers of events for signal and backgrounds respectively, and we have not included any  systematic uncertainties in our analysis. The expected event yields at the HL-LHC after all the cuts above are shown in Table~\ref{tab:events}. It is clear that the significance can reach $5\sigma$ in the cut-based analysis, which implies a great potential for discovery of the signal $H^{\pm\pm} \to W^\pm W^\pm \phi$ at the HL-LHC.

\begin{table}[!t]
\begin{center}
\caption{\label{tab:events}
Number of events in cut-based and BDT analysis for signal and backgrounds at the HL-LHC with 3 ab$^{-1}$ luminosity and for  $M_{H^{\pm\pm}}=400\,\text{GeV}$. The last column shows the significance of signal.}
\vspace{5pt}
\begin{tabular}{ |c|c|c c c c|c||c| }
 \hline
 & Signal & $WZ$ & $WW$ & $t\bar{t}W$ & $WWW$ & Backgrounds & $\sigma$  \\ \hline
\specialcell{Number of events \\ (cut-based)}& 145.56 & 397.54 & 104.17 & 120.00 & 30.42 & 652.12 &5.15  \\ \hline
\specialcell{Number of events \\ (BDT-based)} & 184.56 & 70.00 & 23.00 & 29.30 & 10.48 & 132.78& 10.36 \\
 \hline
\end{tabular}
\end{center}
\end{table}

\subsubsection{BDT improvement}

\begin{figure}[!t]
\makebox[\linewidth][c]{%
\centering
\includegraphics[width=0.5\textwidth]{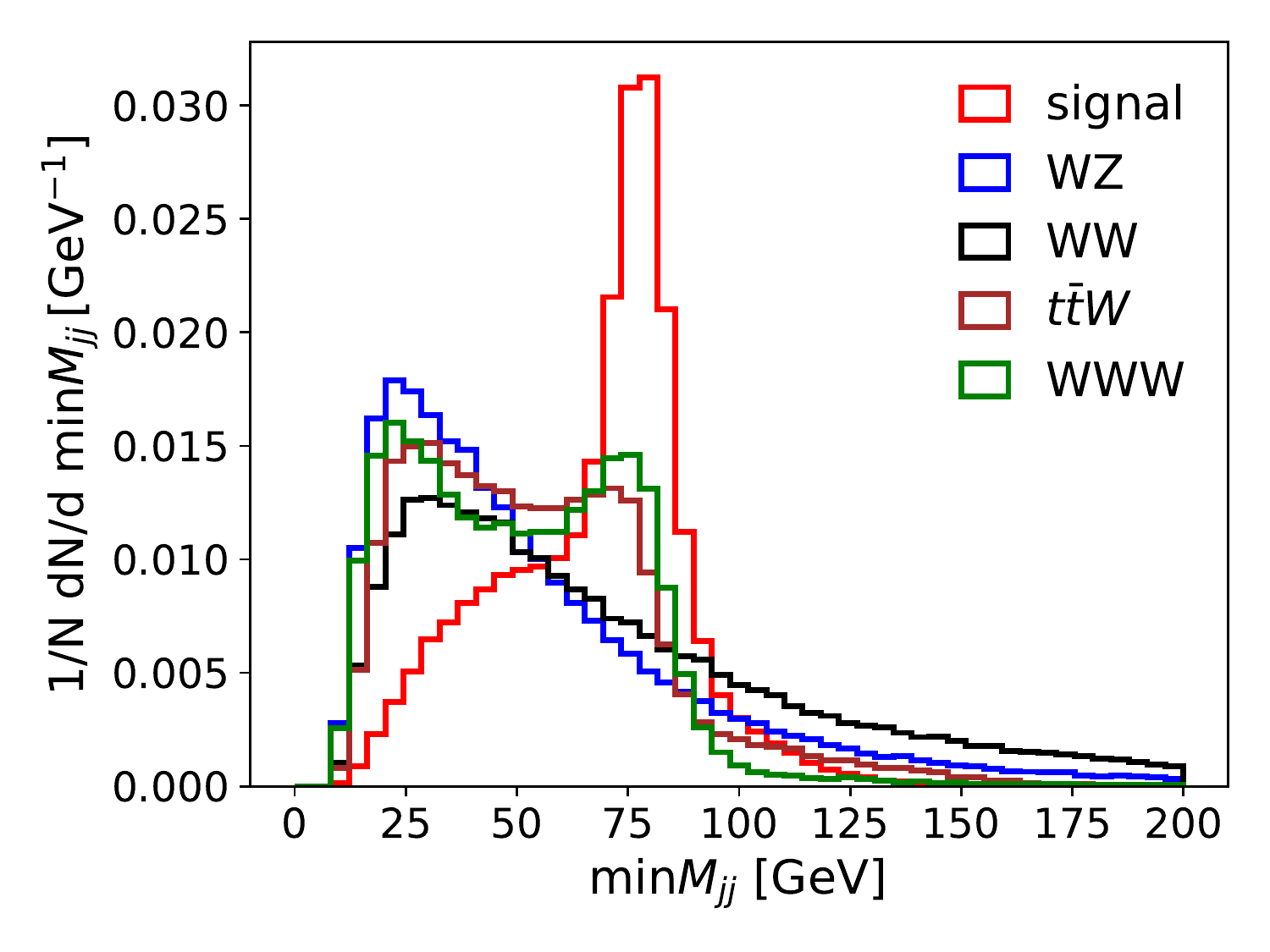}
\includegraphics[width=0.5\textwidth]{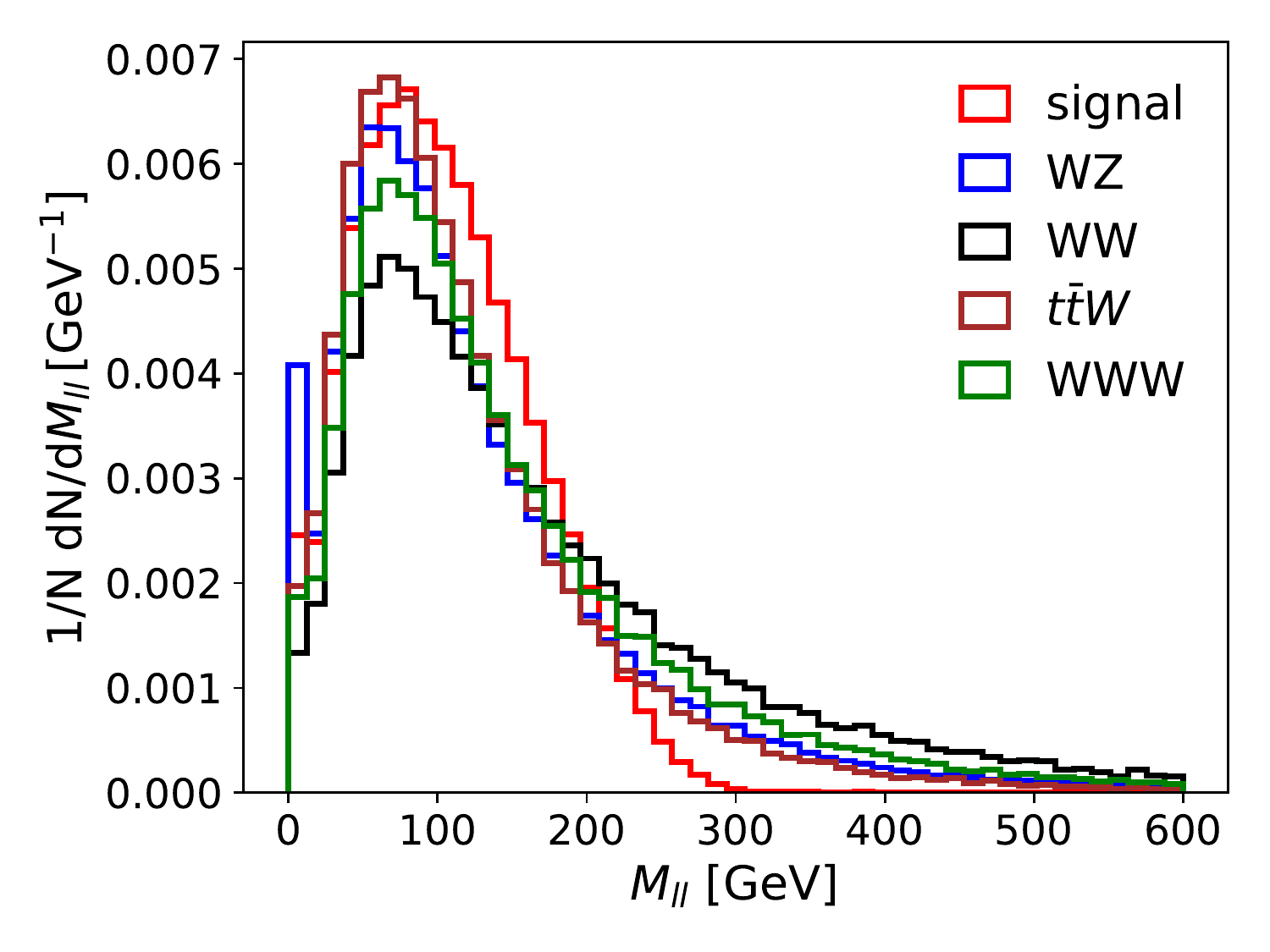}
}
\makebox[\linewidth][c]{%
\centering
\includegraphics[width=0.5\textwidth]{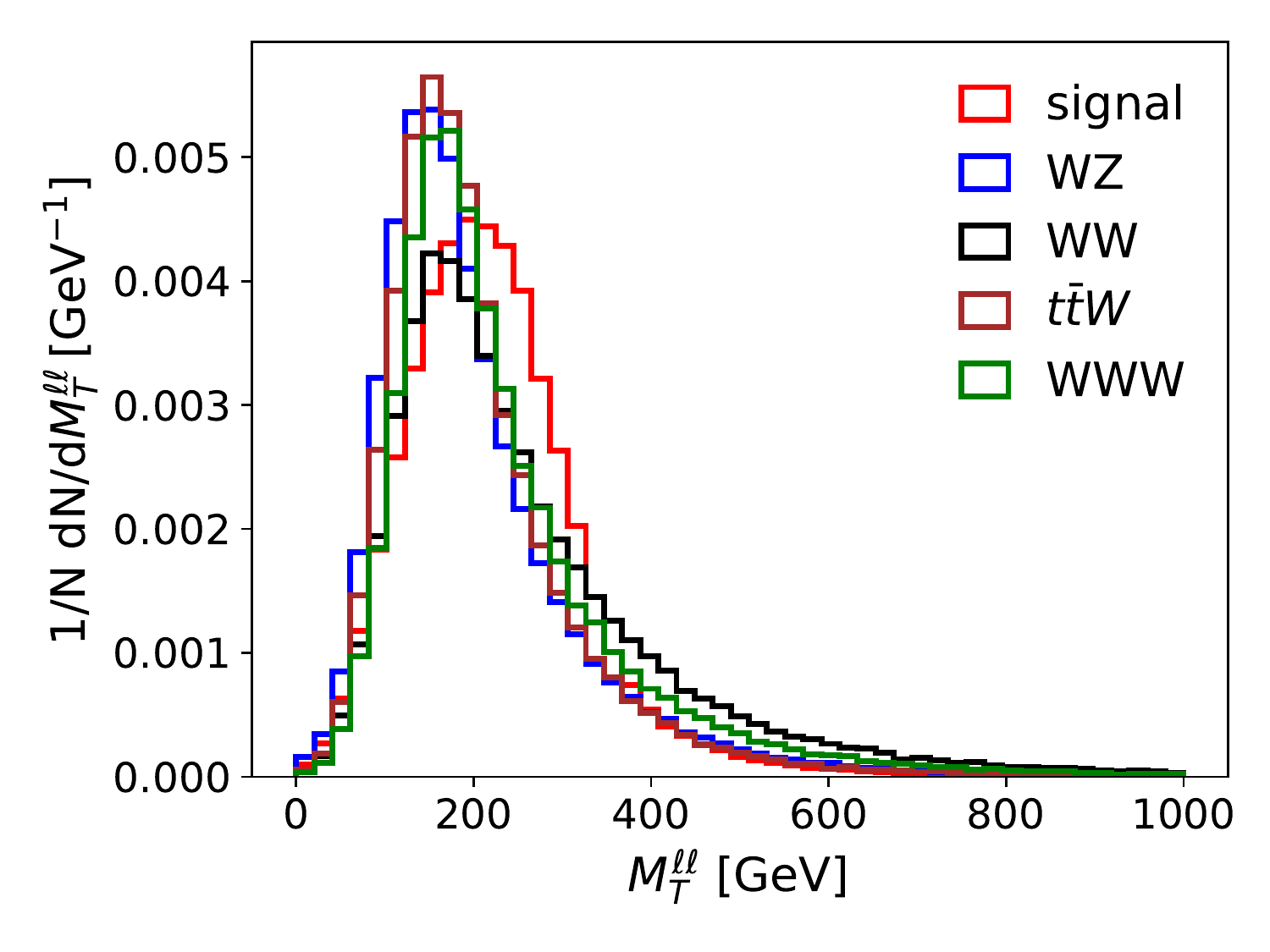}
\includegraphics[width=0.5\textwidth]{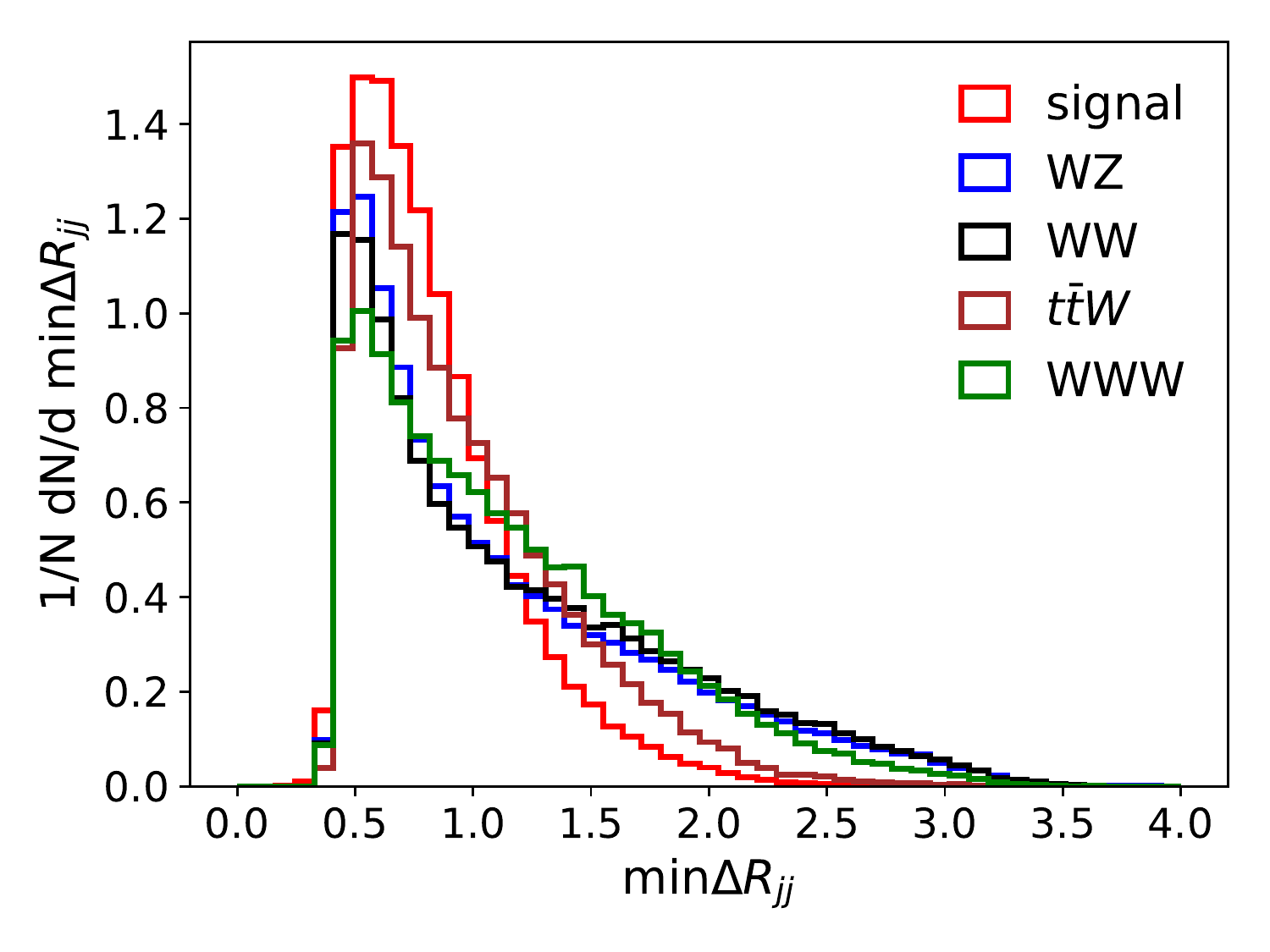}
}
\caption{\label{fig:additional_distributions}
More distributions of variables that are found by BDT to be important for distinguishing signal $W^\pm W^\pm \phi$ from backgrounds $WZ$, $WW$, $t \bar{t} W$, $WWW$: minimum invariant mass min$M_{jj}$ of two jets (upper left), invariant mass $M_{\ell\ell}$ of same-sign dilepton (upper right), transverse mass $M_T^{\ell\ell}$ of leptons and missing energy (lower left), and minimum separation min$\Delta R_{jj}$ of two jets (lower right). }
\end{figure}

\begin{figure}[!t]
\centering
\includegraphics[width=0.45\textwidth]{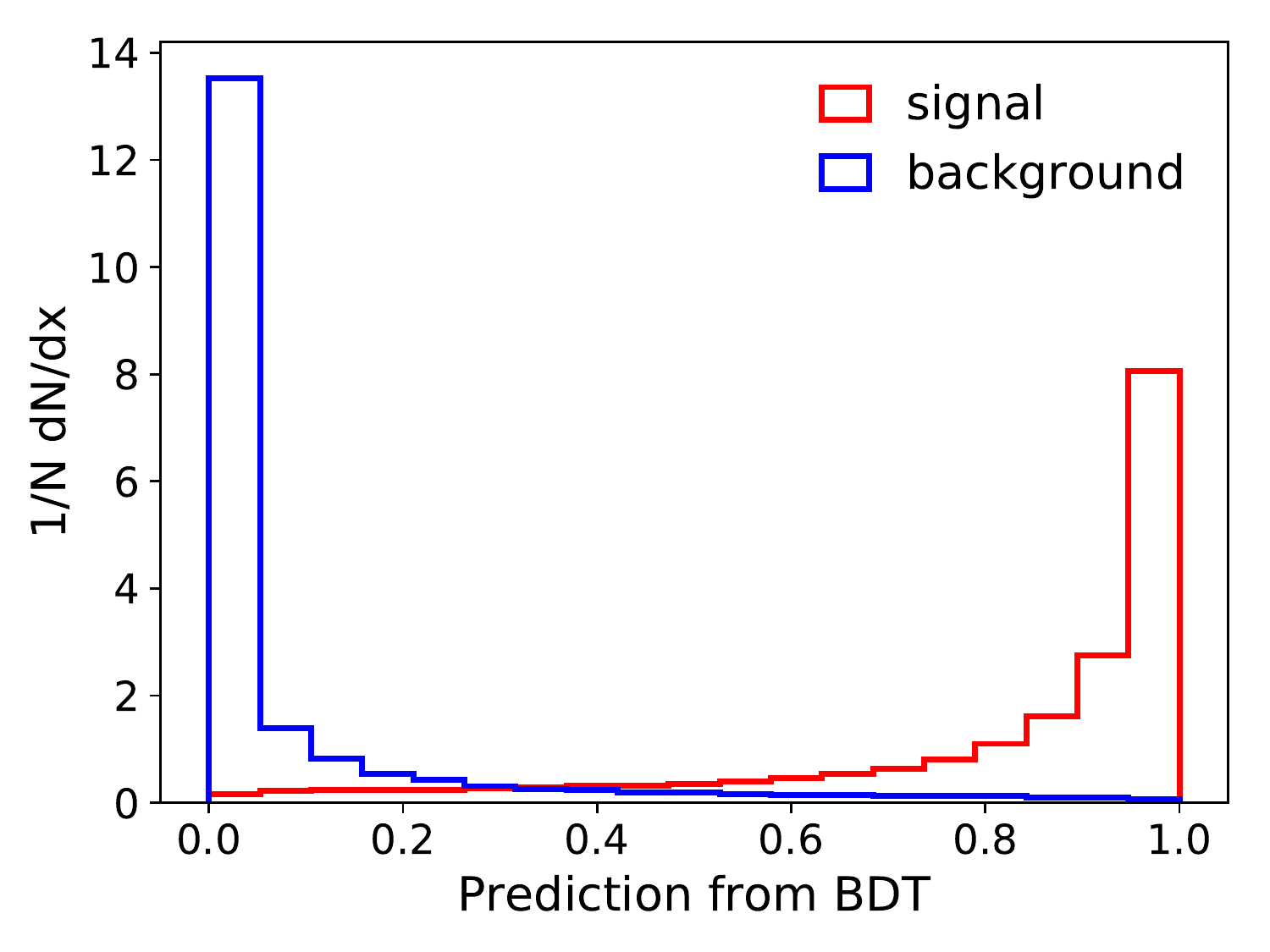} \\
\includegraphics[width=0.45\textwidth]{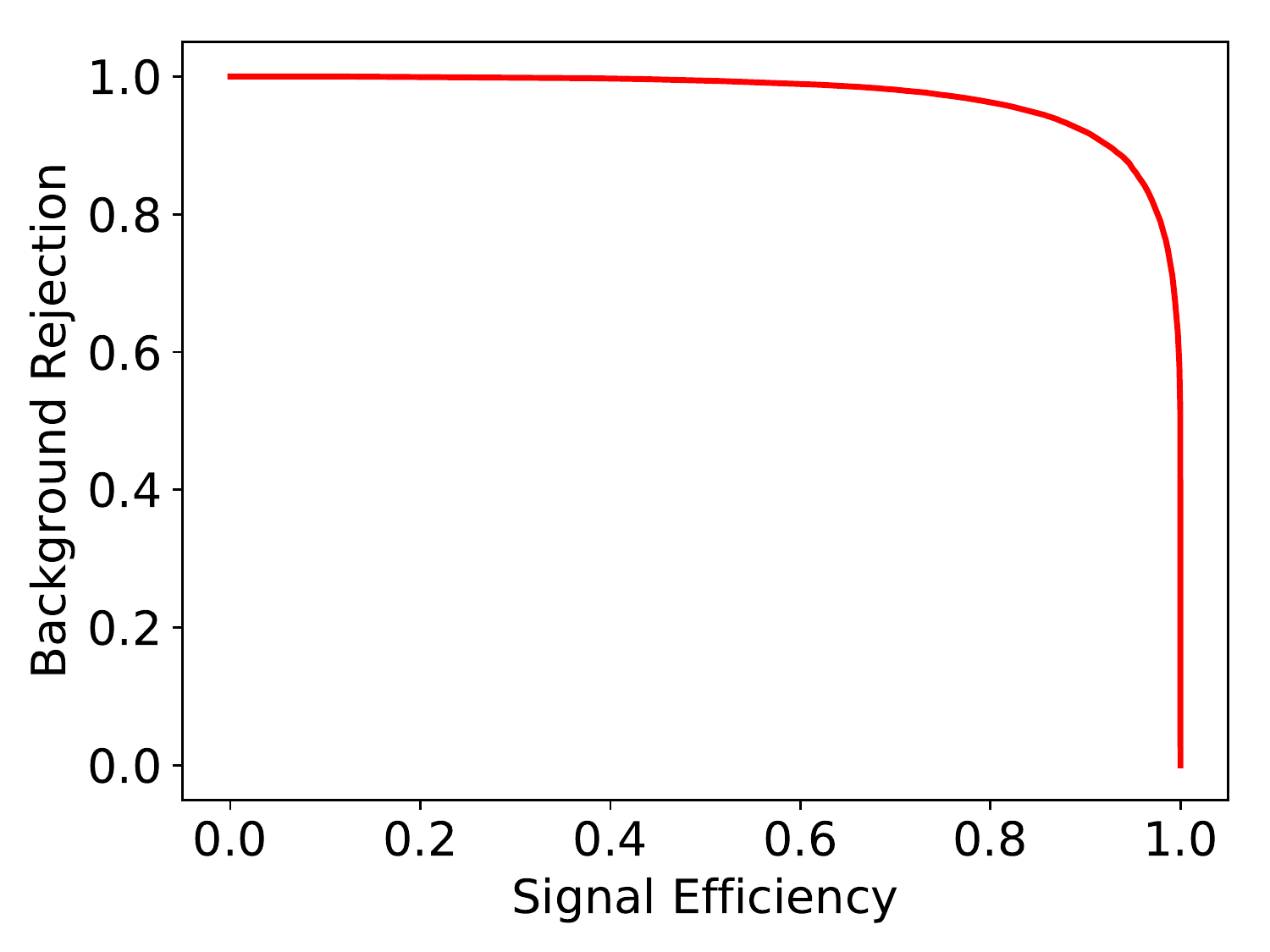}
\includegraphics[width=0.45\textwidth]{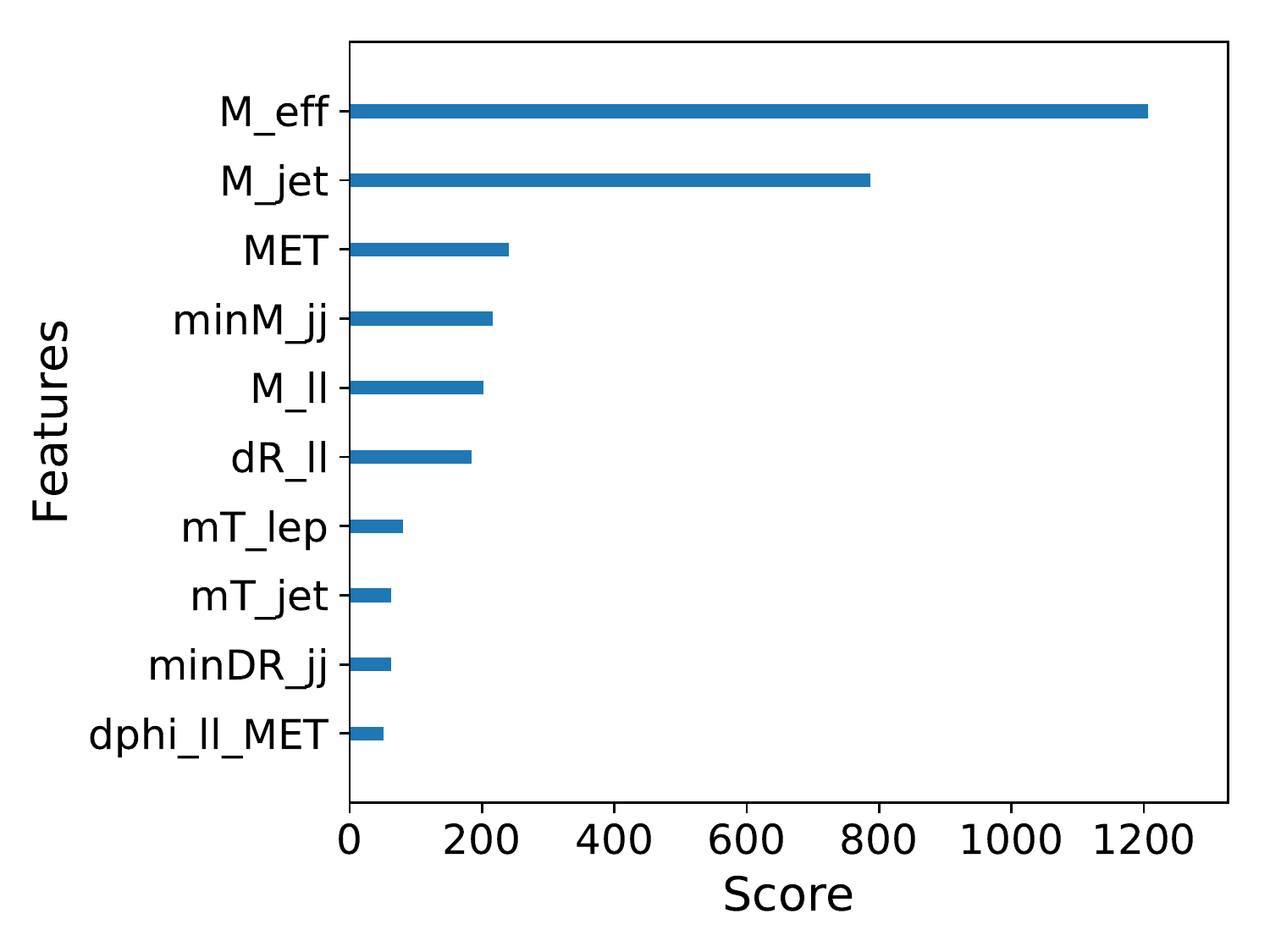}
\caption{\label{fig:BDT}BDT response (upper), ROC curve (lower left) and feature importance (lower right) for the small Yukawa coupling scenario  with $M_{H^{\pm\pm}} = 400$ GeV. In the feature importance plot, the variables from top to bottom are respectively $M_{\rm eff}$, $M_{\rm jets}$, $E_T^{\rm miss}$, ${\rm min} M_{jj}$, $M_{\ell\ell}$, $\Delta R_{\ell\ell}$, $M_T^{\ell\ell}$, $M_T^{\rm jets}$, ${\rm min} \Delta R_{jj}$ and $\Delta \phi(\ell\ell, E_T^{\text{miss}})$.
}
\end{figure}

In order to further control the backgrounds, we adopt the BDT technique. In particular, we use the {\tt XGBoost} package~\cite{Chen:2016btl} to build the BDT. In addition to the variables mentioned above, we also feed the BDT the following variables:
\begin{itemize}
    \item invariant mass $M_{\ell\ell}$ of same-sign dileptons;
    \item transverse mass $M_T^{\ell\ell}$ constructed from leptons and $E_T^{\rm miss}$;
    \item azimuthal angles $\Delta \phi(\ell_1,E_T^{\text{miss}})$ and $\Delta \phi(\ell_2,E_T^{\text{miss}})$ between leptons and $E_T^{\rm miss}$;
    \item azimuthal angle $\Delta \phi(j_1,E_T^{\text{miss}})$ between leading jet and $E_T^{\rm miss}$;
    \item separation $\Delta R_{\ell_1 j_1}$ and $\Delta R_{\ell_2j_1}$ of  leptons and leading jet;
    \item minimum separation $\text{min}\Delta R_{jj}$ of two jets;
    \item minimum separation  $\text{min}\Delta R_{\ell j}$ of leptons and jets;
    \item minimum invariant mass $\text{min}M_{jj}$ of two jets.
\end{itemize}
Some of the distributions, such as those for $\text{min}M_{jj}$, $M_{\ell\ell}$, $M_T^{\ell\ell}$ and $\text{min}\Delta R_{jj}$, are shown in Fig.~\ref{fig:additional_distributions}. We will see in the lower right panel of Fig.~\ref{fig:BDT} that these distributions are also very important for discriminating the signal from backgrounds.

The hyperparameters we used to train BDT are as follows: the learning rate is 0.1, the number of trees is 500, the maximum depth of each tree is 3, the fraction of events to train tree on is 0.6, the fraction of features to train tree on is 0.8, the minimum sum of instance weight needed in a child is 3, and the minimum loss reduction required to make a further partition on a leaf node of the tree is 0.2.

We split the data set into a training set and a testing set to make sure that there is no over-fitting. The BDT responses for our testing set are shown in the upper panel of Fig.~\ref{fig:BDT}. The BDT response close to 1 means the event is more signal-like while the response around 0 means the event is more background-like. We can see that our BDT classifier behaves quite good on the testing set. The receiver operating characteristic curve (ROC curve) of BDT and its feature importance are presented respectively in the lower left and right panels of Fig.~\ref{fig:BDT}. The feature importance is measured by ``gain'', which is defined as the average training loss reduction gained when using a feature for splitting. The importance plot shows the top 10 important variables in the BDT training.  The observables used in the cut-based analysis rank among the top 10 by the BDT, where the most important one is the effective mass $M_{\text{eff}}$, followed by $M_{\text{jets}}$ and $E_T^{\text{miss}}$. In addition, the BDT determines that the distributions $\text{min}M_{jj}$, $M_{\ell\ell}$, $M_T^{\ell\ell}$ and $\text{min}\Delta R_{jj}$ shown in Fig.~\ref{fig:additional_distributions} are also very important.

We choose the BDT cut such that it maximizes the significance of signal. For $M_{H^{\pm\pm}}=400\,\text{GeV}$, the event yields of signal and backgrounds after the BDT cut are reported in Table~\ref{tab:events}.   We can see that the BDT can eliminate backgrounds significantly while keeping most of the signal. The significance can reach 10.36 with the help of BDT, which is improved remarkably in comparison to the cut-based method in Section~\ref{sec:cut}.


\subsubsection{Mass reaches}
\label{sec:mass}

\begin{figure}[htbp]
\centering
\includegraphics[width=0.5\textwidth]{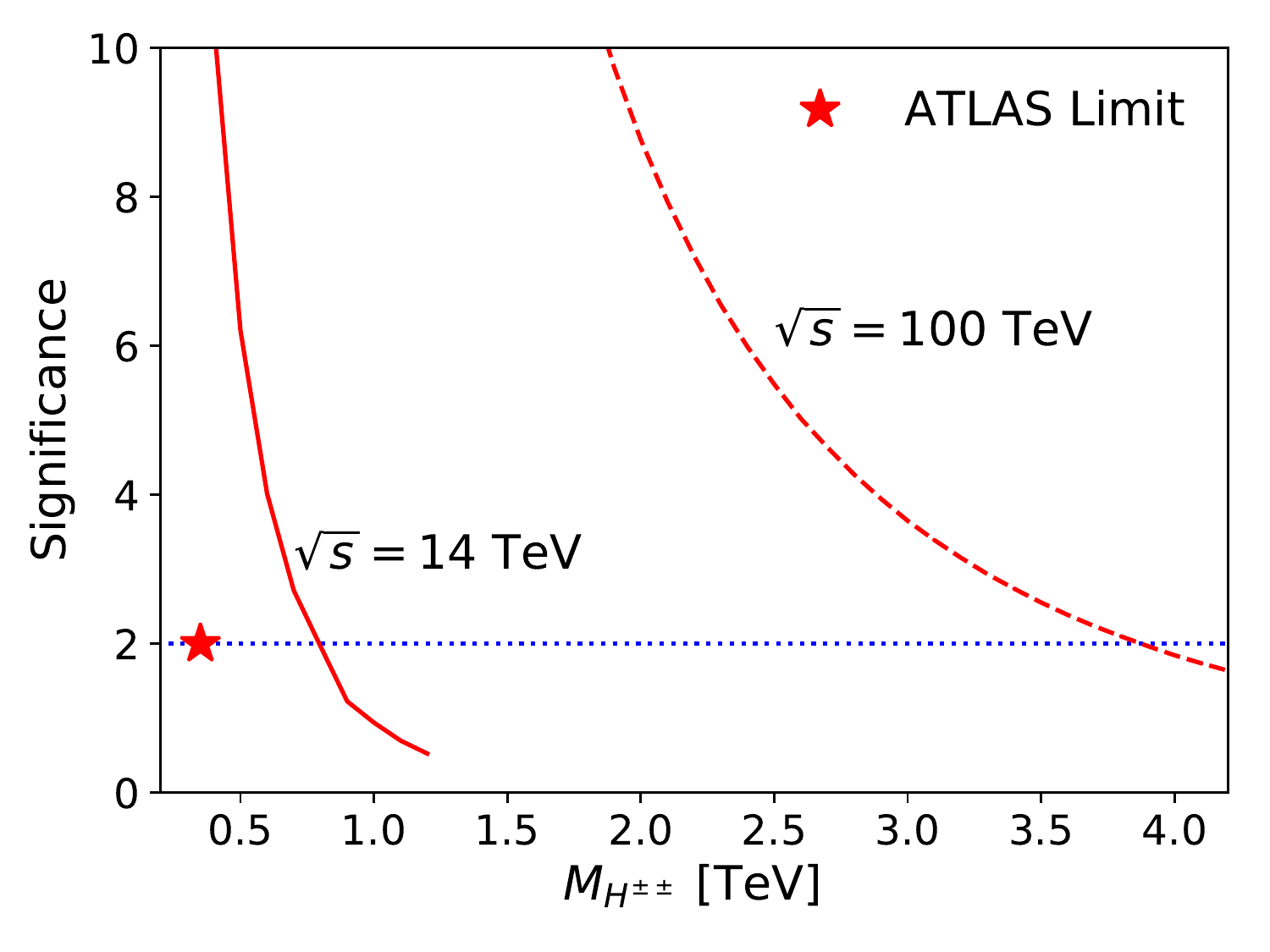}
\caption{\label{fig:significance}
BDT significance as a function of $M_{H^{\pm\pm}}$ at the HL-LHC (solid) and future 100 TeV collider (dashed) for the small Yukawa coupling scenario. The red star is current LHC $2\sigma$ limit on $M_{H^{\pm\pm}}$ in the $W^\pm W^\pm$ channel~\cite{Aad:2021lzu}. }
\label{Hpppairsignificance}
\end{figure}

To explore the discovery potential of $H^{\pm\pm}$ in the small Yukawa coupling scenario at the HL-LHC, we generate event samples for the signal process for $M_{H^{\pm\pm}}$ in the range from $300\,\text{GeV}$ to 1.2 TeV with the step of $100\,\text{GeV}$. We build BDTs for different masses to discriminate the signal from the SM backgrounds and maximize the significance.  The significance as a function of the doubly-charged scalar mass $M_{H^{\pm\pm}}$ is shown in Figure~\ref{fig:significance} as the solid line. It is found that we can reach $M_{H^{\pm\pm}} \simeq 800\,\text{GeV}$  at the $2\sigma$ significance in the $W^\pm W^\pm \phi$ channel for the small Yukawa scenario at the HL-LHC.

At future 100 TeV hadron colliders such as FCC-hh and SPPC, the production cross section of $H^{\pm\pm}$ can be largely enhanced, as shown in Fig.~\ref{fig:cross_section}. Following the same BDT analysis as that at  14 TeV LHC, the significance of signal as a function of $M_{H^{\pm\pm}}$ is presented as the dashed line in Figure~\ref{fig:significance}. Benefiting from the large cross section, the prospect of $M_{H^{\pm\pm}}$ can reach up to 3.8 TeV at the $2\sigma$ sensitivity at the 100 TeV collider.


\subsection{Large Yukawa coupling scenario}
\label{sec:large}

Another case of interest in contrast to the previous one is the large Yukawa coupling scenario. According to the low-energy flavor limits in
Table~\ref{tab:limits}, most elements of the Yukawa coupling matrix $Y_{\alpha\beta}$ are bounded to be small while $Y_{\mu\mu}$ can be of ${\cal O}(1)$ for TeV-scale $H^{\pm\pm}$. Note that the effective coupling between neutrinos and leptonic scalars ($H_1$ and $A_1$) in our model is of order $\lambda_{\alpha \beta} \sim 2\sqrt{2} \, Y_{\alpha \beta} \, \sin\theta$ (cf.~Table~\ref{keyvertices1}); therefore, $Y_{\mu\mu}\sim {\cal O}(1)$ could also be probed at hadron colliders via the VBF process discussed in our previous study~\cite{deGouvea:2019qaz}. For example, a $Y_{\mu \mu }=1.5$ Yukawa coupling leads to an effective coupling $\lambda_{\mu \mu} \sim 0.58$ which is within the $2\sigma$ LHC sensitivity in the VBF mode~\cite{deGouvea:2019qaz}. Although the $Y_{\tau\tau}$ coupling is the least constrained (cf.~Table~\ref{tab:limits}), final states involving taus at the hadron colliders are more difficult to analyze; therefore, we only focus on the muon final states and leave the tau signal for a future work.

After considering the constraints from perturbativity and unitarity in Section~\ref{sec:Limits2}, we found that the $Y_{\mu\mu}$ component can be as high as 1.5 as presented in Fig.~\ref{fig:pert}. This is still consistent with the muon $g-2$ bound given in Table~\ref{tab:limits} for a TeV-scale $H^{\pm\pm}$. In this scenario, the contributions from other Yukawa coupling elements are negligible, and the doubly-charged scalar $H^{\pm\pm}$ decays predominately into a pair of same-sign muons, {\it i.e.} $\mathrm{BR}(H^{\pm\pm} \rightarrow \mu^\pm \mu^\pm) \simeq 100\%$. For large $Y_{\mu\mu}$ the main decay channel for the singly-charged scalar will be $H^\pm \rightarrow \mu^\pm \, \nu$. However, the $H^\pm \rightarrow W^\pm \, \phi$ channel is still feasible and its BR varies from 10\% to 20\% depending on the mass of $H^\pm$, as shown in the lower left panel of Fig.~\ref{fig:br_Hpp}. With the $W$ boson decaying hadronically, the $\phi$ induced signal at the hadron collider emerges from the associated production channel as follows:
\begin{align*}
p p \rightarrow H^{\pm\pm}(\rightarrow \mu^{\pm} \mu^{\pm}) \, H^{\mp}(\rightarrow W^{\mp} \phi) \rightarrow \mu^{\pm} \mu^{\pm}+2\,\text{jets} + E_T^{\text{miss}} \,,
\end{align*}
{\it i.e.}~same-sign muon pair plus two jets from $W$ boson decay plus transverse missing energy from $\phi$. We should mention here that the traditional 3-$\mu$ or 4-$\mu$ channels will still be the discovery mode for this scenario, but our choice of the final state and analysis is useful to determine the mass of leptonic scalar $\phi$ ($H_1/A_1$) as will be shown in Section~\ref{sec:phimass}.

The same-sign dilepton signals are ``smoking-gun'' signals of doubly-charged scalars at the high-energy colliders, and have been searched for at the LEP~\cite{OPAL:2001luy, DELPHI:2002bkf, L3:2003zst}, Tevatron~\cite{CDF:2004teg, CDF:2008vdv, D0:2008qnv, D0:2011eug}, LHC data at 7 TeV~\cite{ATLAS:2011rha, CMS:2011sqa}, 8 TeV~\cite{ATLAS:2014kca, CMS:2016cpz} and 13 TeV~\cite{Aaboud:2017qph, CMS:2017pet}.
For the scenario ${\rm BR} (H^{\pm\pm} \to \mu^\pm \mu^\pm) = 100\%$, the current most stringent lower dilepton limit on $M_{H^{\pm\pm}}$ is from the LHC 13 TeV data, being $846 \, \text{GeV}$~\cite{Aaboud:2017qph}.
For illustration purpose, we use
\begin{eqnarray}
\label{eqn:benchmark}
M_{H^{\pm\pm}} = 900\,\text{GeV} \,, \quad
M_{H^\pm} = 893 \, \text{GeV}
\end{eqnarray}
as our benchmark scenario for the analysis below.

\subsubsection{Analysis and mass reaches}

\begin{figure}[!t]
\centering
\includegraphics[width=0.5\textwidth]{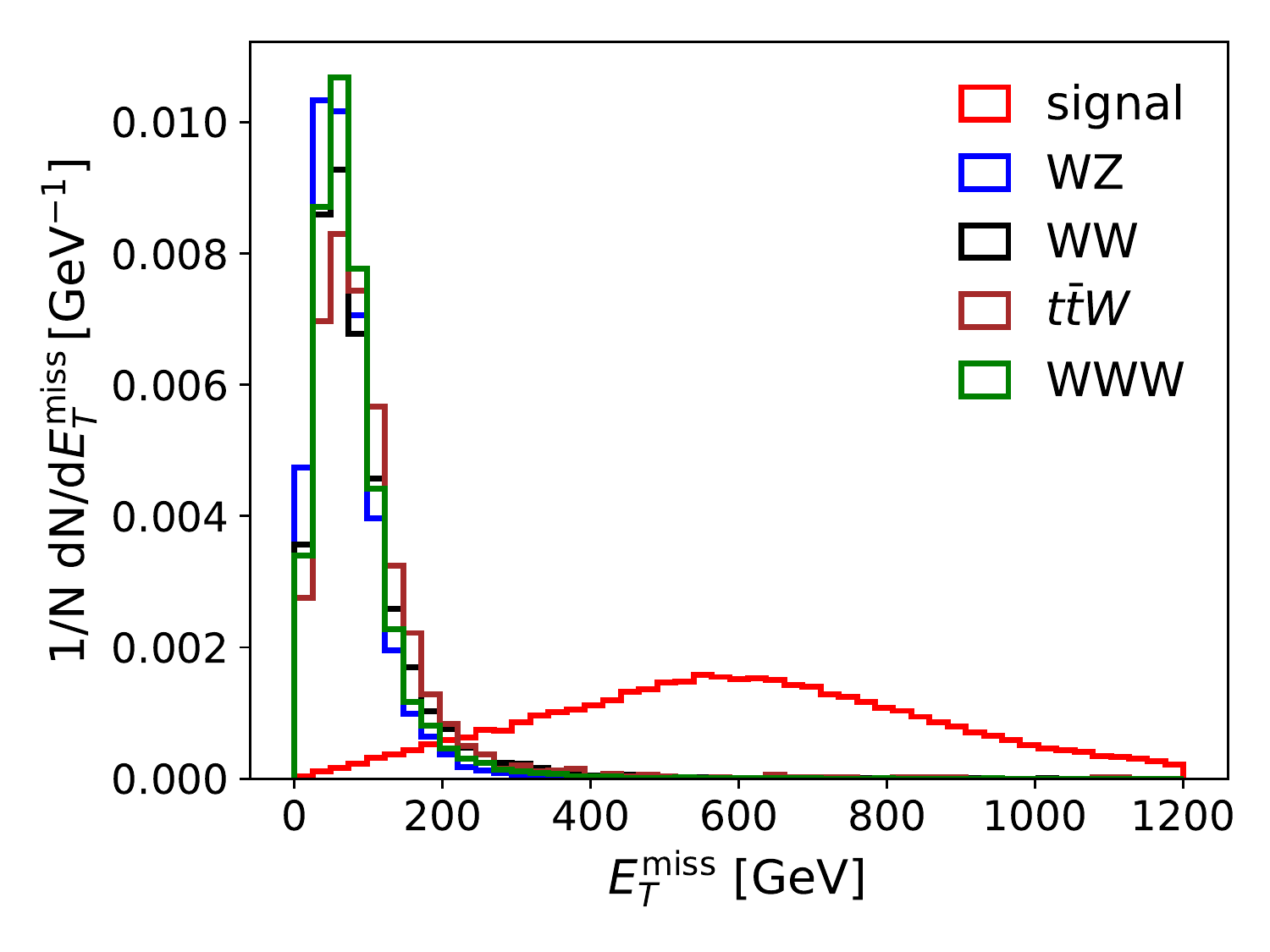} \\
\includegraphics[width=0.49\textwidth]{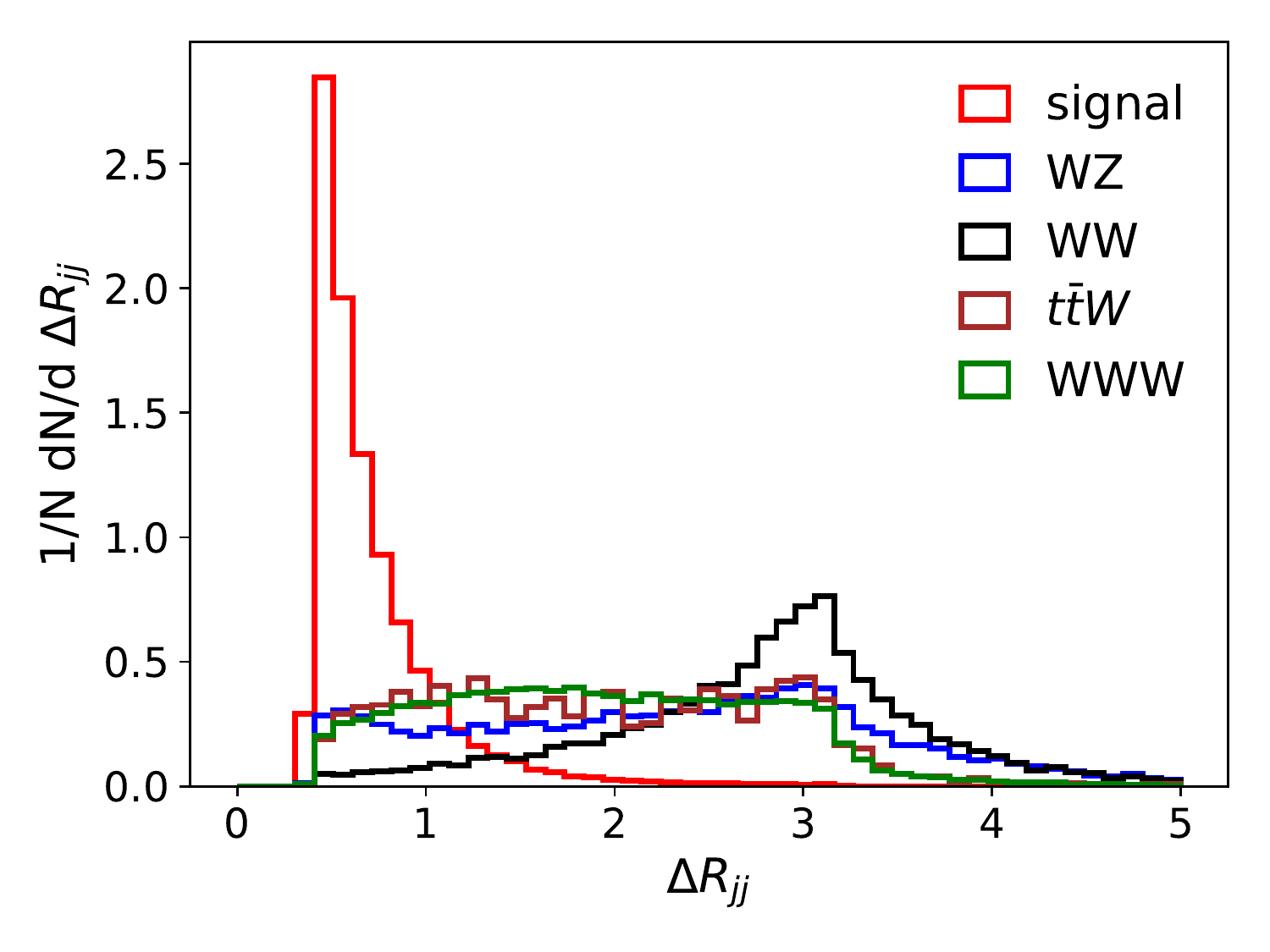}
\includegraphics[width=0.49\textwidth]{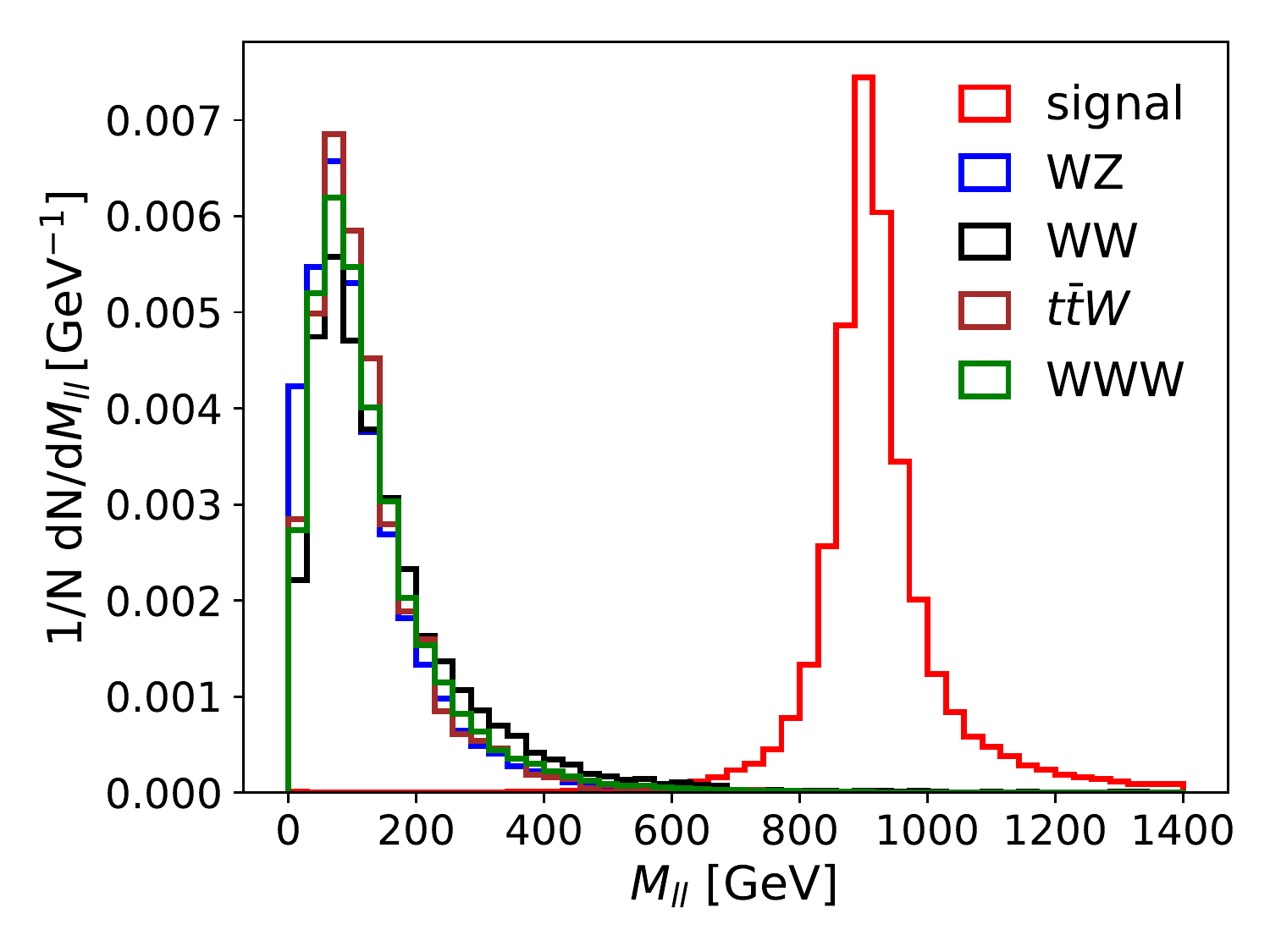}
\caption{\label{fig:ap_distributions}
Distributions of $E_T^{\text{miss}}$ (upper), $\Delta R_{j j}$ (lower left) and $M_{\mu^\pm \mu^\pm}$ (lower right) in associated production $H^{\pm\pm} H^{\mp}$ and the SM backgrounds $WZ$, $WW$, $t \bar{t}W$, $WWW$.}
\end{figure}

The signal samples are generated by using {\tt MadGraph5}. Since the final state is similar to the small Yukawa coupling case, we use the same background samples as in Section~\ref{sec:small}. The muon and jet definitions are also kept unchanged. All the events are required to have two reconstructed same-sign muons and two jets without any $b$-tagged jet. In addition, to further control the backgrounds the following cuts are applied, and the corresponding cut-flows for the cross sections of signal and backgrounds are presented in Table~\ref{tab:cut_flow2}.
\begin{itemize}
\item $\text{min} \Delta R_{\mu j} >0.4$ and $ \Delta R_{\mu^{} \mu^{}} > 0.3$. This is to satisfy the muon isolation criteria.

\item $ E_T^{\text{miss}} > 200 \, \text{GeV}$. Since  $E_T^{\text{miss}}$ in the signal is from the scalar $\phi=H_1,\;A_1$, it tends to have a larger value than the backgrounds with a broader distribution, as shown in the upper panel of Fig.~\ref{fig:ap_distributions}.

\item $\Delta R_{j j} <2 $. The two jets in the signal are from the decay products of a very energetic $W$ boson, so they tend to be more collimated than the backgrounds. With the distributions shown in the lower left panel of Fig.~\ref{fig:ap_distributions}, a small $\Delta R_{j j}$ can help us to reduce the backgrounds.

\item $ 700\,\text{GeV} < M_{\mu^\pm \mu^\pm} < 1100\,\text{GeV}$. Since the same-sign muon pair appears from the decay of the $H^{\pm \pm}$ boson, their Breit–Wigner peak provides a strong discrimination against the SM backgrounds. This can be clearly seen in the lower right panel of Fig.~\ref{fig:ap_distributions}.
\end{itemize}

\begin{table}[tb]
\begin{center}
\caption{
\label{tab:cut_flow2}
Cut-flow of the cross sections for signal and SM backgrounds $WZ$, $WW$, $t \bar{t} W$, $WWW$ at the HL-LHC for the large Yukawa coupling scenario~(\ref{eqn:benchmark}). Backgrounds that are essentially eliminated are denoted by ``$-$''s.}
\vspace{5pt}
\begin{tabular}{ |c|c|c c c c| }
 \hline
 Cut Selection& \specialcell{Signal \\ $[\text{fb}]$} &\specialcell{$WZ$ \\ $[\text{fb}]$} & \specialcell{$WW$\\ $[\text{fb}]$}  & \specialcell{$t\bar{t}W$ \\ $[\text{fb}]$} & \specialcell{$WWW$  \\ $[\text{fb}]$}   \\ \hline
 \hline
$\text{min} \Delta R_{\mu j} >0.4$ and  $ \Delta R_{\mu^{} \mu^{}} > 0.3$ & 0.0059 & 1.7 & 0.81 & 0.044 & 0.27 \\ \hline
$ E_T^{\text{miss}} > 200 \, \text{GeV}$ & 0.0056 & 0.036 & 0.049 & 0.0027& 0.010 \\ \hline
$\Delta R_{j j} <2 $ & 0.0054 & 0.017 & 0.013 & 0.0019 & 0.0082 \\ \hline
 $ 700\,\text{GeV} < M_{\mu^\pm \mu^\pm} < 1100\,\text{GeV}$ & 0.050 & 0.00010 & 0.00015 & $-$ & 0.00019 \\ \hline
\end{tabular}
\end{center}
\end{table}

As a result of very distinct topologies of the signal and backgrounds, the number of background events can be highly suppressed after the cuts, as reported in Table~\ref{tab:cut_flow2}. The expected numbers of events at the HL-LHC are shown in Table~\ref{tab:ap_events}. In the cut-based analysis,   the significance can reach $\sigma = 3.67$ for the benchmark scenario in Eq.~(\ref{eqn:benchmark}).
\begin{table}[!t]
\begin{center}
\caption{\label{tab:ap_events}
Number of events in cut-based and BDT analysis for associated production $H^{\pm\pm} H^{\mp}$ in the benchmark scenario (\ref{eqn:benchmark}) and the SM backgrounds at the HL-LHC with 3 ab$^{-1}$ luminosity. The last column shows the significance of signal. Backgrounds that are essentially eliminated by our cuts are denoted by ``$-$''s.}
\vspace{5pt}
\begin{tabular}{ |c|c|c c c c|c||c| }
 \hline
 & Signal & $WZ$ & $WW$ & $t\bar{t}W$ & $WWW$ & Backgrounds & $\sigma$  \\ \hline
\specialcell{Number of events \\ (Cut-based)}& 14.87 & 0.32 & 0.46 & $-$ & 0.57 & 1.35 & 3.69  \\ \hline
\specialcell{Number of events \\ (BDT-based)} & 19.00 & $-$ & $-$ & $-$ & 0.06 & 0.06& 4.35 \\
 \hline
\end{tabular}
\end{center}
\end{table}

\begin{figure}[!t]
\makebox[\linewidth][c]{%
\centering
\includegraphics[width=0.5\textwidth]{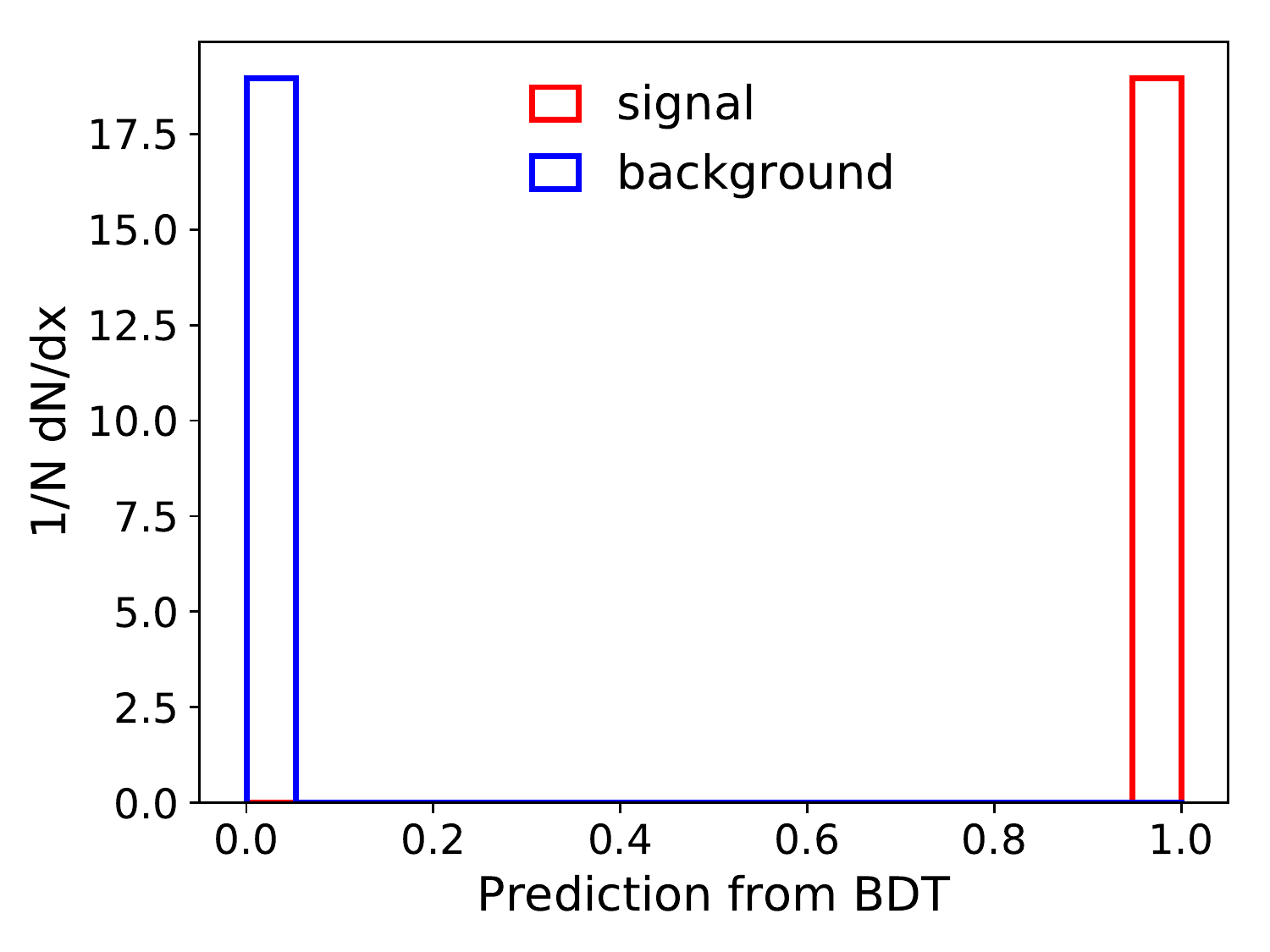}
}
\caption{\label{fig:BDT_M900}
BDT score distribution for the large Yukawa coupling scenario. }
\end{figure}

As in the small Yukawa coupling case in Section~\ref{sec:small}, BDT can help us improve to some extent the sensitivity. In addition to the observables above in cut-and-count analysis, we also use the following observables:
\begin{itemize}
    \item transverse momenta  $p^{}_{T,\,\mu_1}$ and $p^{}_{T,\,\mu_2}$ of the two muons;
    \item effective mass $M_{\text{eff}}$;
    \item invariant mass $M_{jj}$ of two jets;
    \item total transverse momentum $p^{}_{T,\,jj}$ of two jets;
    \item transverse mass $M_T$ constructed from jets and $E_T^{\text{miss}}$;
    \item azimuthal angle $ \Delta \phi(\mu \mu, E_T^{\text{miss}})$ between two muons and $E_T^{\text{miss}}$.
\end{itemize}
The BDT score distribution is presented in Fig.~\ref{fig:BDT_M900}. As expected, the signal is well separated from the backgrounds.  Therefore the BDT can eliminate almost all the background events while keeping most of the signal events. The expected numbers of signal and background events after optimal BDT cuts are collected in the last row of Table~\ref{tab:ap_events}. With the help of  BDT, the sensitivity can reach a higher value at $\sigma = 4.35$.

\begin{figure}[!t]
\makebox[\linewidth][c]{%
\centering
\includegraphics[width=0.5\textwidth]{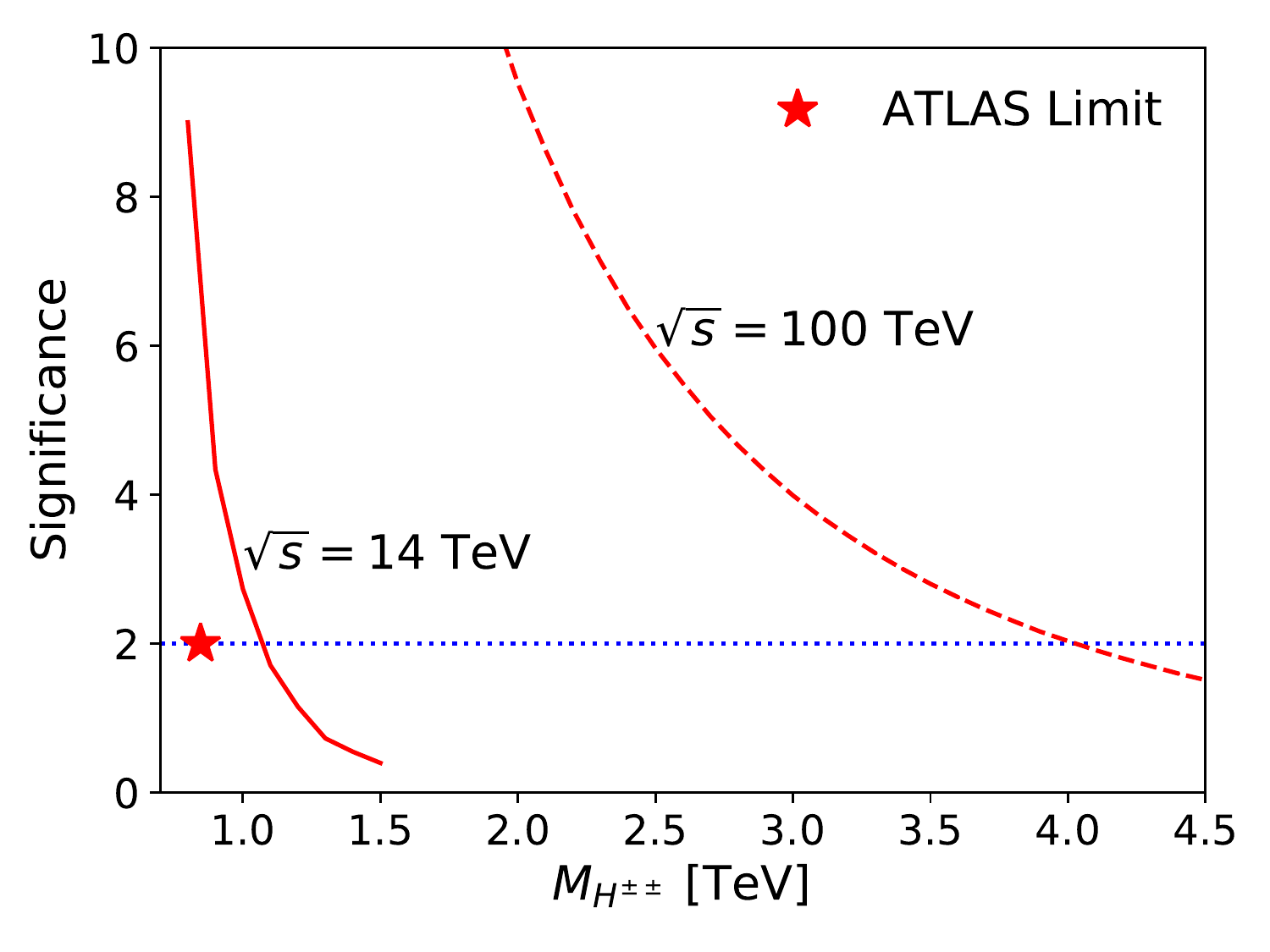}
}
\caption{\label{fig:significance2}
BDT significance as a function of $M_{H^{\pm\pm}}$ at the HL-LHC (solid) and future 100 TeV collider (dashed) for the large Yukawa coupling scenario. The red star indicates the current LHC $2\sigma$ limit on $M_{H^{\pm\pm}}$  with $100\%$ BR into $\mu^\pm \mu^\pm$~\cite{Aaboud:2017qph}.
}
\end{figure}

Since the backgrounds can be highly suppressed 
by the BDT analysis, the significance will be mainly determined by the cross section of signal, which in turn depends on the mass of $H^{\pm\pm}$. We generate our signal samples in the step of $100\,\text{GeV}$ for $M_{H^{\pm\pm}}$ varying from $900\,\text{GeV}$ to $1.5\,\text{TeV}$. The resultant significance at the HL-LHC as a function of $M_{H^{\pm\pm}}$ is shown in Fig.~\ref{fig:significance2} as the solid line. It turns out $H^{\pm\pm}$ can be probed up to 1.1 TeV at the $2\sigma$ sensitivity at the HL-LHC in the large Yukawa coupling scenario. At a future 100 TeV collider, the production cross section $\sigma (pp \to H^{\pm\pm} H^\mp)$ can be enhanced by over one order of magnitude (see Fig.~\ref{fig:cross_section}). The corresponding prospect of $M_{H^{\pm\pm}}$ can reach up to 4 TeV at the $2\sigma$ sensitivity, which is indicated by the dashed line in Fig.~\ref{fig:significance2}.


\subsubsection{Mass determination of the leptonic scalar $\phi$}
\label{sec:phimass}

For the associated production $H^{\pm\pm} H^{\mp}$ in the large Yukawa coupling case, the only missing particles is $\phi=H_1$, $A_1$, which provides a  possibility to measure its mass. However, at the hadron colliders such as LHC, we can at most determine the  transverse momentum of $\phi$ while its longitudinal momentum is completely lost. Therefore the usual method to determine a particle's mass is not applicable here. An alternative approach is to utilizes the transverse mass of a mother particle whose decay products contain a massive invisible daughter particle. To achieve this, we need to modify the definition of transverse mass in Eq.~(\ref{eq:M_T}). In that equation, we do not consider the mass of the missing particles but simply assume the transverse energy of missing particles to be the same as the missing transverse momentum. The modified definition of missing transverse energy is
\begin{eqnarray}
E_T^{\text{miss}} (\tilde{m}) = \sqrt{\tilde{m}^2 + p_{T,\,\text{miss}} ^2} \,,
\end{eqnarray}
where $\tilde{m}$ is the assumed mass of $\phi$, and $p_{T,\,\text{miss}}$ is the missing transverse momentum. Thus the cluster transverse mass $M_T$ can be re-expressed as a function of the assumed mass $\tilde{m}$:
\begin{align}
M_T(\tilde{m})=\left[ \left( \sqrt{M_{\text{jets}}^2 + \bigg| \sum_{j} \overrightarrow{p}^j_T \bigg|^2} + \sqrt{\tilde{m}^2 + p_{T,\,\text{miss}} ^2} \right)^2 - \Bigg| \sum_{j} \overrightarrow{p}^j_T + \overrightarrow{p}_{T,\,\text{miss}} \Bigg|^2 \right]^{1/2}.
\label{eq:realM_T}
\end{align}
As shown in Refs.~\cite{Gripaios:2007is,Barr:2007hy}, the endpoint of $M_T$ distribution will increase with the assumed mass $\tilde{m} $, and a kink will appear at the point of $\tilde{m} = m$ when the assumed mass $\tilde{m}$ is equal to the real mass $m$ of the invisible daughter particle.

\begin{figure}[!t]
\centering
\includegraphics[width=0.6\textwidth]{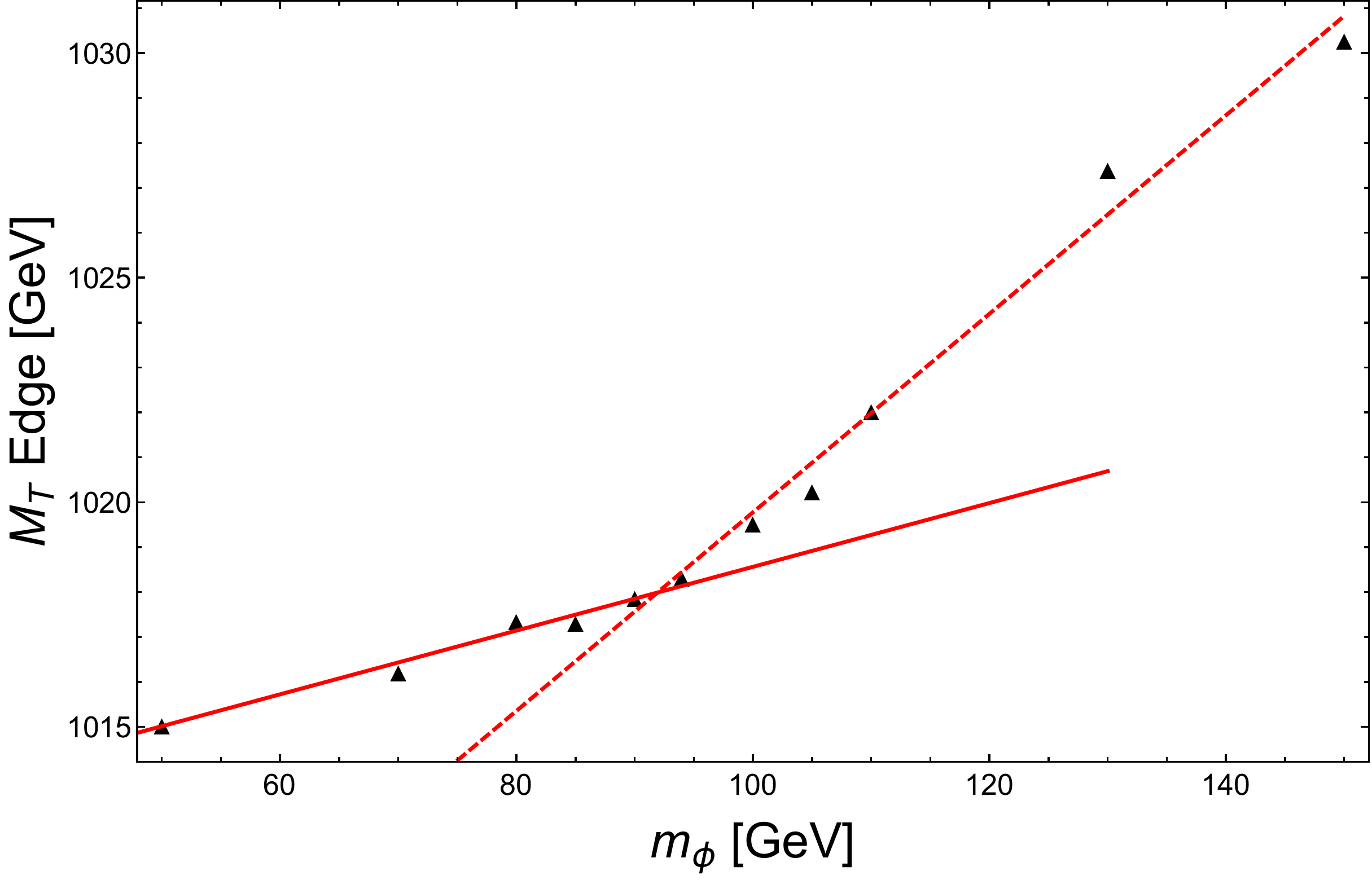}
\caption{\label{fig:edgeVSmphi}
$M_T$ endpoints (black triangles) from  {\tt EdgeFinder} fitting~\cite{Curtin:2011ng} as a function of assumed trial mass $m_{\phi}$. The red straight lines are from linear fittings as an illustration of kink position.
}
\end{figure}

As an explicit example, we choose the scalar mass $m_{\phi} = 89.28\, \text{GeV}$, and the masses of charged scalars are set as in Eq.~(\ref{eqn:benchmark}).
We calculate the transverse mass $M_T$ of the simulated events by Eq.~(\ref{eq:realM_T}) with different choices of $\tilde{m}$, and then use package {\tt EdgeFinder}~\cite{Curtin:2011ng} to find the endpoint of $M_T$ distribution for each $\tilde{m}$ choice. The result is shown in Fig.~\ref{fig:edgeVSmphi}. By fitting the data points, a kink is found at $\tilde{m} = (93.60 \pm 11.43) \, \text{GeV} $. Comparing $\tilde{m}$ at the kink with the real mass $m_{\phi}$, we find that this method provides a great potential for measuring the mass of the invisible light scalar $\phi = H_1,\, A_1$ at the LHC.

We note that the fitting process may be associated with some 
uncertainties for both  $M_T$ edges and $m_\phi$. To test the robustness of fitting result, we smear the $M_T$ edge according to the initial error bars from the {\tt EdgeFinder} package in a Normal distribution. Using 100 points for  trial, we find that the mass determination by the kink yields a result $\tilde{m} = (93.55 \pm 11.41) \, \text{GeV}$. Since the uncertainty range does not change,  we can state that the kink-finding method leads to a rather reliable mass  determination. 
It should be noted that it is difficult to apply the mass determination technique used here to the small Yukawa coupling scenario in Section~\ref{sec:small}, since in that case $\phi$ is from $H^{\pm\pm}$ decay, which leads to the appearance of missing energy from both neutrinos from $W$ boson decay and the invisible scalar $\phi$.


\subsection{Intermediate Yukawa coupling scenario}
\label{sec:medium-yukawa}

\begin{figure}[!hbt]
    \makebox[\linewidth][c]{%
    \centering
    \includegraphics[width=0.5\textwidth]{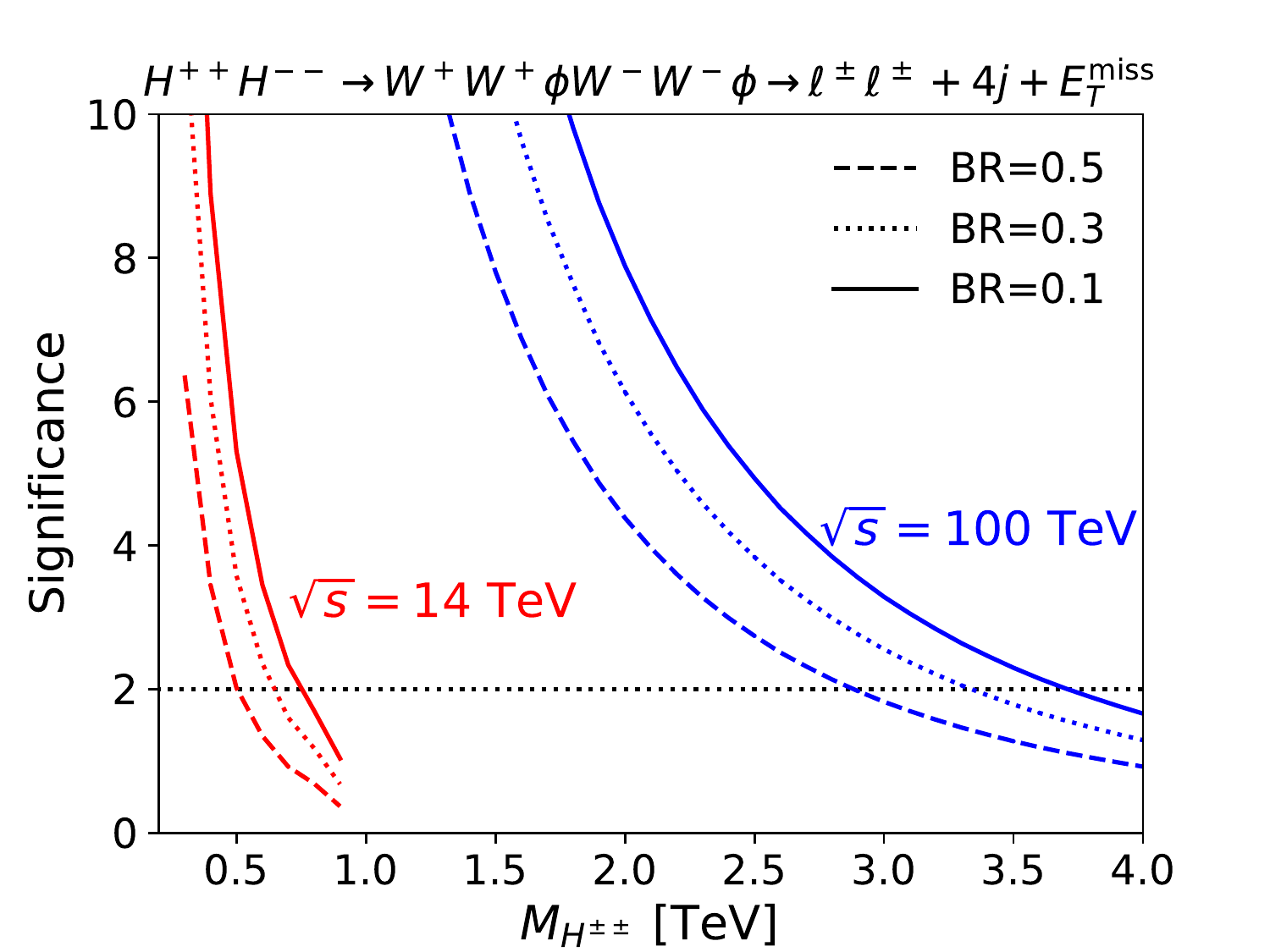}
    \includegraphics[width=0.5\textwidth]{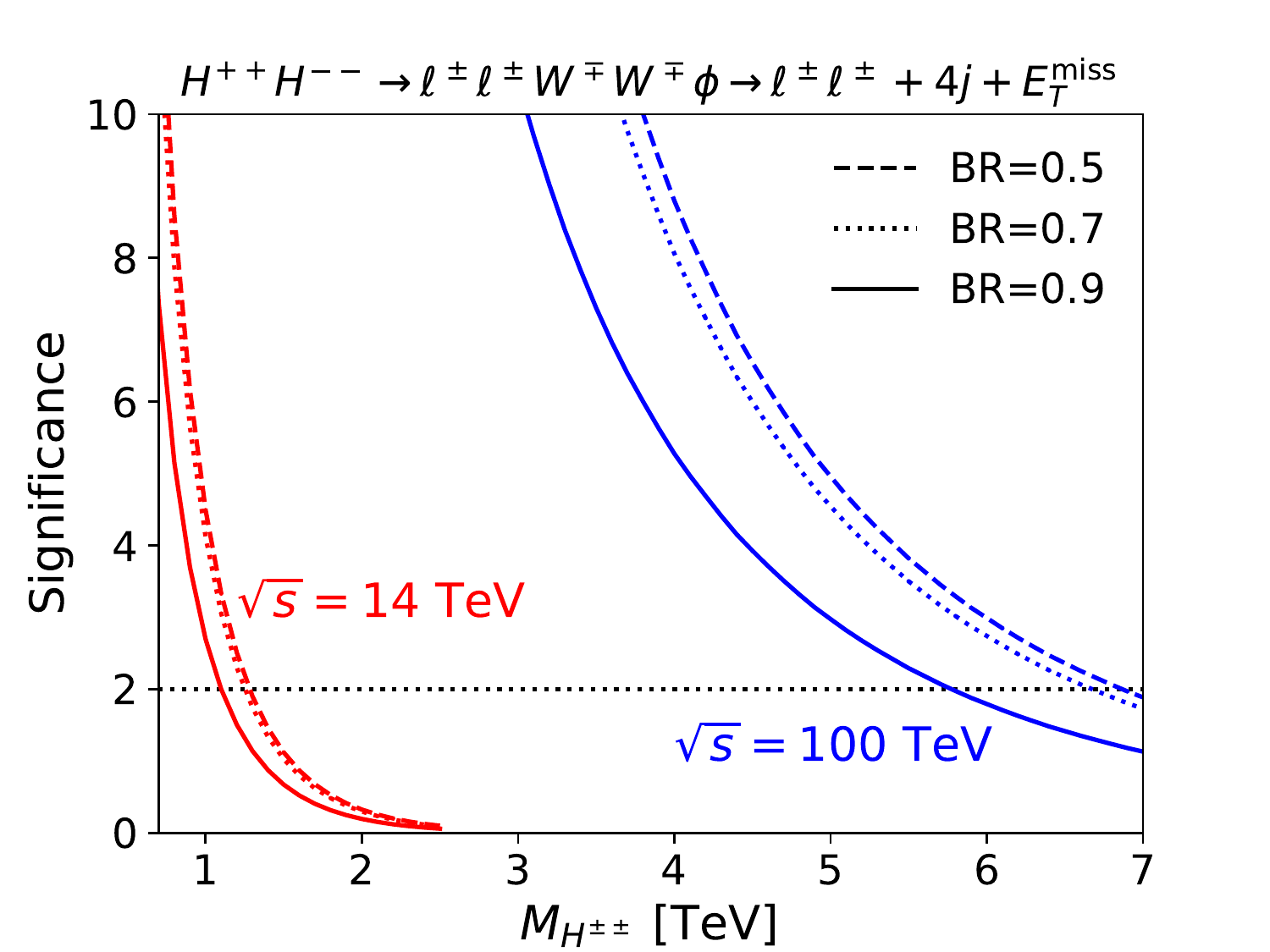}
    }
    \makebox[\linewidth][c]{%
    \centering
    \includegraphics[width=0.5\textwidth]{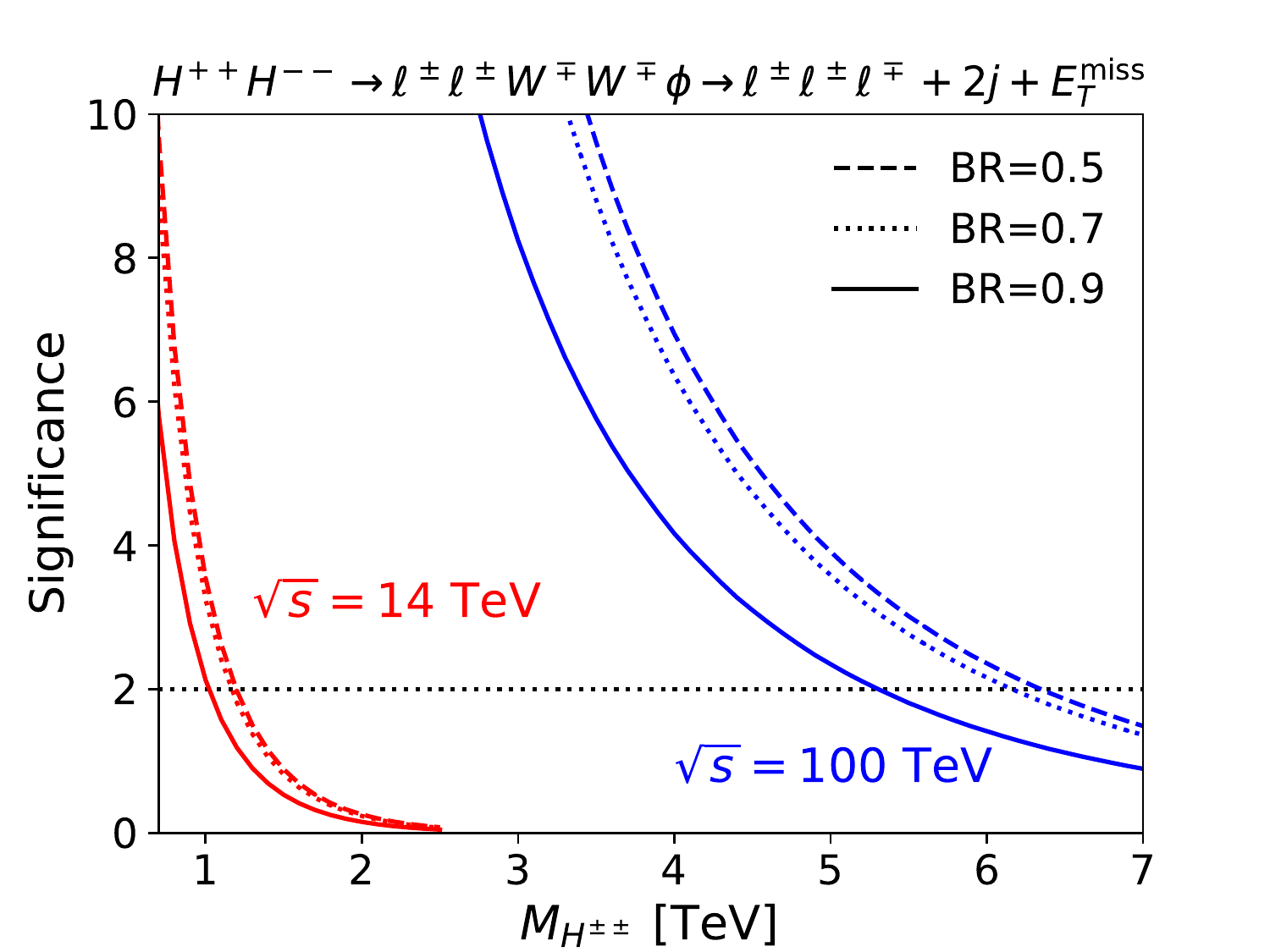}
    \includegraphics[width=0.5\textwidth]{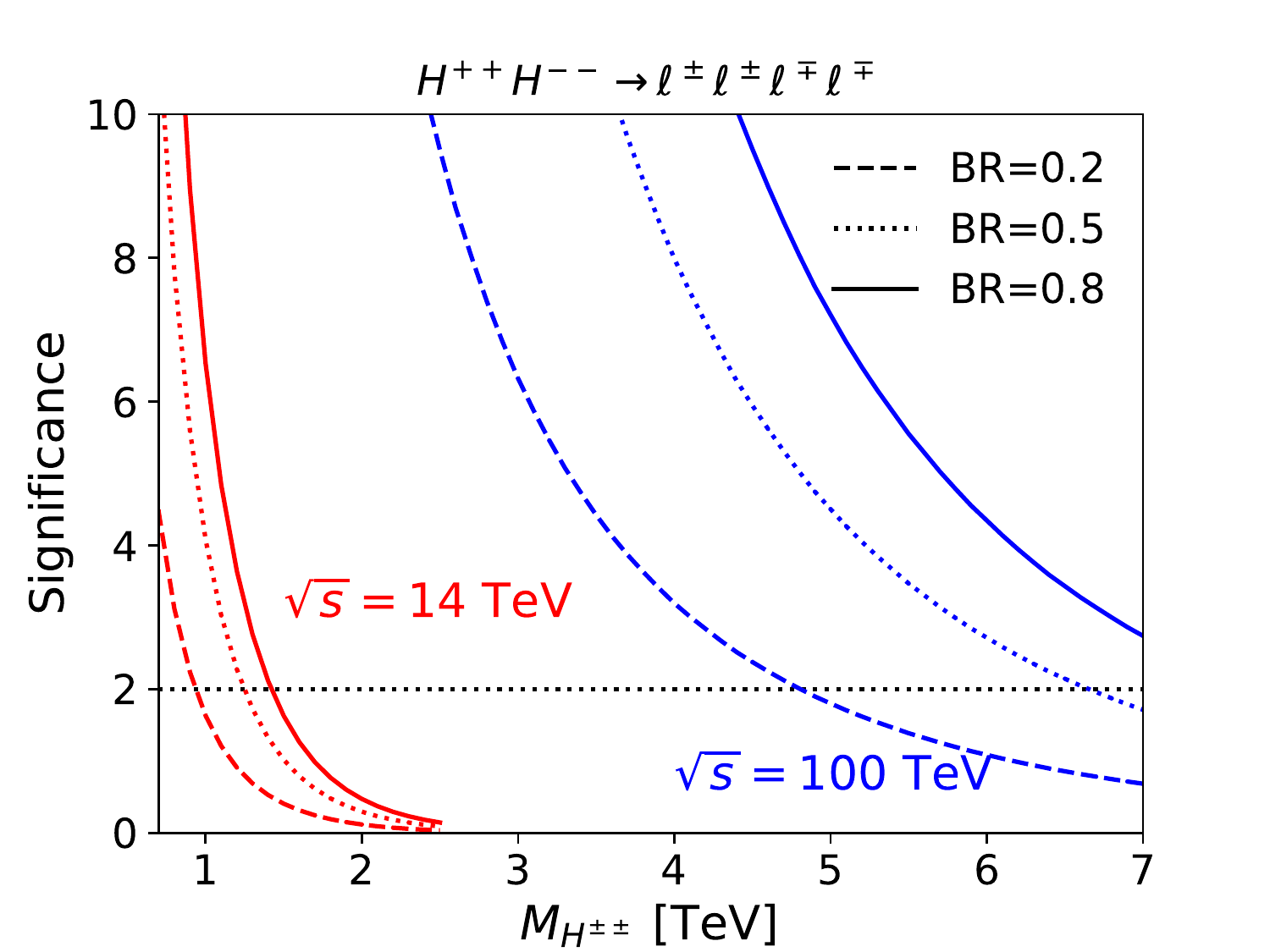}
    }
    \caption{\label{fig:intermediate_yukawa}
    Significance as a function of $M_{H^{\pm\pm}}$ at the HL-LHC (red) and future 100 TeV collider (blue) for the intermediate Yukawa coupling scenario, in the channels of $H^{++} H^{--} \to W^+ W^+ \phi W^- W^- \phi \to \ell^\pm \ell^\pm + 4j + E_T^{\rm miss}$ (top left), $\ell^\pm \ell^\pm W^\mp W^\mp \phi \to \ell^\pm \ell^\pm + 4j + E_T^{\rm miss}$ (top right), $\ell^\pm \ell^\pm W^\mp W^\mp \phi \to \ell^\pm \ell^\pm \ell^\mp + 2j + E_T^{\rm miss}$ (bottom left) and $\ell^+ \ell^+ \ell^- \ell^-$ (bottom right). The ``BR'' in all the legends refers to the leptonic decay branching fraction $\mathrm{BR}(H^{\pm \pm} \rightarrow \ell^\pm \ell^\pm)$ of the doubly-charged scalar. }
\end{figure}

For the completeness of our study, we also investigate the mass reach in the intermediate Yukawa coupling scenario. If the Yukawa coupling is of order ${\cal O}(10^{-2} - 1) $, the branching fraction of leptonic channel $H^{\pm \pm} \rightarrow \ell_\alpha^\pm \ell_\beta^\pm $ could be comparable to the bosonic channel $H^{\pm \pm} \rightarrow W^\pm W^\pm \phi$. Since these two channels make up all the doubly-charged scalar decay, once we fix the branching fraction of one channel, the other one could be easily obtained, thus we could scale the cross section of pair production $pp \to H^{++} H^{--}$ accordingly to estimate the mass reach with different final states.

The first process we consider is the same as that in small Yukawa coupling scenario in Section~\ref{sec:small}, i.e.~with both doubly-charged scalars decaying bosonically, and the same-sign $W$ bosons decaying leptonically.
The final state would be a pair of same-sign leptons plus jets and large missing transverse energy: $ H^{++}(\rightarrow W^+W^+ \phi) \; H^{--}(\rightarrow W^-W^- \phi) \rightarrow \ell^{\pm} \ell^{\pm}+\text{4 jets} + E_T^{\text{miss}} $. Since the branching fraction of the bosonic channel is no longer 100\% for intermediate Yukawa couplings, the mass reach would be undermined by the rising branching fraction of the leptonic decay channel $H^{\pm \pm} \rightarrow \ell^\pm \ell^\pm$. The significance of $H^{\pm\pm}$ in this channel is shown in the top left panel of Fig.~\ref{fig:intermediate_yukawa} as function of $M_{H^{\pm\pm}}$, where the red and blue lines are respectively for the HL-LHC and future 100 TeV collider. As shown in this figure, the doubly-charged scalar can be probed at the $2\sigma$ C.L. with mass below 500 GeV (2.9 TeV) at the HL-LHC (future 100 TeV collider) for $\text{BR}(H^{\pm \pm} \rightarrow \ell^\pm \ell^\pm) = 50 \%$. As the leptonic BR decreases, the mass reach increases, as expected, up to the ones reported in Fig.~\ref{Hpppairsignificance} (corresponding to ${\rm BR} (H^{\pm \pm} \rightarrow \ell^\pm \ell^\pm)$=0).


When the leptonic branching fraction is large enough, it is more likely that one of the pair-produced $H^{\pm \pm}$ decays leptonically and the other one decays bosonically. In this case, the final states with two or three charged leptons are of great interest. The two same-sign leptons can be used to reconstruct the Breit–Wigner peak of the mother doubly-changed scalar, making such signals almost background free.  The corresponding significances of $H^{\pm\pm}$ in the two-lepton channel $H^{++}H^{--}\rightarrow \ell^\pm \ell^\pm W^\mp W^\mp \phi \rightarrow \ell^\pm \ell^\pm + 4j + E_T^{\mathrm{miss}}$ and three-lepton channel $H^{++}H^{--}\rightarrow \ell^\pm \ell^\pm W^\mp W^\mp \phi \rightarrow \ell^\pm \ell^\pm \ell^\mp + 2j + E_T^{\mathrm{miss}}$ are shown respectively in the top right and bottom left panels of Fig.~\ref{fig:intermediate_yukawa}. In the two-lepton channel, the $2\sigma$ sensitivities for $H^{\pm \pm}$ mass are respectively 1.1 TeV at HL-LHC and 5.7 TeV at future 100 TeV collider for $\text{BR}(H^{\pm \pm} \rightarrow \ell^\pm \ell^\pm) < 90 \% $. With the same branching fraction choice, the mass reach of $H^{\pm\pm}$ in the three-lepton final state is slightly lower -- 1 (5.3) TeV at the HL-LHC (future 100 TeV collider).

The last case is the four-lepton final state via the process $H^{++} H^{--} \to \ell^+ \ell^+ \ell^- \ell^-$. Since we have two Breit–Wigner peaks from the two pairs of same-sign leptons, the search of $H^{\pm\pm}$ is same as in the standard Type-II seesaw, and the only limitations are the cross section of pair production and the branching fraction of the leptonic decay channel. The resultant significance of $H^{\pm\pm}$ in this channel is shown in the bottom right panel of Fig.~\ref{fig:intermediate_yukawa}. As shown in this figure, at the $2\sigma$ C.L. the doubly-charged scalar mass can reach respectively 950 GeV and 4.8 TeV at HL-LHC and future 100 TeV collider in the $\ell^+ \ell^+ \ell^- \ell^-$ channel with $\text{BR}(H^{\pm \pm} \rightarrow \ell^\pm \ell^\pm) > 20 \%$.

\section{Discussions and conclusion}
\label{sec:Sum}



In this paper, we have presented a global  $(B-L)$-conserved UV-complete neutrino mass model which contains a scalar triplet $\Delta$ and a singlet $\Phi$ both carrying a $B-L$ charge of +2.
From mixing of the neutral components of $\Delta$ with $\Phi$, this model features new neutrino interactions along with a pair of (light) leptonic scalars $H_1$ and $A_1$, collectively denoted by $\phi$. The light leptonic scalar $\phi$  induces very rich phenomenological consequences.
We list here the main features of the model, allowed  parameter space and the prospects of discovering this model at the HL-LHC and a future 100 TeV collider. Here are the main points:
\begin{itemize}
    \item The  proposed model looks similar to the Type-II seesaw model. But unlike the standard Type-II seesaw model, the neutral component of the triplet $\Delta$ of this model does not acquire any VEV. As a result, there is no Majorana mass term, neutrinos are Dirac fermions, and the $SU(2)_L$ custodial symmetry remains unbroken in this model.

    \item In light of all the low-energy LFV constraints, the coupling $Y_{\mu\mu}$ can be as large as ${\cal O} (1)$ for a TeV-scale $H^{\pm\pm}$ while all other Yukawa couplings are more stringently constrained (see Fig.~\ref{fig:limits} and Table~\ref{tab:limits}).
    Using RGEs, we have also determined the largest values of $\lambda_8$ and $Y_{\alpha\beta}$ at the EW scale in order to keep the theory perturbative all the way to the UV-complete scale, as shown in Fig.~\ref{fig:pert}. It is remarkable that as a good approximation the perturbativity limits can be obtained analytically. We checked also the unitarity constraints for these couplings and found them to be much weaker compared to the perturbativity limits.

    \item
    Originating from the gauge couplings, $H^{\pm\pm}$ and $H^\pm$ can decay into the light leptonic scalar $\phi = H_1,\; A_1$ via $H^{\pm\pm}\rightarrow W^\pm W^\pm \phi$ and $H^\pm\rightarrow W^\pm \phi$. The scalar $\phi$ provides additional sources of missing energy (along with the neutrinos from the decays of $W$ when the leptonic final states are selected) since it decays only into neutrinos, {\it i.e.} $\phi \to \nu\nu$. These new decay channels $H^{\pm\pm}\rightarrow W^\pm W^\pm \phi$ and $H^\pm\rightarrow W^\pm \phi$  dominate for small $Y_{\alpha\beta}$. For ${\cal O}(1)$ values of $Y_{\alpha\beta}$,  $H^{\pm\pm}$ and $H^\pm$  decay primarily into $\ell^\pm \ell^\pm$ and $\ell^\pm \nu$ respectively, while the decay $H^\pm \to W^\pm \phi$ can still occur with a BR of $10\% -20\%$ level, as shown in  the left panels of Fig.~\ref{fig:br_Hpp}, which is used for signal selection in this case.

    \item For our LHC analysis, we utilized the presence of the new source of missing energy from $\phi$ in the decays of $H^{\pm\pm}$ and $H^\pm$, and the BDT analysis can improve significantly the signal significance, in particular for the small Yukawa coupling case.
    At the HL-LHC, we found that for small and large $Y_{\alpha\beta}$, the $2\sigma\ (5\sigma)$  sensitivity reaches for $H^{\pm\pm}$ are respectively 800 (500) GeV and 1.1 (0.8) TeV (see Tables~\ref{tab:events} and \ref{tab:ap_events}), as denoted by the solid lines in Figs.~\ref{fig:significance} and \ref{fig:significance2}. These prospects are well above the current LHC constraints.

    \item At a future 100 TeV collider, the production cross section of $H^{\pm\pm}$ can be enhanced by over one order of magnitude in both pair production and associated production channels (see Fig.~\ref{fig:cross_section}). Therefore the mass reaches of $H^{\pm\pm}$ can be largely improved via the observation of $\phi$ induced signals. For the small and large Yukawa coupling cases, the mass $M_{H^{\pm\pm}}$ can reach up to 3.8 (2.6) TeV and 4 (2.7) TeV respectively at the $2\sigma\ (5\sigma)$ significance (see  Tables~\ref{tab:events} and \ref{tab:ap_events}), as indicated by the dashed lines in Figs.~\ref{fig:significance} and \ref{fig:significance2}.

    \item In the large Yukawa coupling scenario, the missing transverse energy is completely from the invisible light scalar $\phi$ at the parton level in the $pp \to H^{\pm\pm} H^\mp \to \mu^\pm \mu^\pm + 2j + E_T^{\rm miss}$ channel, and the mass $m_{\phi}$ can be determined with 10\% accuracy at the LHC via the transverse mass distributions associated with jets and missing energy. This is demonstrated in Fig.~\ref{fig:edgeVSmphi}.

\item In the intermediate Yukawa coupling case with $|Y_{\alpha\beta}| = {\cal O} (10^{-2} - 1)$, the branching fractions of leptonic $\ell^\pm \ell^\pm$ and bosonic $W^\pm W^\pm \phi$ decays of $H^{\pm\pm}$ are comparable to each other, the doubly-charged scalar $H^{\pm\pm}$ can be searched at the future hadron colliders in the  $H^{++} H^{--} \to \ell^\pm \ell^\pm + 4j + E_T^{\rm miss}$, $\ell^\pm \ell^\pm \ell^\mp + 2j + E_T^{\rm miss}$ and $\ell^+ \ell^+ \ell^- \ell^-$ channels. The corresponding prospects of $H^{\pm\pm}$ depend largely on the leptonic branching fraction of $H^{\pm\pm}$ and the search channels.
For the purpose of studying the leptonic scalar $\phi$ in the final state, the intermediate Yukawa coupling case can be most beneficial, from combining the leptonic and bosonic decay channels.
\end{itemize}

In this paper, we have focused on the light leptonic scalar case with mass $M_h/2 < M_\phi \lesssim {\cal O}(100\, {\rm GeV})$. It should be noted that the analysis in this paper can be generalized to the cases with relatively heavier leptonic scalars $\phi$, say with masses of few hundreds of GeV or even larger. Then the $\phi$-induced signals will depend largely on the mass $M_\phi$.
The light $\phi$ induced signal in this paper can also be compared with the searches of $H^{\pm\pm}$ at future hadron colliders in the standard Type-II seesaw. For instance, the $H^{\pm\pm}$ mass reach has been estimated in the standard Type-II scenario for the LHC and future 100 TeV colliders in Refs.~\cite{Du:2018eaw, Arkani-Hamed:2015vfh}. In a large region of parameter space of Type-II seesaw, the bosonic decay channel $H^{\pm\pm} \to W^\pm W^\pm$ dominates, and the mass reach of $H^{\pm\pm}$ is found to be 1.8 TeV at 5$\sigma$ at the 100 TeV collider, which is smaller than our reach of $\sim$2.6 TeV in both the large and small Yukawa coupling scenarios (cf. the dashed line in Figs.~\ref{fig:significance} and \ref{fig:significance2}).
The better reach in our model is due to the extra source of missing energy via $\phi$. This makes the signal in our model more easily distinguishable from the SM backgrounds.

\section*{Acknowledgments}
We thank Andr\'e de Gouv\^{e}a for discussions and collaboration on the previous related publication~\cite{deGouvea:2019qaz}, which motivated this work. The work of PSBD was supported in part by the U.S. Department  of  Energy  under  Grant  No.   DE-SC0017987, by the Neutrino Theory  Network  Program  and by a Fermilab Intensity Frontier Fellowship.
The work of BD was supported in part by the U.S.~Department of Energy under grant No. DE-SC0010813.
TG would like to acknowledge support from the Department of Atomic Energy, Government of India, for the Regional Centre for Accelerator-based Particle
Physics (RECAPP). TG was partly supported by the U.S. Department of Energy under
grant No. DE-FG02- 95ER40896, the PITT PACC, and by the FAPESP process no. 2019/17182-0.
The work of TH and HQ was supported in part by the U.S.~Department of Energy under grant No.~DE-FG02- 95ER40896 and in part by the PITT PACC.
%
The work of Y.Z.\ is supported by the National Natural Science Foundation of China under Grant No.\  12175039, the 2021 Jiangsu Shuangchuang (Mass Innovation and Entrepreneurship) Talent Program No.\ JSSCBS20210144, and the ``Fundamental Research Funds for the Central Universities''.
This work was partly performed at the Aspen Center for Physics, which is supported by National Science Foundation grant PHY-1607611.
\appendix
\section{Feynman rules}
\label{sec:AppA}

This appendix summarizes all the interaction vertices and their Feynman Rules for the model presented in Section~\ref{sec:Model}. The model contains three CP-even scalars $h,\, H_1,\,H_2$; two CP-odd scalars $A_1,\, A_2$; the singly-charged scalars $H^\pm$; and the doubly-charged scalars $H^{\pm\pm}$. The component $h$ from the $SU(2)_L$-doublet is identified with the 125 GeV SM Higgs boson. In our convention, $H_1$ is lighter than $H_2$, and $A_1$ is lighter than $A_2$. The trilinear and quartic scalar couplings are collected in Tables~\ref{table:couplings:trilinear} and \ref{table:couplings:quartic} respectively, the trilinear and quartic gauge couplings are presented in Tables~\ref{table:couplings:trilinear2} and \ref{table:couplings:quartic2} respectively, and the Yukawa couplings can be found in Table~\ref{table:couplings:Yukawa}.



\begin{table}[!t]
\centering
\caption{Trilinear scalar couplings.}
\label{table:couplings:trilinear}
\vspace{5pt}
\begin{tabular}{ccc}
\hline\hline
Vertices & Couplings \\
\hline
$H_1 H_1 h$, $A_1 A_1 h$ & $-i\, ((\lambda_1+\lambda_4)\sin^2\theta + \lambda_6 \cos^2\theta + \lambda_8 \sin 2\theta) v$ \\ \hline
$H_2 H_2 h$, $A_2 A_2 h$ & $-i\, ((\lambda_1+\lambda_4)\cos^2\theta + \lambda_6 \sin^2\theta - \lambda_8 \sin 2\theta) v$ \\ \hline
$H_1 H_2 h$, $A_1 A_2 h$ & $ \frac12 i \, ((\lambda_1 + \lambda_4 - \lambda_6) \sin 2\theta + 2\lambda_8 \cos 2\theta) v$ \\ \hline
$H^+ H^- h$ & $-i \, (\lambda_1 + \frac12 \lambda_4) v$ \\ \hline
$H^{++} H^{--} h$ & $-i \, \lambda_1 v$ \\ \hline\hline
\end{tabular}
\end{table}

\begin{table}[!t]
    \centering
    \caption{Quartic scalar couplings}
    \label{table:couplings:quartic}
    \vspace{5pt}
    \begin{tabular}{ccc}
    \hline\hline
    Vertices & Couplings \\ \hline
    $h h H_1 H_1$, $h h A_1 A_1$ & $-i\, (\lambda_6 \cos^2\theta - \lambda_8 \sin 2 \theta +(\lambda_1+\lambda_4)\sin^2\theta)$ \\ 
    $h h H_2 H_2$, $h h A_2 A_2$ & $-i (\lambda_1 \cos^2 \theta + \lambda_4 \cos^2 \theta + \lambda_8 \sin 2\theta +\lambda_6 \sin^2\theta)$ \\ 
    $h h H_2 H_1$, $h h A_2 A_1$ & $i\, (\lambda_8 \cos 2 \theta - \frac{1}{2}(\lambda_1+\lambda_4-\lambda_6)\sin 2 \theta)$ \\ 

    $H_1 H_1 H_1 H_1$, $A_1 A_1 A_1 A_1$ & $-6i (\lambda_5\cos^2\theta + \lambda_7 \cos^2\theta \sin^2\theta +(\lambda_2+\lambda_3)\sin^4\theta)$ \\ 
    $H_1 H_1 A_1 A_1$ & $-2i\,(\lambda_5 \cos^4 \theta+\lambda_7\cos^2 \theta \sin^2 \theta + (\lambda_2+\lambda_3)\sin^4 \theta)$ \\ 

    $H_2 H_2 H_2 H_2$, $A_2 A_2 A_2 A_2$ & $-6 i ((\lambda_2+\lambda_3) \cos^4 \theta + \lambda_7 \cos^2\theta \sin^2\theta + \lambda_5 \sin^4\theta)$ \\ 
    $H_2 H_2 A_2 A_2$ & $-2i((\lambda_2 + \lambda_3)\cos^4\theta +\lambda_7 \cos^2\theta\sin^2\theta+\lambda_5\sin^4\theta)$ \\ 

$H_1 H_1 H_1 H_2$, $A_1 A_1 A_1 A_2$ & $\frac{3}{2} i\, (-\lambda_2 - \lambda_3 +\lambda_5 +(\lambda_2 +\lambda_3 + \lambda_5 - \lambda_7)\cos 2 \theta)\sin 2 \theta$ \\ 
$H_1 H_1 H_2 H_2$, $A_1 A_1 A_2 A_2$ & $-\frac{1}{4} i\, (3(\lambda_2+\lambda_3+\lambda_5)+\lambda_7-3(\lambda_2+\lambda_3+\lambda_5-\lambda_7)\cos 4 \theta)$ \\ 
$H_1 H_2 H_2 H_2$, $A_1 A_2 A_2 A_2$ & $-\frac{3}{2} i \, (\lambda_2 + \lambda_3 - \lambda_5 + (\lambda_2 + \lambda_3 +\lambda_5 - \lambda_7)\cos 2\theta)\sin 2\theta$ \\ 

$H_1 H_1 A_2 A_2$, $H_2 H_2 A_1 A_1$ & $-\, \frac{1}{4}i(\lambda_2+\lambda_3+\lambda_5+3\lambda_7-(\lambda_2+\lambda_3+\lambda_5-\lambda_7)\cos4\theta)$ \\ 

$H_1 H_2 A_2 A_2$, $H_2 H_2 A_1 A_2$ & $-\frac{1}{2} i \, (\lambda_2 + \lambda_3 - \lambda_5 + (\lambda_2 + \lambda_3 +\lambda_5 - \lambda_7)\cos 2\theta)\sin 2\theta$ \\ 

$H_1 H_1 A_1 A_2$, $H_1 H_2 A_1 A_1$ & $\frac{1}{2} i\, (-\lambda_2 - \lambda_3 +\lambda_5 +(\lambda_2 +\lambda_3 + \lambda_5 - \lambda_7)\cos 2 \theta)\sin 2 \theta$ \\ 

$H_1 H_2 A_1 A_2$ & $-\frac{1}{2} i \, (\lambda_2 + \lambda_3 + \lambda_5 - \lambda_7) \sin^2 2 \theta$ \\ \hline

    $H^+ H^- hh$ & $-i (\lambda_1 + \frac12 \lambda_4)$ \\ 
    $H^+ H^- H_1 H_1$, $H^+ H^- A_1 A_1$ & $-i\, (\lambda_7 \cos^2\theta+2(\lambda_2+\lambda_3)\sin^2\theta)$ \\ 
    $H^+ H^- H_2 H_2$, $H^+ H^- A_2 A_2$ & $-i(2 (\lambda_2 +\lambda_3) \cos^2\theta + \lambda_7 \sin^2\theta)$ \\ 
    $H^+ H^- H_1 H_2$, $H^+ H^- A_1 A_2$ & $-i(\lambda_2 + \lambda_3 - \frac{1}{2} \lambda_7)\sin 2 \theta$ \\ 

    $H^+ H^+ H^- H^-$ & $-2i (2\lambda_2+\lambda_3)$ \\ \hline


    $H^{++} H^{--} hh$ & $-i \lambda_1$ \\ 

    $H^{++} H^{--} H_1 H_1$, $H^{++} H^{--} A_1 A_1$ & $-i\,(\lambda_7 \cos^2\theta+2\lambda_2\sin^2\theta)$ \\ 
    $H^{++} H^{--} H_2 H_2$, $H^{++} H^{--} A_2 A_2$ & $-i(2 \lambda_2 \cos^2\theta + \lambda_7 \sin^2\theta)$ \\ 
    $H^{++} H^{--} H_1 H_2$, $H^{++} H^{--} A_1 A_2$ & $-i\, (\lambda_2 - \frac{1}{2} \lambda_7)\sin 2\theta$ \\ 

    $H^{++} H^- H^- H_1$,  $H^{++} H^- H^- H_2$ & $\sqrt{2}i\lambda_3 \sin\theta$ \\ 
    $H^{++} H^- H^- A_1$, $H^{++} H^- H^- A_2$ & $\sqrt{2}\lambda_3 \sin\theta$ \\ 

    $H^{++} H^{--} H^+ H^-$ & $-2i(\lambda_2 + \lambda_3)$ \\ 
    $H^{++} H^{++}H^{--}H^{--}$ & $-4i(\lambda_2+\lambda_3)$ \\ 

\hline\hline
    \end{tabular}
\end{table}


\begin{table}[!t]
    \centering
    \caption{Trilinear gauge couplings. Here $p_1,\, p_2$ are the momenta of the first and second particles in the vertices.}
    \label{table:couplings:trilinear2}
    \vspace{5pt}
    \begin{tabular}{ccc}
    \hline\hline
    Vertices & Couplings \\ \hline
     $A_1 H_1 Z_\mu$ & $\dfrac{g_L}{c_W} \, (p_1 - p_2)_{\mu} \, \sin^2 \theta$  \\ 
     $A_2 H_2 Z_\mu $ &  $\dfrac{g_L}{c_W} \, (p_1 - p_2)_{\mu} \, \cos^2 \theta$  \\
 $A_1 H_2 Z_\mu$, $A_2 H_1 Z_\mu$ & $- \dfrac{g_L}{2\,c_W} \, (p_1 - p_2)_{\mu} \, \sin 2 \theta$  \\ 
 \hline

  $H^+ H^- \gamma_\mu$ & $i \, e (p_1 - p_2)_\mu$ \\ 
 $H^+ H^- Z_\mu$ &  $-i\, e \dfrac{s_W}{c_W} (p_1 - p_2)_\mu$ \\  
  $H^+ H_1 W^-_\mu$  & $-i\, \dfrac{g_L}{\sqrt{2}} \, (p_1 -p_2)_\mu \, \sin \theta$ \\ 
 $H^+ H_2 W^-_\mu$ & $i\, \dfrac{g_L}{\sqrt{2}} \, (p_1 -p_2)_\mu \, \cos \theta$ \\ 
  $H^+ A_1 W^-_\mu$ & $- \dfrac{g_L}{\sqrt{2}} \, (p_1 -p_2)_\mu \, \sin \theta$ \\ 
 $H^+ A_2 W^-_\mu$  & $\dfrac{g_L}{\sqrt{2}} \, (p_1 - p_2)_\mu \, \cos \theta$ \\ \hline

 $H^{++} H^{--} \gamma_\mu$ & $2i \, e  (p_1 - p_2)_\mu$ \\ 
 $H^{++} H^{--} Z_\mu$ &  $i\, e \dfrac{c^2_W - s^2_W}{c_W s_W} (p_1 -p_2)_\mu$ \\ 
 $H^{++} H^- W^-_\mu$ & $-i \, g_L (p_1 - p_2)_\mu$ \\ 
 \hline\hline
    \end{tabular}
\end{table}


\begin{table}[!t]
    \centering
    \caption{Quartic gauge couplings.}
    \vspace{5pt}
    \label{table:couplings:quartic2}
    \scalebox{0.9}{%
    \begin{tabular}{ccc}
    \hline\hline
    Vertices & Couplings \\ \hline
    $H_1 H_1 Z_\mu Z_\nu$, $A_1 A_1 Z_\mu Z_\nu$ & $2i \, \dfrac{g_L^2}{c_W^2} \sin^2 \theta \, g_{\mu \nu}$ \\ 
    $H_2 H_2 Z_\mu Z_\nu$, $A_2 A_2 Z_\mu Z_\nu$ & $2i \, \dfrac{g_L^2}{c_W^2} \cos^2 \theta \, g_{\mu \nu}$ \\ 
    $H_1 H_2 Z_\mu Z_\nu$, $A_1 A_2 Z_\mu Z_\nu$ &  $-i \, \dfrac{g_L^2}{c^2_W} \sin 2\theta \, g_{\mu \nu}$  \\  \hline


$H_1 H_1 W^+_\mu W^-_\nu$, $A_1 A_1 W^-_\mu W^-_\nu$ & $i \, g_L^2 \, \sin^2\theta \, g_{\mu \nu}$ \\ 
$H_2 H_2 W^+_\mu W^-_\nu$, $A_2 A_2 W^+_\mu W^-_\nu$ & $i \, g_L^2 \,  \cos^2 \theta  \, g_{\mu \nu}$ \\ 
$H_1 H_2 W^+_\mu W^-_\nu$, $A_1 A_2 W^+_\mu W^-_\nu$ & $- \frac12 i \, g_L^2 \, \sin 2\theta \, g_{\mu \nu}$ \\ \hline\hline


    $H^+ H^- \gamma_\mu \gamma_\nu$ & $2i\, e^2 g_{\mu \nu}$ \\ 
    $H^{+} H^{-} Z_\mu Z_\nu$ & $2i \, e^2\, \dfrac{s_W^2}{c_W^2} \, g_{\mu \nu}$ \\ 
    $H^{+} H^{-} Z_\mu \gamma_\nu$ & $-2i \, e^2\, \dfrac{s_W}{c_W } g_{\mu \nu}$ \\ \hline

    $H^+ H^- W^+_\mu W^-_\nu$ & $2i \, g_L^2  \, g_{\mu \nu}$ \\ \hline

    $H^+ H_1 W^-_\mu \gamma_\nu$ & $-i \, \dfrac{e^2}{\sqrt{2} s_W} \, \sin \theta \, g_{\mu \nu}$ \\ 
    $H^+ H_2 W^-_\mu \gamma_\nu$ & $i\, \dfrac{e^2}{\sqrt{2} s_W} \, \cos \theta \, g_{\mu \nu}$ \\ 
    $H^+ A_1 W^-_\mu \gamma_\nu$ & $-\dfrac{e^2}{\sqrt{2} s_W} \, \sin \theta \, g_{\mu \nu}$ \\ 
    $H^+ A_2 W^-_\mu \gamma_\nu$ &  $\dfrac{e^2}{\sqrt{2} s_W} \, \cos \theta \, g_{\mu \nu}$ \\ \hline

    $H^+ H_1 W^-_\mu Z_\nu$ & $i\, \dfrac{e^2}{\sqrt{2}\, c_W} \left(2+\dfrac{c_W^2}{s_W^2}\right) \sin\theta \, g_{\mu \nu}$ \\ 
    $H^+ H_2 W^-_\mu Z_\nu$ & $-i \, \dfrac{e^2}{\sqrt{2}\, c_W} \left(2+\dfrac{c_W^2}{s_W^2}\right) \cos\theta \, g_{\mu \nu}$ \\ 
    $H^+ A_1 W^-_\mu Z_\nu$ & $\dfrac{e^2}{\sqrt{2}\, c_W} \left(2+\dfrac{c_W^2}{s_W^2}\right) \sin\theta \, g_{\mu \nu}$ \\ 
    $H^+ A_2 W^-_\mu Z_\nu$ &  $-\dfrac{e^2}{\sqrt{2}\, c_W} \left(2+\dfrac{c_W^2}{s_W^2}\right) \cos\theta \, g_{\mu \nu}$ \\ \hline\hline

$H^{++} H^{--} \gamma_\mu \gamma_\nu$ & $8i\, e^2 g_{\mu \nu}$ \\ 
$H^{++} H^{--} Z_\mu Z_\nu$ & $2i \, g_L^2 \dfrac{(c_W^2 - s_W^2)^2}{c_W^2} \, g_{\mu \nu}$ \\ 
$H^{++} H^{--} Z_\mu \gamma_\nu$ & $4i \, e^2\, \dfrac{c_W^2 -s_W^2}{c_W \, s_W } g_{\mu \nu}$ \\ \hline

$H^{++} H^{--} W^+_\mu W^-_\nu$ & $i \, g_L^2  \, g_{\mu \nu}$ \\ \hline

$H^{++} H_1 W^-_\mu W^-_\nu$ & $\sqrt{2}i g_L^2 \, \sin \theta \, g_{\mu \nu}$ \\ 
$H^{++} H_2 W^-_\mu W^-_\nu$ & $-\sqrt{2}i g_L^2 \, \cos \theta \, g_{\mu \nu}$ \\ 
$H^{++} A_1 W^-_\mu W^-_\nu$ & $\sqrt{2} g_L^2 \, \sin \theta \, g_{\mu \nu}$ \\ 
$H^{++} A_2 W^-_\mu W^-_\nu$ & $-\sqrt{2} g_L^2 \, \cos \theta \, g_{\mu \nu}$ \\ \hline

$H^{++} H^- W^-_\mu \gamma_\nu$ & $-3i\, \dfrac{e^2}{s_W} \, g_{\mu \nu}$ \\ 
$H^{++} H^- W^-_\mu Z_\nu$ &  $i \, \dfrac{e^2}{ c_W} \left(2+\dfrac{c_W^2}{s_W^2}\right) \, g_{\mu \nu}$ \\ 
\hline\hline
    \end{tabular}}
\end{table}



\begin{table}[hbtp]
\centering
\caption{Yukawa couplings.}
\vspace{5pt}
\label{table:couplings:Yukawa}
\begin{tabular}{lc}
\hline\hline
Vertices & Couplings \\ \hline
$H^{++} l^-_\alpha l^-_\beta$ & $2 i \, Y_{\alpha\beta} \, P_L$ \\ \hline
$H^{+} l^-_\alpha v_\beta$ & $\sqrt{2} i\, Y_{\alpha\beta} \, P_L$ \\
\hline
$H_2  \nu_\alpha \nu_\beta$ & $-\sqrt{2}i \,  Y_{\alpha\beta} \, P_L \, \cos \theta$ \\
\hline
$H_1  \nu_\alpha \nu_\beta$ & $- \sqrt{2}i \, Y_{\alpha\beta} \, P_L \, \sin \theta$  \\
\hline
$A_2 \nu_\alpha \nu_\beta$ & $\sqrt{2} \,  Y_{\alpha\beta} \, P_L \, \cos \theta$ \\
\hline
$A_1 \nu_\alpha \nu_\beta$ &  $\sqrt{2} \, Y_{\alpha\beta} \, P_L \, \sin \theta$ \\
\hline
\end{tabular}
\end{table}

\section{The functions $G$ and ${\cal F}$}
\label{sec:F}
For the decays in Eq.~\eqref{eqn:Width:HpWH2}, the function $G(x,y)$ is given by
\begin{eqnarray}
G(x,y) & =  & \frac{1}{12y} \Bigg\{ 2(-1+x)^3-9(-1+x^2)y+6(-1+x)y^2 \nonumber \\
&& +6(1+x-y)y\sqrt{-\lambda(x,y)}\left[\mathrm{arctan}\left(\frac{-1+x-y}{\sqrt{-\lambda(x,y)}}\right) +\mathrm{arctan}\left(\frac{-1+x+y}{\sqrt{-\lambda(x,y)}}\right) \right] \nonumber \\
&& -3y \Big[ 1+(x-y)^2-2y \Big] \mathrm{log}x \Bigg\} \,.
\end{eqnarray}

For the decays in Eq.~(\ref{eqn:decayWWphi}), the function ${\cal F}$ is defined as
\begin{eqnarray}
F &  =  & 4 + \frac{1}{2} (x-2)^2 \nonumber \\
&& + \frac{1}{2 (y-u)^2} \left[ (y-1)^2 - 2r (y+1) + r^2 \right]\left[ (y-1)^2 - 2w (y+1) + w^2 \right] \nonumber \\
&& + \frac{1}{2 (z-u)^2} \left[ (z-1)^2 - 2r (z+1) + r^2 \right]\left[ (z-1)^2 - 2w (z+1) + w^2 \right] \nonumber \\
&& + \frac{1}{(y-u)(z-u)} \left[ (y-r)(z-w) + (y+z+r+w-3) \right]  \nonumber \\
&& \times \left[ (z-r)(y-w) + (y+z+r+w-3) \right]  \nonumber \\
&& - \frac{1}{y-u} \left[ (x-2) (y-r-1)(y-w+1) \right. \nonumber \\
&& \left. + 2(y-r-1)^2 + 2(z-r-1)^2 + 2(x-2)(z-r-1) - 8r \right] \nonumber \\
&& - \frac{1}{z-u} \left[ (x-2) (z-r-1)(z-w+1) \right. \nonumber \\
&& \left. + 2(y-r-1)^2 + 2(z-r-1)^2 + 2(x-2)(y-r-1) - 8r \right] \,,
\end{eqnarray}
where we have defined
\begin{eqnarray}
x \equiv \frac{m_{12}^2}{M_W^2} \,, \quad
y \equiv \frac{m_{23}^2}{M_W^2} \,, \quad
z \equiv \frac{m_{13}^2}{M_W^2} \,, \quad
r \equiv \frac{m_{\phi}^2}{M_W^2} \,, \quad
u \equiv \frac{M_{H^\pm}^2}{M_W^2} \,, \quad
w \equiv \frac{M_{H^{\pm\pm}}^2}{M_W^2} \,.
\end{eqnarray}

\section{One-loop RGEs}
\label{sec:AppB}

In this appendix, we list the $\beta$-functions for all the one-loop RGEs for the gauge couplings, quartic couplings and Yukawa couplings in our model. These were obtained using the {\tt PyR@TE} package~\cite{Lyonnet:2013dna, Sartore:2020gou}. For simplicity, we keep only the Yukawa coupling $Y_{\mu\mu}$ in the matrix $Y_{\alpha\beta}$. The gauge coupling $g_Y$ is normalized to be $g_1 = \sqrt{3/5} g_Y$~\cite{Arason:1991ic}.
\begin{align}
\label{eqn:bgS}
(4\pi)^2\beta_{g_{S}} = & \ - 7 g_{S}^{3} \,, \\
\label{eqn:bgL}
(4\pi)^2 \beta_{g_{L}} = & \ - \frac{5}{2} g_{L}^{3} \,, \\
\label{eqn:bg1}
(4\pi)^2\beta_{g_{1}} = & \ +\frac{47}{6} g_{1}^{3} \,, \\
(4\pi)^2\beta_{\lambda} = & \
\frac{3}{2} \left( 3 g_{L}^{4} + 2 g_{1}^{2} g_{L}^{2} + g_{1}^{4}  \right)
+6 \lambda^{2} + 12 \lambda_{1}^{2}  +5 \lambda_{4}^{2} + 4 \lambda_{6}^{2} +8 \lambda_{8}^{2}  + 12 \lambda_{1} \lambda_{4} \nonumber \\
& - 24 y_t^4  - 3 \lambda \left( 3 g_{L}^{2} + g_{1}^{2} \right)  +12 \lambda y_t^2
 \,, \\
(4\pi)^2\beta_{\lambda_{1}} =
& \ 3 \left( 2 g_{L}^{4} - 2 g_{1}^{2} g_{L}^{2} + g_{1}^{4}  \right)
+ 4 \lambda_{1}^{2} + \lambda_{4}^{2}
+3 \lambda \lambda_{1} + \lambda \lambda_{4} +16 \lambda_{1} \lambda_{2} + 12 \lambda_{1} \lambda_{3} \nonumber \\
& + 6 \lambda_{2} \lambda_{4}
 + 2 \lambda_{3} \lambda_{4} + 2 \lambda_{6} \lambda_{7}
- \frac{3}{2} \lambda_1 \left( 11 g_{L}^{2} +5 g_{1}^{2} \right) \nonumber \\
&+ 2\lambda_1 \left( 3 y_t^2 + 2 \left| Y_{\mu\mu} \right|^2 \right) \,, \\
(4\pi)^2\beta_{\lambda_{2}} =
& \ 3 \left( 5 g_{L}^{4} - 4 g_{1}^{2} g_{L}^{2} + 2 g_{1}^{4} \right)
+2 \lambda_{1}^{2}	+28 \lambda_{2}^{2} + 6 \lambda_{3}^{2} + \lambda_{7}^{2} +2 \lambda_{1} \lambda_{4} +24 \lambda_{2} \lambda_{3} \nonumber \\
& - 12 \lambda_2 \left( 2 g_{L}^{2}  + g_{1}^{2} \right) 			+ 8 \lambda_{2} \left|  Y_{\mu\mu} \right|^2 \,, \\
(4\pi)^2\beta_{\lambda_{3}} =
& \ - 6 g_L^2 \left( g_{L}^{2} - 4 g_{1}^{2} \right)
+18 \lambda_{3}^{2} + \lambda_{4}^{2} 	+24 \lambda_{2} \lambda_{3} - 16 \left| Y_{\mu\mu} \right|^4 \nonumber \\
&- 12 \lambda_{3} \left( 2 g_{L}^{2} + g_{1}^{2} \right) + 8 \lambda_{3} \left| Y_{\mu\mu} \right|^2    \,, \\
(4\pi)^2\beta_{\lambda_{4}} =
& \ 12 g_{1}^{2} g_{L}^{2} +4 \lambda_{4}^{2} + 8 \lambda_{8}^{2}
+\lambda \lambda_{4} +8 \lambda_{1} \lambda_{4}  + 4 \lambda_{2} \lambda_{4} +8 \lambda_{3} \lambda_{4} \nonumber \\
& - \frac{3}{2} \lambda_4 \left( 11 g_{L}^{2} + 5 g_{1}^{2} \right) + 2\lambda_4 \left( 3 y_t^2 + 2 \left| Y_{\mu\mu} \right|^2 \right)  \,, \\
(4\pi)^2\beta_{\lambda_{5}} =
& \ 20 \lambda_{5}^{2} + 2 \lambda_{6}^{2} + 3 \lambda_{7}^{2} \,, \\
(4\pi)^2\beta_{\lambda_{6}} =
& \ 4 \lambda_{6}^{2} + 12 \lambda_{8}^{2}
+3 \lambda \lambda_{6} +6 \lambda_{1} \lambda_{7} + 8 \lambda_{5} \lambda_{6} + 3 \lambda_{4} \lambda_{7}
\nonumber \\
& - \frac{3}{2} \lambda_{6} \left( 3 g_{L}^{2} + g_{1}^{2} \right)  +6 \lambda_{6} y_t^2 \,, \\
(4\pi)^2\beta_{\lambda_{7}} =
& \ 4 \lambda_{7}^{2}  + 4 \lambda_{8}^{2} +4 \lambda_{1} \lambda_{6} + 16 \lambda_{2} \lambda_{7} + 12 \lambda_{3} \lambda_{7} +2 \lambda_{4} \lambda_{6} +8 \lambda_{5} \lambda_{7}  \nonumber \\
& - 6 \lambda_{7} \left( 2 g_{L}^{2} +g_{1}^{2} \right) + 4 \lambda_{7} \left| Y_{\mu\mu} \right|^2 \,, \\
(4\pi)^2\beta_{\lambda_{8}} =
& \ \lambda \lambda_{8} +4 \lambda_{1} \lambda_{8} + 6 \lambda_{4} \lambda_{8} +4 \lambda_{6} \lambda_{8} +2 \lambda_{7} \lambda_{8} \nonumber \\
& - \frac{3}{2} \lambda_{8} \left( 7 g_{L}^{2} +3 g_{1}^{2} \right)
+ 2 \lambda_{8} \left( 3 y_t^2 + \left| Y_{\mu\mu} \right|^2 \right)	\,, \\
\label{eqn:beta:yt}
(4\pi)^2\beta_{y_{t}} =
& \ \frac{9}{2} y_t^3 - y_t \left( 8 g_{S}^{2} + \frac{9}{4} g_{L}^{2} + \frac{17}{12} g_{1}^{2} \right) \,, \\
\label{eqn:beta:Ymumu}
(4\pi)^2\beta_{Y_{\mu\mu}} = & \ 8 \left| Y_{\mu\mu} \right|^2 Y_{\mu\mu} - \frac32 Y_{\mu\mu} \left( 3 g_L^2 + g_1^2 \right) \,.
\end{align}

\section{Analytical perturbativity limits}
\label{sec:AppC}
For the gauge couplings $g_{i}$, it is trivial to get the analytical one-loop expressions for the couplings, which turn out to be
\begin{eqnarray}
\alpha_i (\mu) = \frac{ \alpha_i (v) }{ 1 - \frac{b_i}{2\pi} \alpha_i (v) \log (\mu/v) } \,,
\end{eqnarray}
with $\alpha_3 = g_S^2/4\pi$, $\alpha_2 = g_L^2/4\pi$, $\alpha_1 = g_1^2/4\pi$ for the $SU(3)_c$, $SU(2)_L$ and $U(1)_Y$ couplings respectively, and $b_3 = -7$, $b_2 = -5/2$, $b_1 = 47/6$ [cf. Eqs.~(\ref{eqn:bgS})-(\ref{eqn:bg1})].
For the SM top-quark Yukawa coupling $y_t$, let us first consider only the $y_t^3$ and $g_S^2 y_t$ terms on the RHS of Eq.~(\ref{eqn:beta:yt}), {\it i.e.}:
\begin{eqnarray}
\label{eqn:yt}
(4\pi)^2 \frac{\rm d}{{\rm d} t} y_{t} & = & \frac{9}{2} y_t^3  - 8 g_{S}^{2} y_t \,.
\end{eqnarray}
To implement the running of $g_S$, we rewrite the equation above to be in the form of
\begin{eqnarray}
8\pi^2 \left[ \frac{1}{y_t^2} \frac{\rm d}{{\rm d} t} y_{t}^2
 + \frac{8}{b_3} \frac{1}{\alpha_3} \frac{\rm d}{{\rm d} t} \alpha_3 \right] & = & \frac{9}{2} y_t^3 \, , \nonumber \\
 {\rm or,} \quad
8\pi^2 \frac{\rm d}{{\rm d} t} \log \left( y_t^2 \alpha_3^{8/b_3} \right)
& = & \frac{9}{2} y_t^2 \,.
\end{eqnarray}
Then we can obtain the analytical running of $y_t$:
\begin{eqnarray}
y_t^2 (\mu) \simeq y_t^2 (v)
\left( \frac{\alpha_3 (v)}{ \alpha_3(\mu)} \right)^{8/b_3}
\left[ 1 - \frac{9}{16\pi^2} y_t^2(v) \alpha_3^{8/b_3}(v)
\int_0^t {\rm d} t' \, \alpha_3^{-8/b_3} (t') \right]^{-1} \,.
\end{eqnarray}
If we include also the $g_L^2 y_t$ and $g_1^2 y_t$ terms in Eq.~(\ref{eqn:beta:yt}), it is straightforward to get the full analytical one-loop solution for $y_t$:
\begin{eqnarray}
y_t^2 (\mu) = y_t^2 (v)
\left( \frac{E_\alpha (v)}{ E_\alpha(\mu)} \right)
\left[ 1 - \frac{9}{16\pi^2} y_t^2(v) E_\alpha (v)
\int_0^t {\rm d} t' \, E_\alpha^{-1} (t') \right]^{-1} \,,
\end{eqnarray}
where the function
\begin{eqnarray}
E_\alpha (\mu) = \alpha_3^{8/b_3} (\mu) \alpha_2^{9/4b_2} (\mu) \alpha_1^{17/12b_1} (\mu) \,.
\end{eqnarray}

In the one-loop RGE of $Y_{\mu\mu}$ in Eq.~(\ref{eqn:beta:Ymumu}), if we consider only the $Y_{\mu\mu}^3$ term on the RHS, it is trivial to obtain
\begin{eqnarray}
\alpha_{\mu} (\mu) = \frac{ \alpha_{\mu} (v) }{ 1 - \frac{4}{\pi} \alpha_{\mu} (v)t } \,,
\end{eqnarray}
where $\alpha_{\mu} \equiv Y_{\mu\mu}^2/4\pi$. It is clear that the coupling $Y_{\mu\mu}$ will blow up when the $t$ parameter approaches the value of
\begin{eqnarray}
t_c = \log \left( \frac{\mu_c}{v} \right) = \frac{\pi^2}{Y_{\mu\mu}^2(v)} \,.
\end{eqnarray}
With an initial value of $Y_{\mu\mu} (v) = 1.5$, we can get the critical value of $t_c \simeq 4.39$. As in Eq.~(\ref{eqn:yt}), we can first include the gauge coupling $g_L$, then
\begin{eqnarray}
Y_{\mu\mu}^2 (\mu) \simeq Y_{\mu\mu}^2 (v)
\left( \frac{\alpha_2 (v)}{ \alpha_2 (\mu)} \right)^{9/2b_2}
\left[ 1 - \frac{1}{\pi^2} Y_{\mu\mu}^2(v) \alpha_2^{9/2b_2}(v)
\int_0^t {\rm d} t' \, \alpha_2^{-9/2b_2} (t') \right]^{-1} \,.
\end{eqnarray}
In this case, the coupling $g_L$ becomes divergent when the parameter $t_c = 4.62$. If we have all the terms on the RHS of Eq.~(\ref{eqn:beta:Ymumu}), it turns out that
\begin{eqnarray}
Y_{\mu\mu}^2 (\mu) & = & Y_{\mu\mu}^2 (v)
\left( \frac{\alpha_2 (v)}{ \alpha_2 (\mu)} \right)^{9/2b_2}
\left( \frac{\alpha_1 (v)}{ \alpha_1 (\mu)} \right)^{3/2b_1} \nonumber \\
&& \times \left[ 1 - \frac{1}{\pi^2} Y_{\mu\mu}^2(v) \alpha_2^{9/2b_2}(v) \alpha_1^{3/2b_1}(v)
\int_0^t {\rm d} t' \, \alpha_2^{-9/2b_2} (t') \alpha_1^{-3/2b_1} (t') \right]^{-1} \,.
\end{eqnarray}
In this case, the critical value $t_c = 4.67$.

We also show the analytical solution of $\lambda_8(\mu)$ below:
\begin{equation}
    \lambda_8(\mu)=\lambda_8(v) \exp \left\{ \frac{1}{4\pi^2}\int^\mu_v E_8(\mu) d\mu \right\} \,,
\end{equation}
where
\begin{eqnarray}
\label{eqn:E8}
    E_8(\mu) &=&
    3y_t(v)^2 \left( 1-\frac{\mu b_3 \alpha_3(v)}{2\pi} \right)^{{8}/{b_3}}
    -\alpha_\mu(v) \left( 1-\frac{4\mu\alpha_\mu(v)}{\pi} \right)^{-1} \\ \nonumber
    &-&\frac{21}{2}{\alpha_2(v)} \left( 1-\frac{\mu b_2\alpha_2(v)}{2\pi} \right)^{-1}
    -\frac{9}{2}{\alpha_1(v)} \left( 1+\frac{\mu b_1\alpha_1(v)}{2\pi} \right)^{-1} \,.
\end{eqnarray}
These results agree well with the full numerical results shown in Fig.~\ref{fig:pert}.

\section{Unitarity limits}
\label{sec:AppD}
Following the analysis for the Type-II seesaw model~\cite{Arhrib:2011uy}, the unitarity bounds in our model can be found by diagonalizing the sub-matrices ${\cal M}_i$ which correspond to the coefficients for $2\leftrightarrow 2$ scalar scattering processes. Writing the scalar multiplets explicitly as
\begin{eqnarray}
H = \left( \begin{matrix}
h^\pm \\ \frac{1}{\sqrt2} ( h + iZ_1 )
\end{matrix} \right) \,, \quad
\Delta =
\begin{pmatrix} \frac{1}{\sqrt{2}}\delta^+ & \delta^{++} \\ \frac{1}{\sqrt2} ( \xi + iZ_2 ) & -\frac{1}{\sqrt{2}}\delta^+ \end{pmatrix} \,, \quad
\Phi = \frac{1}{\sqrt2} ( s + iZ_3 ) \,,
\end{eqnarray}
the sub-matrices for the initial and final states
($h\xi$, $hs$, $Z_1 Z_2$, $Z_1 Z_3$,
$h Z_2$, $h Z_3$, $\xi Z_1$, $s Z_1$,
$h^+ \delta^-$, $\delta^+ h^-$) and ($\xi s$, $Z_2 Z_3$, $\xi Z_3$, $s Z_3$) respectively are
\begin{eqnarray}
{\cal M}_1 &=&
\left(\begin{matrix}
\lambda_{14} & -\lambda_8  & 0 & \lambda_8 & 0 & 0 & 0 & 0 & \frac{\lambda_4}{2\sqrt2} & \frac{\lambda_4}{2\sqrt2} \\
-\lambda_8 & \lambda_6  & -\lambda_8 & 0 & 0 & 0 & 0 & 0 & -\frac{\lambda_8}{\sqrt2} & -\frac{\lambda_8}{\sqrt2} \\
0 & -\lambda_8  & \lambda_{14} & \lambda_8 & 0 & 0 & 0 & 0 & \frac{\lambda_4}{2\sqrt2} & \frac{\lambda_4}{2\sqrt2} \\
\lambda_8 & 0 & \lambda_8 & \lambda_6 & 0 & 0 & 0 & 0 & \frac{\lambda_8}{\sqrt2} & \frac{\lambda_8}{\sqrt2} \\
0 & 0 & 0 & 0 & \lambda_{14} & -\lambda_8 & 0 & -\lambda_8 & -\frac{i\lambda_4}{2\sqrt2} & -\frac{i\lambda_4}{2\sqrt2}  \\
0 & 0 & 0 & 0 & -\lambda_8 & \lambda_{6} & \lambda_8 & 0 & \frac{i\lambda_8}{\sqrt2} & \frac{i\lambda_8}{\sqrt2} \\
0 & 0 & 0 & 0 & 0 & \lambda_8 & \lambda_{14} & \lambda_8 & \frac{i\lambda_4}{2\sqrt2} & \frac{i\lambda_4}{2\sqrt2} \\
0 & 0 & 0 & 0 & -\lambda_8 & 0 & \lambda_8 & \lambda_{6} & \frac{i\lambda_8}{\sqrt2} & -\frac{i\lambda_8}{\sqrt2} \\
\frac{\lambda_4}{2\sqrt2} & -\frac{\lambda_8}{\sqrt2} & \frac{\lambda_4}{2\sqrt2} & \frac{\lambda_8}{\sqrt2} & \frac{i\lambda_4}{2\sqrt2} &  -\frac{i\lambda_8}{\sqrt2} & -\frac{i\lambda_4}{2\sqrt2} & -\frac{i\lambda_8}{\sqrt2} & \lambda'_{14} & 0 \\
\frac{\lambda_4}{2\sqrt2} & -\frac{\lambda_8}{\sqrt2} & \frac{\lambda_4}{2\sqrt2} & \frac{\lambda_8}{\sqrt2} & -\frac{i\lambda_4}{2\sqrt2} &  \frac{i\lambda_8}{\sqrt2} & \frac{i\lambda_4}{2\sqrt2} & \frac{i\lambda_8}{\sqrt2} & 0 & \lambda'_{14}
\end{matrix}\right) \,, \\
{\cal M}_2 &=&
\left(\begin{matrix}
\lambda_7 & 0 & 0 & 0 \\
0 & \lambda_7 & 0 & 0 \\
0 & 0 & \lambda_7 & 0 \\
0 & 0 & 0 & \lambda_7
\end{matrix}\right) \,,
\end{eqnarray}
where we have defined the combinations of quartic couplings:
\begin{eqnarray}
\lambda_{ij} \equiv \lambda_i + \lambda_j \,, \qquad
\lambda'_{ij} \equiv \lambda_i + \frac12 \lambda_j \,.
\end{eqnarray}
The eigenvalues are
\begin{eqnarray}
\label{eqn:values1}
\lambda_{1,\, 6,\, 7} \,, \quad
\lambda_1 + \lambda_4 \,, \quad
\lambda_{146}^\pm \,,
\end{eqnarray}
with
\begin{eqnarray}
\lambda_{146}^\pm \equiv
\frac14 \left[ (2\lambda_1+3\lambda_4+2\lambda_6) \pm \sqrt{(2\lambda_1+3\lambda_4-2\lambda_6)^2 + 96\lambda_8^2} \right] \,.
\end{eqnarray}

For the states ($\frac{1}{\sqrt2} hh$, $\frac{1}{\sqrt2} \xi\xi$, $\frac{1}{\sqrt2} ss$, $\frac{1}{\sqrt2} Z_1 Z_1$, $\frac{1}{\sqrt2} Z_2 Z_2$, $\frac{1}{\sqrt2} Z_3 Z_3$, $h^+ h^-$, $\delta^+ \delta^-$, $\delta^{++} \delta^{--}$) with factor of ${1}/{\sqrt2}$ accounting for the identical particles, the sub-matrix is
\begin{equation}
{\cal M}_3 =
\left(\begin{matrix}
\frac{3\lambda}{4} & \frac{\lambda_{14}}{2} & \frac{\lambda_6}{2} & \frac{\lambda}{4} & \frac{\lambda_{14}}{2} & \frac{\lambda_6}{2} & \frac{\lambda}{2\sqrt2} & \frac{\lambda'_{14}}{\sqrt2} & \frac{\lambda_1}{\sqrt2} \\
\frac{\lambda_{14}}{2} & 3\lambda_{23} & \frac{\lambda_7}{2} & \frac{\lambda_{14}}{2} & \lambda_{23} & \frac{\lambda_7}{2} & \frac{\lambda_1}{\sqrt2} & \sqrt2 \lambda_{23} & \sqrt2\lambda_2  \\
\frac{\lambda_6}{2} & \frac{\lambda_7}{2} & 3\lambda_5 &  \frac{\lambda_6}{2} &  \frac{\lambda_7}{2} & \lambda_5 & \frac{\lambda_6}{\sqrt2} & \frac{\lambda_7}{\sqrt2} & \frac{\lambda_7}{\sqrt2} \\
\frac{\lambda}{4} & \frac{\lambda_{14}}{2} & \frac{\lambda_{6}}{2} & \frac{3\lambda}{4} & \frac{\lambda_{14}}{2} & \frac{\lambda_{6}}{2} & \frac{\lambda}{2\sqrt2} & \frac{\lambda'_{14}}{\sqrt2} & \frac{\lambda_1}{\sqrt2} \\
\frac{\lambda_{14}}{2} & \lambda_{23} & \frac{\lambda_7}{2} & \frac{\lambda_{14}}{2} & 3\lambda_{23} & \frac{\lambda_7}{2} & \frac{\lambda_1}{\sqrt2} & \sqrt2 \lambda_{23} & \sqrt2\lambda_2 \\
\frac{\lambda_6}{2} & \frac{\lambda_7}{2} & \lambda_5 &  \frac{\lambda_6}{2} &  \frac{\lambda_7}{2} & 3\lambda_5 & \frac{\lambda_6}{\sqrt2} & \frac{\lambda_7}{\sqrt2} & \frac{\lambda_7}{\sqrt2} \\
\frac{\lambda}{2\sqrt2} & \frac{\lambda_1}{\sqrt2} & \frac{\lambda_6}{\sqrt2} & \frac{\lambda}{2\sqrt2} & \frac{\lambda_1}{\sqrt2} & \frac{\lambda_6}{\sqrt2} & \lambda & \lambda'_{14} & \lambda_{14} \\
\frac{\lambda'_{14}}{\sqrt2} & \sqrt2 \lambda_{23} & \frac{\lambda_7}{\sqrt2} & \frac{\lambda'_{14}}{\sqrt2} & \sqrt2 \lambda_{23} & \frac{\lambda_7}{\sqrt2} & \lambda'_{14} & 4\lambda'_{23} & 2\lambda_{23} \\
\frac{\lambda_1}{\sqrt2} & \sqrt2\lambda_2 & \frac{\lambda_7}{\sqrt2} & \frac{\lambda_1}{\sqrt2} & \sqrt2\lambda_2 & \frac{\lambda_7}{\sqrt2} & \lambda_{14} & 2\lambda_{23} & 4\lambda_{23}
\end{matrix}\right) \,,
\end{equation}
and the eigenvalues are
\begin{eqnarray}
\label{eqn:values3}
\frac12 \lambda \,, \quad
2\lambda_{2,\, 5} \,, \quad
2(\lambda_2 + \lambda_3) \,, \quad
\lambda_{023} \,, \quad x_{1,\, 2,\, 3} \,,
\end{eqnarray}
with
\begin{eqnarray}
\lambda_{023} \equiv
\frac14 \left[ (\lambda+4\lambda_2+8\lambda_3) \pm \sqrt{(\lambda-4\lambda_2-8\lambda_3)^2 + 16\lambda_4^2} \right]
\end{eqnarray}
and $x_{1,\, 2,\, 3}$ are the roots of the equation
\begin{eqnarray}
&& x^3 -2x^2 \left( 3\lambda + 16 \lambda_2 + 12\lambda_3 + 8\lambda_5 \right) \nonumber \\
&& + 8x \left[ 6\lambda ( 4\lambda_2 + 3\lambda_3 + 2\lambda_5 )
-3 ( 2\lambda_1 + \lambda_4 )^2 + 64 \lambda_2 \lambda_5 + 48 \lambda_3 \lambda_5
-4 \lambda_6^2  -6 \lambda_7^2 \right] \nonumber \\
&& + 32 \left[ 9\lambda \lambda_7^2
+12\lambda_5 ( -2 \lambda (4\lambda_2+3\lambda_3) + (2\lambda_1+\lambda_4)^2 )
+ 8 \lambda_6^2 (4\lambda_2+3\lambda_3)
-12 \lambda_6\lambda_7 (2\lambda_1+\lambda_4) \right] = 0 \,. \nonumber \\ &&
\end{eqnarray}

The sub-matrix for the states ($h Z_1$, $\xi Z_2$, $s Z_3$) is
\begin{equation}
{\cal M}_4 =
\left(\begin{matrix}
\frac12 \lambda & 0 & 0 \\
0 & 2 (\lambda_2 + \lambda_3) & 0 \\
0 & 0 & 2\lambda_5
\end{matrix}\right) \,,
\end{equation}
whose eigenvalues are
\begin{eqnarray}
\label{eqn:values4}
\frac12 \lambda \,, \quad
 2 (\lambda_2 + \lambda_3) \,, \quad
 2\lambda_5 \,.
\end{eqnarray}

The sub-matrix for ($h h^+$, $\xi h^+$, $s h^+$, $Z_1 h^+$, $Z_2 h^+$, $Z_3 h^+$, $h \delta^+$, $\xi \delta^+$, $s \delta^+$, $Z_1 \delta^+$, $Z_2 \delta^+$, $Z_3 \delta^+$, $\delta^{++} h^-$, $\delta^{++} \delta^-$) is
\vspace{-10pt}
\begin{center}
\scriptsize
\begin{equation}
{\cal M}_5 =
\left(\begin{matrix}
\frac{\lambda}{2} & 0 & 0 & 0 & 0 & 0 &
0 & \frac{\lambda_4}{2\sqrt2} & -\frac{\lambda_8}{\sqrt2} & 0 & \frac{i\lambda_4}{2\sqrt2} & -\frac{i\lambda_8}{\sqrt2} &
0 & -\frac{\lambda_4}{2} \\
0 & \lambda_1 & 0 & 0 & 0 & 0 &
0 & 0 & 0 & 0 & 0 & 0 & -\frac{\lambda_8}{\sqrt2} & -\frac{\lambda_8}{\sqrt2} \\
0 & 0 & \lambda_6 & 0 & 0 & 0 &
-\frac{\lambda_8}{\sqrt2} & 0 & 0 & -\frac{i\lambda_8}{\sqrt2} & 0 & 0 &
\sqrt2\lambda_8 & 0 \\
0 & 0 & 0 & \frac{\lambda}{2} & 0 & 0 &
0 & -\frac{i\lambda_4}{2\sqrt2} & -\frac{i\lambda_8}{\sqrt2} & 0 & \frac{\lambda_4}{2\sqrt2} & \frac{\lambda_8}{\sqrt2} &
0 & \frac{i\lambda_4}{2} \\
0 & 0 & 0 & 0 & \lambda_1 & 0 &
\frac{i\lambda_4}{2\sqrt2} & 0 & 0 & \frac{\lambda_4}{2\sqrt2} & 0 & 0 &
0 & 0 \\
0 & 0 & 0 & 0 & 0 & \lambda_6 &
-\frac{i\lambda_8}{\sqrt2} & 0 & 0 & \frac{\lambda_8}{\sqrt2} & 0 & 0 &
\sqrt2 i \lambda_8 & 0 \\
0 & \frac{\lambda_4}{2\sqrt2} & -\frac{\lambda_8}{\sqrt2} & 0 & -\frac{i\lambda_4}{2\sqrt2} & \frac{i\lambda_8}{\sqrt2}
& \lambda'_{14} & 0 & 0 & 0 & 0 & 0 &
-\frac{\lambda_4}{2} & 0  \\
\frac{\lambda_4}{2\sqrt2} & 0 & 0 & \frac{i\lambda_4}{2\sqrt2} & 0 & 0 &
0 & 2\lambda_{23} & 0 & 0 & 0 & 0 &
0 & -\sqrt2 \lambda_3 \\
-\frac{\lambda_8}{\sqrt2} & 0 & 0 & \frac{i\lambda_8}{\sqrt2} & 0 & 0 &
0 & 0 & \lambda_7 & 0 & 0 & 0 & 0 & 0 \\
0 & \frac{i\lambda_4}{2\sqrt2} & \frac{i\lambda_8}{\sqrt2} & 0 & \frac{\lambda_4}{2\sqrt2} & \frac{\lambda_8}{\sqrt2}
& 0 & 0 & 0 & \lambda'_{14} & 0 & 0 &
\frac{i\lambda_4}{2} & 0  \\
-\frac{i\lambda_4}{2\sqrt2} & 0 & 0 & \frac{\lambda_4}{2\sqrt2} & 0 & 0 &
0 & 0 & 0 & 0 & 2\lambda_{23} & 0 &
0 & \sqrt2 i\lambda_3 \\
\frac{i\lambda_8}{\sqrt2} & 0 & 0 & \frac{\lambda_8}{\sqrt2} & 0 & 0 &
0 & 0 & 0 & 0 & 0 & \lambda_7 & 0 & 0 \\
0 & 0 & \sqrt2 \lambda_8 & 0 & 0 & -\sqrt2 i \lambda_8 &
-\frac{\lambda_4}{2} & 0 & 0 & -\frac{i\lambda_4}{2} & 0 & 0 &
\lambda_{14} & 0  \\
-\frac{\lambda_4}{2} & 0 & 0 & -\frac{i\lambda_4}{2} & 0 & 0 &
 0 & -\sqrt2 \lambda_3 & 0 & 0 & -\sqrt2 i \lambda_3 & 0 &
0 & 2\lambda_{23}
\end{matrix}\right) \,,
\end{equation}
\end{center}
and the eigenvalues are
\begin{eqnarray}
\label{eqn:values5}
\lambda_{1} \,, \quad
2\lambda_{2,\, 5,\, 6,\, 7} \,, \quad
\lambda_{1} + \lambda_4 \,, \quad
\lambda_1 - \frac12 \lambda_4 \,, \quad
2(\lambda_2 + \lambda_3) \,, \quad
\lambda_{023}^\pm \,, \quad
\lambda_{078}^\pm \,, \quad \lambda_{146}^\pm \,,
\end{eqnarray}
with
\begin{eqnarray}
\lambda_{078}^\pm \equiv
\frac14 \left[ (\lambda+2\lambda_7) \pm \sqrt{(\lambda-2\lambda_7)^2 + 32\lambda_8^2} \right] \,.
\end{eqnarray}

Finally, the sub-matrix for ($\frac{1}{\sqrt2} h^+ h^+$,  $\frac{1}{\sqrt2} \delta^+ \delta^+$, $h^+ \delta^+$, $\delta^{++} h$, $\delta^{++} \xi$, $\delta^{++} s$, $\delta^{++} Z_1$, $\delta^{++} Z_2$, $\delta^{++} Z_3$) is
\begin{equation}
{\cal M}_6 =
\left(\begin{matrix}
\frac{\lambda}{2} & 0 & 0 & 0 & 0 & \lambda_8 & 0 & 0 & i\lambda_8 \\
0 & 2\lambda'_{23} & 0 & 0 & 0 & 0 & 0 & 0 & 0 \\
0 & 0 & \lambda'_{14} & -\frac{\lambda_4}{2} & 0 & 0 &
\frac{i\lambda_4}{2} & 0 & 0 \\
0 & 0 & -\frac{\lambda_4}{2} & \lambda_1 & 0 & 0 &
0 & 0 & 0 \\
0 & 0 & 0 & 0 & 2\lambda_{2} & 0 & 0 & 0 & 0 \\
\lambda_8 & 0 & 0 & 0 & 0 & \lambda_{7} & 0 & 0 & 0 \\
0 & 0 & -\frac{i\lambda_4}{2} & 0 & 0 & 0 & \lambda_{1} & 0 & 0 \\
0 & 0 & 0 & 0 & 0 & 0 & 0 & 2\lambda_{2} & 0 \\
-i\lambda_8 & 0 & 0 & 0 & 0 & 0 & 0 & 0 & \lambda_{7} \\
\end{matrix}\right) \,,
\end{equation}
and the eigenvalues are
\begin{eqnarray}
\label{eqn:values6}
\lambda_{1,\, 7} \,, \quad
2\lambda_{2} \,, \quad
2\lambda_2 + \lambda_3 \,, \quad
\lambda_{1} + \lambda_4 \,, \quad
\lambda_1 - \frac12 \lambda_4 \,, \quad
\lambda_{078}^\pm \,.
\end{eqnarray}

To implement the unitarity bounds, we can set all the eigenvalues in Eqs.~(\ref{eqn:values1}), (\ref{eqn:values3}), (\ref{eqn:values4}), (\ref{eqn:values5}) and (\ref{eqn:values6}) to be smaller than $8\pi$. As a comparison to the perturbativity bounds, we set the quartic couplings to be the benchmark values,
\begin{eqnarray}
\lambda_{1} = 0.1 \,, \quad
\lambda_{4} = - 1 \,, \quad
\lambda_{2,\, 3,\, 5,\, 6,\, 7} = 0 \,,
\end{eqnarray}
and check the unitarity bounds on $\lambda_8$. It turns out for this specific benchmark scenario, only the following bounds are relevant to $\lambda_8$:
\begin{eqnarray}
\left| \lambda_{146}^\pm \right| \leq 8\pi \,, \quad
\left| \lambda_{078}^\pm \right| \leq 8\pi \,.
\end{eqnarray}
Among the four constraints, the most stringent one is from $\lambda_{146}^{-}$, which leads to
\begin{eqnarray}
\lambda_8 < 10.0 \, ,
\end{eqnarray}
which is much weaker than the perturbativity bound discussed in Section~\ref{sec:Limits2}.

\bibliographystyle{JHEP}
\bibliography{ref}

\providecommand{\href}[2]{#2}\begingroup\raggedright\begin{thebibliography}{100}

\bibitem{Zyla:2020zbs}
{\bf Particle Data Group} {\bf Collaboration}, P.~A. Zyla {\em et~al.}, {\it
  {Review of Particle Physics}},  {\em PTEP} {\bf 2020} (2020), no.~8 083C01.

\bibitem{Bilenky:2020wjn}
S.~M. Bilenky, {\it {Neutrinos: Majorana or Dirac?}},  {\em Universe} {\bf 6}
  (2020), no.~9 134.

\bibitem{Dev:2019qno}
P.~S.~B. Dev {\em et~al.}, {\it {Neutrino Non-Standard Interactions: A Status
  Report}},  {\em SciPost Phys. Proc.} {\bf 2} (2019) 001,
  [\href{http://www.arxiv.org/abs/1907.00991}{{\tt 1907.00991}}].

\bibitem{Lattanzi:2014mia}
M.~Lattanzi, R.~A. Lineros, and M.~Taoso, {\it {Connecting neutrino physics
  with dark matter}},  {\em New J. Phys.} {\bf 16} (2014), no.~12 125012,
  [\href{http://www.arxiv.org/abs/1406.0004}{{\tt 1406.0004}}].

\bibitem{Hagedorn:2017wjy}
C.~Hagedorn, R.~N. Mohapatra, E.~Molinaro, C.~C. Nishi, and S.~T. Petcov, {\it
  {CP Violation in the Lepton Sector and Implications for Leptogenesis}},  {\em
  Int. J. Mod. Phys. A} {\bf 33} (2018), no.~05n06 1842006,
  [\href{http://www.arxiv.org/abs/1711.02866}{{\tt 1711.02866}}].

\bibitem{Berryman:2018ogk}
J.~M. Berryman, A.~de~Gouv\^ea, K.~J. Kelly, and Y.~Zhang, {\it
  {Lepton-Number-Charged Scalars and Neutrino Beamstrahlung}},  {\em Phys. Rev.
  D} {\bf 97} (2018), no.~7 075030,
  [\href{http://www.arxiv.org/abs/1802.00009}{{\tt 1802.00009}}].

\bibitem{deGouvea:2019qaz}
A.~de~Gouv\^ea, P.~S.~B. Dev, B.~Dutta, T.~Ghosh, T.~Han, and Y.~Zhang, {\it
  {Leptonic Scalars at the LHC}},  {\em JHEP} {\bf 07} (2020) 142,
  [\href{http://www.arxiv.org/abs/1910.01132}{{\tt 1910.01132}}].

\bibitem{Kreisch:2019yzn}
C.~D. Kreisch, F.-Y. Cyr-Racine, and O.~Dor\'e, {\it {Neutrino puzzle:
  Anomalies, interactions, and cosmological tensions}},  {\em Phys. Rev. D}
  {\bf 101} (2020), no.~12 123505,
  [\href{http://www.arxiv.org/abs/1902.00534}{{\tt 1902.00534}}].

\bibitem{Blinov:2019gcj}
N.~Blinov, K.~J. Kelly, G.~Z. Krnjaic, and S.~D. McDermott, {\it {Constraining
  the Self-Interacting Neutrino Interpretation of the Hubble Tension}},  {\em
  Phys. Rev. Lett.} {\bf 123} (2019), no.~19 191102,
  [\href{http://www.arxiv.org/abs/1905.02727}{{\tt 1905.02727}}].

\bibitem{DeGouvea:2019wpf}
A.~De~Gouv\^ea, M.~Sen, W.~Tangarife, and Y.~Zhang, {\it {Dodelson-Widrow
  Mechanism in the Presence of Self-Interacting Neutrinos}},  {\em Phys. Rev.
  Lett.} {\bf 124} (2020), no.~8 081802,
  [\href{http://www.arxiv.org/abs/1910.04901}{{\tt 1910.04901}}].

\bibitem{Lyu:2020lps}
K.-F. Lyu, E.~Stamou, and L.-T. Wang, {\it {Self-interacting neutrinos:
  Solution to Hubble tension versus experimental constraints}},  {\em Phys.
  Rev. D} {\bf 103} (2021), no.~1 015004,
  [\href{http://www.arxiv.org/abs/2004.10868}{{\tt 2004.10868}}].

\bibitem{Kelly:2020aks}
K.~J. Kelly, M.~Sen, and Y.~Zhang, {\it {Intimate Relationship between Sterile
  Neutrino Dark Matter and \ensuremath{\Delta}Neff}},  {\em Phys. Rev. Lett.}
  {\bf 127} (2021), no.~4 041101,
  [\href{http://www.arxiv.org/abs/2011.02487}{{\tt 2011.02487}}].

\bibitem{Das:2020xke}
A.~Das and S.~Ghosh, {\it {Flavor-specific interaction favors strong neutrino
  self-coupling in the early universe}},  {\em JCAP} {\bf 07} (2021) 038,
  [\href{http://www.arxiv.org/abs/2011.12315}{{\tt 2011.12315}}].

\bibitem{FCC:2018vvp}
{\bf FCC} {\bf Collaboration}, A.~Abada {\em et~al.}, {\it {FCC-hh: The Hadron
  Collider}: {Future Circular Collider Conceptual Design Report Volume 3}},
  {\em Eur. Phys. J. ST} {\bf 228} (2019), no.~4 755--1107.

\bibitem{Tang:2015qga}
J.~Tang {\em et~al.}, {\it {Concept for a Future Super Proton-Proton
  Collider}},  \href{http://www.arxiv.org/abs/1507.03224}{{\tt 1507.03224}}.

\bibitem{Konetschny:1977bn}
W.~Konetschny and W.~Kummer, {\it {Nonconservation of Total Lepton Number with
  Scalar Bosons}},  {\em Phys. Lett. B} {\bf 70} (1977) 433--435.

\bibitem{Magg:1980ut}
M.~Magg and C.~Wetterich, {\it {Neutrino Mass Problem and Gauge Hierarchy}},
  {\em Phys. Lett. B} {\bf 94} (1980) 61--64.

\bibitem{Schechter:1980gr}
J.~Schechter and J.~W.~F. Valle, {\it {Neutrino Masses in SU(2) x U(1)
  Theories}},  {\em Phys. Rev. D} {\bf 22} (1980) 2227.

\bibitem{Cheng:1980qt}
T.~P. Cheng and L.-F. Li, {\it {Neutrino Masses, Mixings and Oscillations in
  SU(2) x U(1) Models of Electroweak Interactions}},  {\em Phys. Rev. D} {\bf
  22} (1980) 2860.

\bibitem{Mohapatra:1980yp}
R.~N. Mohapatra and G.~Senjanovic, {\it {Neutrino Masses and Mixings in Gauge
  Models with Spontaneous Parity Violation}},  {\em Phys. Rev. D} {\bf 23}
  (1981) 165.

\bibitem{Lazarides:1980nt}
G.~Lazarides, Q.~Shafi, and C.~Wetterich, {\it {Proton Lifetime and Fermion
  Masses in an SO(10) Model}},  {\em Nucl. Phys. B} {\bf 181} (1981) 287--300.

\bibitem{Barrie:2021mwi}
N.~D. Barrie, C.~Han, and H.~Murayama, {\it {Affleck-Dine Leptogenesis from
  Higgs Inflation}},  \href{http://www.arxiv.org/abs/2106.03381}{{\tt
  2106.03381}}.

\bibitem{Pospelov:2007mp}
M.~Pospelov, A.~Ritz, and M.~B. Voloshin, {\it {Secluded WIMP Dark Matter}},
  {\em Phys. Lett. B} {\bf 662} (2008) 53--61,
  [\href{http://www.arxiv.org/abs/0711.4866}{{\tt 0711.4866}}].

\bibitem{Kelly:2019wow}
K.~J. Kelly and Y.~Zhang, {\it {Mononeutrino at DUNE: New Signals from
  Neutrinophilic Thermal Dark Matter}},  {\em Phys. Rev. D} {\bf 99} (2019),
  no.~5 055034, [\href{http://www.arxiv.org/abs/1901.01259}{{\tt 1901.01259}}].

\bibitem{Du:2020avz}
Y.~Du, F.~Huang, H.-L. Li, and J.-H. Yu, {\it {Freeze-in Dark Matter from
  Secret Neutrino Interactions}},  {\em JHEP} {\bf 12} (2020) 207,
  [\href{http://www.arxiv.org/abs/2005.01717}{{\tt 2005.01717}}].

\bibitem{Roe:2004na}
B.~P. Roe, H.-J. Yang, J.~Zhu, Y.~Liu, I.~Stancu, and G.~McGregor, {\it
  {Boosted decision trees, an alternative to artificial neural networks}},
  {\em Nucl. Instrum. Meth. A} {\bf 543} (2005), no.~2-3 577--584,
  [\href{http://www.arxiv.org/abs/physics/0408124}{{\tt physics/0408124}}].

\bibitem{Planck:2018vyg}
{\bf Planck} {\bf Collaboration}, N.~Aghanim {\em et~al.}, {\it {Planck 2018
  results. VI. Cosmological parameters}},  {\em Astron. Astrophys.} {\bf 641}
  (2020) A6, [\href{http://www.arxiv.org/abs/1807.06209}{{\tt 1807.06209}}].

\bibitem{KATRIN:2019yun}
{\bf KATRIN} {\bf Collaboration}, M.~Aker {\em et~al.}, {\it {Improved Upper
  Limit on the Neutrino Mass from a Direct Kinematic Method by KATRIN}},  {\em
  Phys. Rev. Lett.} {\bf 123} (2019), no.~22 221802,
  [\href{http://www.arxiv.org/abs/1909.06048}{{\tt 1909.06048}}].

\bibitem{ATLAS:2012yve}
{\bf ATLAS} {\bf Collaboration}, G.~Aad {\em et~al.}, {\it {Observation of a
  new particle in the search for the Standard Model Higgs boson with the ATLAS
  detector at the LHC}},  {\em Phys. Lett. B} {\bf 716} (2012) 1--29,
  [\href{http://www.arxiv.org/abs/1207.7214}{{\tt 1207.7214}}].

\bibitem{CMS:2012qbp}
{\bf CMS} {\bf Collaboration}, S.~Chatrchyan {\em et~al.}, {\it {Observation of
  a New Boson at a Mass of 125 GeV with the CMS Experiment at the LHC}},  {\em
  Phys. Lett. B} {\bf 716} (2012) 30--61,
  [\href{http://www.arxiv.org/abs/1207.7235}{{\tt 1207.7235}}].

\bibitem{Chakrabarti:1998qy}
S.~Chakrabarti, D.~Choudhury, R.~M. Godbole, and B.~Mukhopadhyaya, {\it
  {Observing doubly charged Higgs bosons in photon-photon collisions}},  {\em
  Phys. Lett. B} {\bf 434} (1998) 347--353,
  [\href{http://www.arxiv.org/abs/hep-ph/9804297}{{\tt hep-ph/9804297}}].

\bibitem{Chun:2003ej}
E.~J. Chun, K.~Y. Lee, and S.~C. Park, {\it {Testing Higgs triplet model and
  neutrino mass patterns}},  {\em Phys. Lett. B} {\bf 566} (2003) 142--151,
  [\href{http://www.arxiv.org/abs/hep-ph/0304069}{{\tt hep-ph/0304069}}].

\bibitem{Akeroyd:2005gt}
A.~G. Akeroyd and M.~Aoki, {\it {Single and pair production of doubly charged
  Higgs bosons at hadron colliders}},  {\em Phys. Rev. D} {\bf 72} (2005)
  035011, [\href{http://www.arxiv.org/abs/hep-ph/0506176}{{\tt
  hep-ph/0506176}}].

\bibitem{FileviezPerez:2008jbu}
P.~Fileviez~Perez, T.~Han, G.-y. Huang, T.~Li, and K.~Wang, {\it {Neutrino
  Masses and the CERN LHC: Testing Type II Seesaw}},  {\em Phys. Rev. D} {\bf
  78} (2008) 015018, [\href{http://www.arxiv.org/abs/0805.3536}{{\tt
  0805.3536}}].

\bibitem{delAguila:2008cj}
F.~del Aguila and J.~A. Aguilar-Saavedra, {\it {Distinguishing seesaw models at
  LHC with multi-lepton signals}},  {\em Nucl. Phys. B} {\bf 813} (2009)
  22--90, [\href{http://www.arxiv.org/abs/0808.2468}{{\tt 0808.2468}}].

\bibitem{Akeroyd:2011zza}
A.~G. Akeroyd and H.~Sugiyama, {\it {Production of doubly charged scalars from
  the decay of singly charged scalars in the Higgs Triplet Model}},  {\em Phys.
  Rev. D} {\bf 84} (2011) 035010,
  [\href{http://www.arxiv.org/abs/1105.2209}{{\tt 1105.2209}}].

\bibitem{Melfo:2011nx}
A.~Melfo, M.~Nemevsek, F.~Nesti, G.~Senjanovic, and Y.~Zhang, {\it {Type II
  Seesaw at LHC: The Roadmap}},  {\em Phys. Rev. D} {\bf 85} (2012) 055018,
  [\href{http://www.arxiv.org/abs/1108.4416}{{\tt 1108.4416}}].

\bibitem{Aoki:2011pz}
M.~Aoki, S.~Kanemura, and K.~Yagyu, {\it {Testing the Higgs triplet model with
  the mass difference at the LHC}},  {\em Phys. Rev. D} {\bf 85} (2012) 055007,
  [\href{http://www.arxiv.org/abs/1110.4625}{{\tt 1110.4625}}].

\bibitem{Chiang:2012dk}
C.-W. Chiang, T.~Nomura, and K.~Tsumura, {\it {Search for doubly charged Higgs
  bosons using the same-sign diboson mode at the LHC}},  {\em Phys. Rev. D}
  {\bf 85} (2012) 095023, [\href{http://www.arxiv.org/abs/1202.2014}{{\tt
  1202.2014}}].

\bibitem{Han:2015hba}
Z.-L. Han, R.~Ding, and Y.~Liao, {\it {LHC Phenomenology of Type II Seesaw:
  Nondegenerate Case}},  {\em Phys. Rev. D} {\bf 91} (2015) 093006,
  [\href{http://www.arxiv.org/abs/1502.05242}{{\tt 1502.05242}}].

\bibitem{Babu:2016rcr}
K.~S. Babu and S.~Jana, {\it {Probing Doubly Charged Higgs Bosons at the LHC
  through Photon Initiated Processes}},  {\em Phys. Rev. D} {\bf 95} (2017),
  no.~5 055020, [\href{http://www.arxiv.org/abs/1612.09224}{{\tt 1612.09224}}].

\bibitem{Ghosh:2017pxl}
D.~K. Ghosh, N.~Ghosh, I.~Saha, and A.~Shaw, {\it {Revisiting the high-scale
  validity of the type II seesaw model with novel LHC signature}},  {\em Phys.
  Rev. D} {\bf 97} (2018), no.~11 115022,
  [\href{http://www.arxiv.org/abs/1711.06062}{{\tt 1711.06062}}].

\bibitem{Dev:2018kpa}
P.~S.~B. Dev and Y.~Zhang, {\it {Displaced vertex signatures of doubly charged
  scalars in the type-II seesaw and its left-right extensions}},  {\em JHEP}
  {\bf 10} (2018) 199, [\href{http://www.arxiv.org/abs/1808.00943}{{\tt
  1808.00943}}].

\bibitem{Du:2018eaw}
Y.~Du, A.~Dunbrack, M.~J. Ramsey-Musolf, and J.-H. Yu, {\it {Type-II Seesaw
  Scalar Triplet Model at a 100 TeV $pp$ Collider: Discovery and Higgs Portal
  Coupling Determination}},  {\em JHEP} {\bf 01} (2019) 101,
  [\href{http://www.arxiv.org/abs/1810.09450}{{\tt 1810.09450}}].

\bibitem{Antusch:2018svb}
S.~Antusch, O.~Fischer, A.~Hammad, and C.~Scherb, {\it {Low scale type II
  seesaw: Present constraints and prospects for displaced vertex searches}},
  {\em JHEP} {\bf 02} (2019) 157,
  [\href{http://www.arxiv.org/abs/1811.03476}{{\tt 1811.03476}}].

\bibitem{Primulando:2019evb}
R.~Primulando, J.~Julio, and P.~Uttayarat, {\it {Scalar phenomenology in
  type-II seesaw model}},  {\em JHEP} {\bf 08} (2019) 024,
  [\href{http://www.arxiv.org/abs/1903.02493}{{\tt 1903.02493}}].

\bibitem{deMelo:2019asm}
T.~B. de~Melo, F.~S. Queiroz, and Y.~Villamizar, {\it {Doubly Charged Scalar at
  the High-Luminosity and High-Energy LHC}},  {\em Int. J. Mod. Phys. A} {\bf
  34} (2019), no.~27 1950157, [\href{http://www.arxiv.org/abs/1909.07429}{{\tt
  1909.07429}}].

\bibitem{Padhan:2019jlc}
R.~Padhan, D.~Das, M.~Mitra, and A.~Kumar~Nayak, {\it {Probing doubly and
  singly charged Higgs bosons at the $pp$ collider HE-LHC}},  {\em Phys. Rev.
  D} {\bf 101} (2020), no.~7 075050,
  [\href{http://www.arxiv.org/abs/1909.10495}{{\tt 1909.10495}}].

\bibitem{Ashanujjaman:2021txz}
S.~Ashanujjaman and K.~Ghosh, {\it {Revisiting Type-II see-saw: Present Limits
  and Future Prospects at LHC}},
  \href{http://www.arxiv.org/abs/2108.10952}{{\tt 2108.10952}}.

\bibitem{Peskin:1990zt}
M.~E. Peskin and T.~Takeuchi, {\it {A New constraint on a strongly interacting
  Higgs sector}},  {\em Phys. Rev. Lett.} {\bf 65} (1990) 964--967.

\bibitem{Peskin:1991sw}
M.~E. Peskin and T.~Takeuchi, {\it {Estimation of oblique electroweak
  corrections}},  {\em Phys. Rev. D} {\bf 46} (1992) 381--409.

\bibitem{Kanemura:2012rs}
S.~Kanemura and K.~Yagyu, {\it {Radiative corrections to electroweak parameters
  in the Higgs triplet model and implication with the recent Higgs boson
  searches}},  {\em Phys. Rev. D} {\bf 85} (2012) 115009,
  [\href{http://www.arxiv.org/abs/1201.6287}{{\tt 1201.6287}}].

\bibitem{Chun:2012jw}
E.~J. Chun, H.~M. Lee, and P.~Sharma, {\it {Vacuum Stability, Perturbativity,
  EWPD and Higgs-to-diphoton rate in Type II Seesaw Models}},  {\em JHEP} {\bf
  11} (2012) 106, [\href{http://www.arxiv.org/abs/1209.1303}{{\tt 1209.1303}}].

\bibitem{Aaboud:2017qph}
{\bf ATLAS} {\bf Collaboration}, M.~Aaboud {\em et~al.}, {\it {Search for
  doubly charged Higgs boson production in multi-lepton final states with the
  ATLAS detector using proton\textendash{}proton collisions at
  $\sqrt{s}=13\,\text {TeV}$}},  {\em Eur. Phys. J. C} {\bf 78} (2018), no.~3
  199, [\href{http://www.arxiv.org/abs/1710.09748}{{\tt 1710.09748}}].

\bibitem{CMS:2017pet}
{\bf CMS} {\bf Collaboration}, {\it {A search for doubly-charged Higgs boson
  production in three and four lepton final states at
  $\sqrt{s}=13~\mathrm{TeV}$}}, .

\bibitem{ATLAS:2018ceg}
{\bf ATLAS} {\bf Collaboration}, M.~Aaboud {\em et~al.}, {\it {Search for
  doubly charged scalar bosons decaying into same-sign $W$ boson pairs with the
  ATLAS detector}},  {\em Eur. Phys. J. C} {\bf 79} (2019), no.~1 58,
  [\href{http://www.arxiv.org/abs/1808.01899}{{\tt 1808.01899}}].

\bibitem{Aad:2021lzu}
{\bf ATLAS} {\bf Collaboration}, G.~Aad {\em et~al.}, {\it {Search for doubly
  and singly charged Higgs bosons decaying into vector bosons in multi-lepton
  final states with the ATLAS detector using proton-proton collisions at
  $\sqrt{s}$ = 13 TeV}},  \href{http://www.arxiv.org/abs/2101.11961}{{\tt
  2101.11961}}.

\bibitem{Amhis:2016xyh}
{\bf HFLAV} {\bf Collaboration}, Y.~Amhis {\em et~al.}, {\it {Averages of
  $b$-hadron, $c$-hadron, and $\tau$-lepton properties as of summer 2016}},
  {\em Eur. Phys. J. C} {\bf 77} (2017), no.~12 895,
  [\href{http://www.arxiv.org/abs/1612.07233}{{\tt 1612.07233}}].

\bibitem{Hanneke:2008tm}
D.~Hanneke, S.~Fogwell, and G.~Gabrielse, {\it {New Measurement of the Electron
  Magnetic Moment and the Fine Structure Constant}},  {\em Phys. Rev. Lett.}
  {\bf 100} (2008) 120801, [\href{http://www.arxiv.org/abs/0801.1134}{{\tt
  0801.1134}}].

\bibitem{Bennett:2006fi}
{\bf Muon g-2} {\bf Collaboration}, G.~W. Bennett {\em et~al.}, {\it {Final
  Report of the Muon E821 Anomalous Magnetic Moment Measurement at BNL}},  {\em
  Phys. Rev. D} {\bf 73} (2006) 072003,
  [\href{http://www.arxiv.org/abs/hep-ex/0602035}{{\tt hep-ex/0602035}}].

\bibitem{Muong-2:2021ojo}
{\bf Muon g-2} {\bf Collaboration}, B.~Abi {\em et~al.}, {\it {Measurement of
  the Positive Muon Anomalous Magnetic Moment to 0.46 ppm}},  {\em Phys. Rev.
  Lett.} {\bf 126} (2021), no.~14 141801,
  [\href{http://www.arxiv.org/abs/2104.03281}{{\tt 2104.03281}}].

\bibitem{Willmann:1998gd}
L.~Willmann {\em et~al.}, {\it {New bounds from searching for muonium to
  anti-muonium conversion}},  {\em Phys. Rev. Lett.} {\bf 82} (1999) 49--52,
  [\href{http://www.arxiv.org/abs/hep-ex/9807011}{{\tt hep-ex/9807011}}].

\bibitem{Abdallah:2005ph}
{\bf DELPHI} {\bf Collaboration}, J.~Abdallah {\em et~al.}, {\it {Measurement
  and interpretation of fermion-pair production at LEP energies above the Z
  resonance}},  {\em Eur. Phys. J. C} {\bf 45} (2006) 589--632,
  [\href{http://www.arxiv.org/abs/hep-ex/0512012}{{\tt hep-ex/0512012}}].

\bibitem{Lindner:2016bgg}
M.~Lindner, M.~Platscher, and F.~S. Queiroz, {\it {A Call for New Physics : The
  Muon Anomalous Magnetic Moment and Lepton Flavor Violation}},  {\em Phys.
  Rept.} {\bf 731} (2018) 1--82,
  [\href{http://www.arxiv.org/abs/1610.06587}{{\tt 1610.06587}}].

\bibitem{Aoyama:2020ynm}
T.~Aoyama {\em et~al.}, {\it {The anomalous magnetic moment of the muon in the
  Standard Model}},  {\em Phys. Rept.} {\bf 887} (2020) 1--166,
  [\href{http://www.arxiv.org/abs/2006.04822}{{\tt 2006.04822}}].

\bibitem{Dev:2017ftk}
P.~S.~B. Dev, R.~N. Mohapatra, and Y.~Zhang, {\it {Lepton Flavor Violation
  Induced by a Neutral Scalar at Future Lepton Colliders}},  {\em Phys. Rev.
  Lett.} {\bf 120} (2018), no.~22 221804,
  [\href{http://www.arxiv.org/abs/1711.08430}{{\tt 1711.08430}}].

\bibitem{BhupalDev:2018vpr}
P.~S. Bhupal~Dev, R.~N. Mohapatra, and Y.~Zhang, {\it {Probing TeV scale origin
  of neutrino mass at future lepton colliders via neutral and doubly-charged
  scalars}},  {\em Phys. Rev. D} {\bf 98} (2018), no.~7 075028,
  [\href{http://www.arxiv.org/abs/1803.11167}{{\tt 1803.11167}}].

\bibitem{Li:2018cod}
T.~Li and M.~A. Schmidt, {\it {Sensitivity of future lepton colliders to the
  search for charged lepton flavor violation}},  {\em Phys. Rev. D} {\bf 99}
  (2019), no.~5 055038, [\href{http://www.arxiv.org/abs/1809.07924}{{\tt
  1809.07924}}].

\bibitem{Evans:2019xer}
J.~A. Evans, P.~Tanedo, and M.~Zakeri, {\it {Exotic Lepton-Flavor Violating
  Higgs Decays}},  {\em JHEP} {\bf 01} (2020) 028,
  [\href{http://www.arxiv.org/abs/1910.07533}{{\tt 1910.07533}}].

\bibitem{Iguro:2020rby}
S.~Iguro, Y.~Omura, and M.~Takeuchi, {\it {Probing $\mu\tau$ flavor-violating
  solutions for the muon $g-2$ anomaly at Belle II}},  {\em JHEP} {\bf 09}
  (2020) 144, [\href{http://www.arxiv.org/abs/2002.12728}{{\tt 2002.12728}}].

\bibitem{Li:2021lnz}
T.~Li, M.~A. Schmidt, C.-Y. Yao, and M.~Yuan, {\it {Charged lepton flavor
  violation in light of the muon magnetic moment anomaly and colliders}},
  \href{http://www.arxiv.org/abs/2104.04494}{{\tt 2104.04494}}.

\bibitem{Hou:2021qmf}
W.-S. Hou and G.~Kumar, {\it {Charged lepton flavor violation in light of Muon
  $g-2$}},  \href{http://www.arxiv.org/abs/2107.14114}{{\tt 2107.14114}}.

\bibitem{Capdevilla:2020qel}
R.~Capdevilla, D.~Curtin, Y.~Kahn, and G.~Krnjaic, {\it {Discovering the
  physics of $(g-2)_\mu$ at future muon colliders}},  {\em Phys. Rev. D} {\bf
  103} (2021), no.~7 075028, [\href{http://www.arxiv.org/abs/2006.16277}{{\tt
  2006.16277}}].

\bibitem{Buttazzo:2020eyl}
D.~Buttazzo and P.~Paradisi, {\it {Probing the muon g-2 anomaly at a Muon
  Collider}},  \href{http://www.arxiv.org/abs/2012.02769}{{\tt 2012.02769}}.

\bibitem{Yin:2020afe}
W.~Yin and M.~Yamaguchi, {\it {Muon $g-2$ at multi-TeV muon collider}},
  \href{http://www.arxiv.org/abs/2012.03928}{{\tt 2012.03928}}.

\bibitem{Capdevilla:2021rwo}
R.~Capdevilla, D.~Curtin, Y.~Kahn, and G.~Krnjaic, {\it {A No-Lose Theorem for
  Discovering the New Physics of $(g-2)_\mu$ at Muon Colliders}},
  \href{http://www.arxiv.org/abs/2101.10334}{{\tt 2101.10334}}.

\bibitem{Haghighat:2021djz}
G.~Haghighat and M.~Mohammadi~Najafabadi, {\it {Search for
  lepton-flavor-violating ALPs at a future muon collider and utilization of
  polarization-induced effects}},
  \href{http://www.arxiv.org/abs/2106.00505}{{\tt 2106.00505}}.

\bibitem{Perez:1992hc}
M.~A. Perez and M.~A. Soriano, {\it {Flavor changing decays of the Z and
  Z-prime gauge bosons in left-right symmetric models}},  {\em Phys. Rev. D}
  {\bf 46} (1992) 284--289.

\bibitem{Nemevsek:2016enw}
M.~Nemev\v{s}ek, F.~Nesti, and J.~C. Vasquez, {\it {Majorana Higgses at
  colliders}},  {\em JHEP} {\bf 04} (2017) 114,
  [\href{http://www.arxiv.org/abs/1612.06840}{{\tt 1612.06840}}].

\bibitem{Fusaoka:1998vc}
H.~Fusaoka and Y.~Koide, {\it {Updated estimate of running quark masses}},
  {\em Phys. Rev. D} {\bf 57} (1998) 3986--4001,
  [\href{http://www.arxiv.org/abs/hep-ph/9712201}{{\tt hep-ph/9712201}}].

\bibitem{Xing:2007fb}
Z.-z. Xing, H.~Zhang, and S.~Zhou, {\it {Updated Values of Running Quark and
  Lepton Masses}},  {\em Phys. Rev. D} {\bf 77} (2008) 113016,
  [\href{http://www.arxiv.org/abs/0712.1419}{{\tt 0712.1419}}].

\bibitem{Xing:2011aa}
Z.-z. Xing, H.~Zhang, and S.~Zhou, {\it {Impacts of the Higgs mass on vacuum
  stability, running fermion masses and two-body Higgs decays}},  {\em Phys.
  Rev. D} {\bf 86} (2012) 013013,
  [\href{http://www.arxiv.org/abs/1112.3112}{{\tt 1112.3112}}].

\bibitem{Antusch:2013jca}
S.~Antusch and V.~Maurer, {\it {Running quark and lepton parameters at various
  scales}},  {\em JHEP} {\bf 11} (2013) 115,
  [\href{http://www.arxiv.org/abs/1306.6879}{{\tt 1306.6879}}].

\bibitem{Huang:2020hdv}
G.-y. Huang and S.~Zhou, {\it {Precise Values of Running Quark and Lepton
  Masses in the Standard Model}},  {\em Phys. Rev. D} {\bf 103} (2021), no.~1
  016010, [\href{http://www.arxiv.org/abs/2009.04851}{{\tt 2009.04851}}].

\bibitem{Arkani-Hamed:2015vfh}
N.~Arkani-Hamed, T.~Han, M.~Mangano, and L.-T. Wang, {\it {Physics
  opportunities of a 100 TeV proton\textendash{}proton collider}},  {\em Phys.
  Rept.} {\bf 652} (2016) 1--49,
  [\href{http://www.arxiv.org/abs/1511.06495}{{\tt 1511.06495}}].

\bibitem{Alloul:2013bka}
A.~Alloul, N.~D. Christensen, C.~Degrande, C.~Duhr, and B.~Fuks, {\it
  {FeynRules 2.0 - A complete toolbox for tree-level phenomenology}},  {\em
  Comput. Phys. Commun.} {\bf 185} (2014) 2250--2300,
  [\href{http://www.arxiv.org/abs/1310.1921}{{\tt 1310.1921}}].

\bibitem{Alwall:2014hca}
J.~Alwall, R.~Frederix, S.~Frixione, V.~Hirschi, F.~Maltoni, O.~Mattelaer,
  H.~S. Shao, T.~Stelzer, P.~Torrielli, and M.~Zaro, {\it {The automated
  computation of tree-level and next-to-leading order differential cross
  sections, and their matching to parton shower simulations}},  {\em JHEP} {\bf
  07} (2014) 079, [\href{http://www.arxiv.org/abs/1405.0301}{{\tt 1405.0301}}].

\bibitem{Artoisenet:2012st}
P.~Artoisenet, R.~Frederix, O.~Mattelaer, and R.~Rietkerk, {\it {Automatic
  spin-entangled decays of heavy resonances in Monte Carlo simulations}},  {\em
  JHEP} {\bf 03} (2013) 015, [\href{http://www.arxiv.org/abs/1212.3460}{{\tt
  1212.3460}}].

\bibitem{Muhlleitner:2003me}
M.~Muhlleitner and M.~Spira, {\it {A Note on doubly charged Higgs pair
  production at hadron colliders}},  {\em Phys. Rev. D} {\bf 68} (2003) 117701,
  [\href{http://www.arxiv.org/abs/hep-ph/0305288}{{\tt hep-ph/0305288}}].

\bibitem{Sjostrand:2007gs}
T.~Sjostrand, S.~Mrenna, and P.~Z. Skands, {\it {A Brief Introduction to PYTHIA
  8.1}},  {\em Comput. Phys. Commun.} {\bf 178} (2008) 852--867,
  [\href{http://www.arxiv.org/abs/0710.3820}{{\tt 0710.3820}}].

\bibitem{Cacciari:2011ma}
M.~Cacciari, G.~P. Salam, and G.~Soyez, {\it {FastJet User Manual}},  {\em Eur.
  Phys. J. C} {\bf 72} (2012) 1896,
  [\href{http://www.arxiv.org/abs/1111.6097}{{\tt 1111.6097}}].

\bibitem{Cacciari:2008gp}
M.~Cacciari, G.~P. Salam, and G.~Soyez, {\it {The anti-$k_t$ jet clustering
  algorithm}},  {\em JHEP} {\bf 04} (2008) 063,
  [\href{http://www.arxiv.org/abs/0802.1189}{{\tt 0802.1189}}].

\bibitem{deFavereau:2013fsa}
{\bf DELPHES 3} {\bf Collaboration}, J.~de~Favereau, C.~Delaere, P.~Demin,
  A.~Giammanco, V.~Lema\^\i{}tre, A.~Mertens, and M.~Selvaggi, {\it {DELPHES 3,
  A modular framework for fast simulation of a generic collider experiment}},
  {\em JHEP} {\bf 02} (2014) 057,
  [\href{http://www.arxiv.org/abs/1307.6346}{{\tt 1307.6346}}].

\bibitem{Barger:1987re}
V.~D. Barger, T.~Han, and J.~Ohnemus, {\it {HEAVY LEPTONS AT HADRON
  SUPERCOLLIDERS}},  {\em Phys. Rev. D} {\bf 37} (1988) 1174.

\bibitem{ATLAS:2020srl}
{\bf ATLAS} {\bf Collaboration}, {\it {Search for squarks and gluinos in final
  states with an isolated lepton, jets, and missing transverse momentum at
  $\sqrt{s}=13$ TeV with the ATLAS detector}}, .

\bibitem{ATLAS:2020ghe}
{\bf ATLAS} {\bf Collaboration}, {\it {Search for new phenomena in final states
  with large jet multiplicities and missing transverse momentum using
  $\sqrt{s}=13~\mathrm{TeV}$ proton-proton collisions recorded by ATLAS in Run
  2 of the LHC}}, .

\bibitem{Chen:2016btl}
T.~Chen and C.~Guestrin, {\it {XGBoost: A Scalable Tree Boosting System}},
  \href{http://www.arxiv.org/abs/1603.02754}{{\tt 1603.02754}}.

\bibitem{OPAL:2001luy}
{\bf OPAL} {\bf Collaboration}, G.~Abbiendi {\em et~al.}, {\it {Search for
  doubly charged Higgs bosons with the OPAL detector at LEP}},  {\em Phys.
  Lett. B} {\bf 526} (2002) 221--232,
  [\href{http://www.arxiv.org/abs/hep-ex/0111059}{{\tt hep-ex/0111059}}].

\bibitem{DELPHI:2002bkf}
{\bf DELPHI} {\bf Collaboration}, J.~Abdallah {\em et~al.}, {\it {Search for
  doubly charged Higgs bosons at LEP-2}},  {\em Phys. Lett. B} {\bf 552} (2003)
  127--137, [\href{http://www.arxiv.org/abs/hep-ex/0303026}{{\tt
  hep-ex/0303026}}].

\bibitem{L3:2003zst}
{\bf L3} {\bf Collaboration}, P.~Achard {\em et~al.}, {\it {Search for doubly
  charged Higgs bosons at LEP}},  {\em Phys. Lett. B} {\bf 576} (2003) 18--28,
  [\href{http://www.arxiv.org/abs/hep-ex/0309076}{{\tt hep-ex/0309076}}].

\bibitem{CDF:2004teg}
{\bf CDF} {\bf Collaboration}, D.~Acosta {\em et~al.}, {\it {Search for
  doubly-charged Higgs bosons decaying to dileptons in $p\bar{p}$ collisions at
  $\sqrt{s} = 1.96$ TeV}},  {\em Phys. Rev. Lett.} {\bf 93} (2004) 221802,
  [\href{http://www.arxiv.org/abs/hep-ex/0406073}{{\tt hep-ex/0406073}}].

\bibitem{CDF:2008vdv}
{\bf CDF} {\bf Collaboration}, T.~Aaltonen {\em et~al.}, {\it {Search for
  Doubly Charged Higgs Bosons with Lepton-Flavor-Violating Decays involving Tau
  Leptons}},  {\em Phys. Rev. Lett.} {\bf 101} (2008) 121801,
  [\href{http://www.arxiv.org/abs/0808.2161}{{\tt 0808.2161}}].

\bibitem{D0:2008qnv}
{\bf D0} {\bf Collaboration}, V.~M. Abazov {\em et~al.}, {\it {Search for pair
  production of doubly-charged Higgs bosons in the $H^{++} H^{--} \to \mu^{+}
  \mu^{+} \mu^{-} \mu^{-}$ final state at D0}},  {\em Phys. Rev. Lett.} {\bf
  101} (2008) 071803, [\href{http://www.arxiv.org/abs/0803.1534}{{\tt
  0803.1534}}].

\bibitem{D0:2011eug}
{\bf D0} {\bf Collaboration}, V.~M. Abazov {\em et~al.}, {\it {Search for
  doubly-charged Higgs boson pair production in $p\bar {p}$ collisions at
  $\sqrt{s} = 1.96$ TeV}},  {\em Phys. Rev. Lett.} {\bf 108} (2012) 021801,
  [\href{http://www.arxiv.org/abs/1106.4250}{{\tt 1106.4250}}].

\bibitem{ATLAS:2011rha}
{\bf ATLAS} {\bf Collaboration}, {\it {Search for Doubly Charged Higgs Boson
  Production in Like-sign Muon Pairs in pp Collisions at
  \ensuremath{\sqrt{}}s=7 TeV}}, .

\bibitem{CMS:2011sqa}
{\bf CMS} {\bf Collaboration}, {\it {Inclusive search for doubly charged higgs
  in leptonic final states at sqrt s=7 TeV}}, .

\bibitem{ATLAS:2014kca}
{\bf ATLAS} {\bf Collaboration}, G.~Aad {\em et~al.}, {\it {Search for
  anomalous production of prompt same-sign lepton pairs and pair-produced
  doubly charged Higgs bosons with $ \sqrt{s}=8 $ TeV $pp$ collisions using the
  ATLAS detector}},  {\em JHEP} {\bf 03} (2015) 041,
  [\href{http://www.arxiv.org/abs/1412.0237}{{\tt 1412.0237}}].

\bibitem{CMS:2016cpz}
{\bf CMS} {\bf Collaboration}, {\it {Search for a doubly-charged Higgs boson
  with $\sqrt{s}=8~\mathrm{TeV}$ $pp$ collisions at the CMS experiment}}, .

\bibitem{Gripaios:2007is}
B.~Gripaios, {\it {Transverse observables and mass determination at hadron
  colliders}},  {\em JHEP} {\bf 02} (2008) 053,
  [\href{http://www.arxiv.org/abs/0709.2740}{{\tt 0709.2740}}].

\bibitem{Barr:2007hy}
A.~J. Barr, B.~Gripaios, and C.~G. Lester, {\it {Weighing Wimps with Kinks at
  Colliders: Invisible Particle Mass Measurements from Endpoints}},  {\em JHEP}
  {\bf 02} (2008) 014, [\href{http://www.arxiv.org/abs/0711.4008}{{\tt
  0711.4008}}].

\bibitem{Curtin:2011ng}
D.~Curtin, {\it {Mixing It Up With MT2: Unbiased Mass Measurements at Hadron
  Colliders}},  {\em Phys. Rev. D} {\bf 85} (2012) 075004,
  [\href{http://www.arxiv.org/abs/1112.1095}{{\tt 1112.1095}}].

\bibitem{Lyonnet:2013dna}
F.~Lyonnet, I.~Schienbein, F.~Staub, and A.~Wingerter, {\it {PyR@TE:
  Renormalization Group Equations for General Gauge Theories}},  {\em Comput.
  Phys. Commun.} {\bf 185} (2014) 1130--1152,
  [\href{http://www.arxiv.org/abs/1309.7030}{{\tt 1309.7030}}].

\bibitem{Sartore:2020gou}
L.~Sartore and I.~Schienbein, {\it {PyR@TE 3}},  {\em Comput. Phys. Commun.}
  {\bf 261} (2021) 107819, [\href{http://www.arxiv.org/abs/2007.12700}{{\tt
  2007.12700}}].

\bibitem{Arason:1991ic}
H.~Arason, D.~J. Castano, B.~Keszthelyi, S.~Mikaelian, E.~J. Piard, P.~Ramond,
  and B.~D. Wright, {\it {Renormalization group study of the standard model and
  its extensions. 1. The Standard model}},  {\em Phys. Rev. D} {\bf 46} (1992)
  3945--3965.

\bibitem{Arhrib:2011uy}
A.~Arhrib, R.~Benbrik, M.~Chabab, G.~Moultaka, M.~C. Peyranere, L.~Rahili, and
  J.~Ramadan, {\it {The Higgs Potential in the Type II Seesaw Model}},  {\em
  Phys. Rev.} {\bf D84} (2011) 095005,
  [\href{http://www.arxiv.org/abs/1105.1925}{{\tt 1105.1925}}].

\end{thebibliography}\endgroup

\end{document}